\documentclass[ aps,  twocolumn, nofootinbib, showkeys, notitlepage, superscriptaddress]{revtex4-1}

\usepackage{graphicx}
\usepackage{dcolumn}
\usepackage{bm}
\usepackage{amsmath}
\usepackage{float}
\usepackage{multirow}
\usepackage{slashed}
\usepackage{xcolor}
\usepackage{physics}
\usepackage{multirow}
\usepackage{gensymb}
\usepackage[colorlinks=true, pdfstartview=FitV, bookmarks=true, bookmarksnumbered=true, breaklinks]{hyperref}
\usepackage{mathtools,braket}

\usepackage{lipsum}  
\usepackage{color}
\definecolor{blue}{rgb}{0.0, 0.0, 1.0}
\definecolor{red}{rgb}{1.0, 0.0, 0.0}
\definecolor{royalblue}{rgb}{0.0, 0.14, 0.4}
\hypersetup{linkcolor=red, citecolor=blue, urlcolor=blue}

\def\orcid#1{\kern .08em\href{https://orcid.org/#1}{\includegraphics[keepaspectratio,width=0.7em]{ORCID_iD.png}}}
\usepackage{calligra}
\DeclareMathAlphabet{\mathcalligra}{T1}{calligra}{m}{n}

\def\la{\langle}
\def\ra{\rangle}

\def\al{\alpha}

\def\be{\begin{equation}}
\def\ee{\end{equation}}
\def\bea{\begin{eqnarray}}
\def\eea{\end{eqnarray}}

\def\la{\langle}
\def\ra{\rangle}

\def\al{\alpha}

\def\be{\begin{equation}}
\def\ee{\end{equation}}
\def\bea{\begin{eqnarray}}
\def\eea{\end{eqnarray}}
\usepackage{amsfonts}
\usepackage{amssymb}
\usepackage{amsmath}
\usepackage{graphicx,color}
\usepackage{dcolumn}
\usepackage{epsfig}
\usepackage{bm}
\usepackage{bbm}
\usepackage{ulem}
\usepackage{subcaption}
\usepackage{ulem,slashed,tensor,braket}
\usepackage{tabularx}

\usepackage{braket}

\usepackage{color}

\begin{document}
\preprint{KNU-ANL-02/2024}
\title{\Large Dynamical Model of $J/\psi$ photo-production on the nucleon}
\author{S. Sakinah}
\email{ssakinahf@knu.ac.kr}
\affiliation{Department of  Physics, Kyungpook National University, Daegu 41566, South Korea}
\author{T.-S. H. Lee}
\email{tshlee@anl.gov}
\affiliation{Physics Division, Argonne National Laboratory, Argonne, Illinois 60439, USA}
\author{H. M. Choi}
\email{homyoung@knu.ac.kr}
\affiliation{Department of  Physics, Kyungpook National University, Daegu 41566, South Korea}

\date{\today}
\begin{abstract}
A dynamical model based on  a phenomenological charm quark-nucleon ($c$-N) potential $v_{cN}$ and the
Pomeron-exchange mechanism is constructed to investigate the  $J/\psi$ photo-production  on the nucleon
from threshold to invariant mass $W=300$ GeV.  The $J/\psi$-N potential, $V_{J/\psi N}(r)$, is constructed  by folding $v_{cN}$ 
into the  wavefunction $\Phi_{J/\psi}(c\bar{c})$ of $J/\psi$ within a Constituent Quark Model (CQM) of  Ref.~\cite{SEFH13}.
A photo-production amplitude is also generated  by  $v_{cN}$ by a $c\bar{c}$-loop integration over the $\gamma\rightarrow c\bar{c}$
vertex function and  $\Phi_{J/\psi}(c\bar{c})$. No  commonly used Vector Meson Dominance assumption is used to define
this photo-production amplitude which is needed to describe the data near the threshold.
The $c$-N potential $v_{cN}(r)$ is parameterized in a form such that the predicted $V_{J/\psi N}(r)$ at large distances has the same
Yukawa potential form extracted from a Lattice QCD (LQCD) calculation of Ref.~\cite{KS10b} . 
The  parameters of  $v_{cN}$  are determined by fitting the total cross section data  of JLab by performing 
calculations  that include $J/\psi$-N  final state interactions (FSI). The resulting differential cross sections $d\sigma/dt$ are found  in 
good agreements with  the data. It is shown that the FSI effects dominate the cross section in the very near-threshold region, allowing for sensitive 
testing of the predicted $J/\psi$-N scattering amplitudes. By imposing the constraints of $J/\psi$-N potential extracted from the LQCD  calculation of Ref.~\cite{KS10b},
we have obtained  three $J/\psi$-N potentials which fit the JLab  data  equally well.
The resulting $J/\psi$-N scattering lengths are in the range of $a=(-0.05$ fm $\sim$ $-0.25$ fm). 
With the determined $v_{cN}(r)$ and the wavefunctions generated from the same CQM, the constructed model is used to  predict
the cross  sections  of photo-production of $\eta_c(1S)$  and $\psi(2S)$ mesons for future  experimental tests.
\end{abstract}
\pacs{ 13.60.Le,  14.20.Gk}



\maketitle

\section{Introduction}

It is well recognized~\cite{lso23} that the information on
 the interactions between the $J/\psi$ meson  and the nucleon(N)
 can improve our understanding 
of the roles of gluons ($g$) in determining the structure of hadrons and
hadron-hadron interactions. 
In addition, a  model of the $J/\psi$-N interaction is needed to understand the nucleon
resonances $N^*(P_c)$  reported by the LHCb
collaboration~\cite{LHCb-15,LHCb-16a,LHCb-19,LHCb-21a}.
It is also needed to extract the gluonic distributions in nuclei, and to study the
existence of nuclei with hidden charms~\cite{BD88a,BSD90,GLM00,BSFS06,WL12}.

\begin{figure}[b]
\centering
\includegraphics[width=0.6\columnwidth,angle=0]{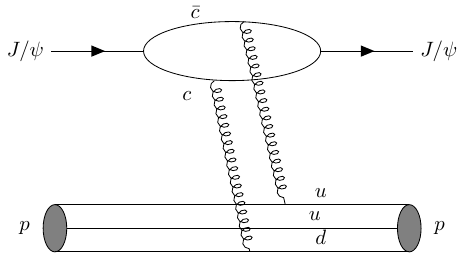}
\caption{Two gluon exchange mechanism of 
$\gamma + N  \rightarrow  J/\psi +N$ reaction.}  
\label{fig:fig-lqcd}
\end{figure}

The leading $J/\psi$-N interaction
is the two-gluon exchange mechanism, as illustrated in Fig.~\ref{fig:fig-lqcd}.
Higher order multi-gluon exchange  effects can not be  neglected in the non-perturbative region.
By using continuum and lattice studies at low energies
and the heavy quark effective field theory 
and perturbative QCD (PQCD) at high energies, the $J/\psi$-N interaction can be estimated.

Several attempts had been made to 
determine the $J/\psi$-N interactions.
Peskin~\cite{Peskin79}  applied the operator product expansion to 
evaluate the strength of the color field emitted by heavy $q\bar{q}$ systems, and suggested~\cite{BP79}
that the van der Waals force induced by the color field of $J/\psi$ on nucleons can generate an attractive short-range
$J/\psi$-N interaction.
The results of Peskin were used by Luke, Manohar, and Savage~\cite{LMS92} to predict, using the effective field theory
method, the $J/\psi$-N forward scattering amplitude which was then used to get an estimation that $J/\psi$ can have
a few MeV/nucleon attraction in nuclear matter.
The $J/\psi$-N forward scattering amplitude of Ref.~\cite{LMS92} was further investigated by Brodsky and Miller~\cite{BM97a}
to derive a $J/\psi$-N potential which gives a $J/\psi$-N scattering length of $-0.24$~fm.
The result of Peskin was also used by Kaidalov and Volkovitsky~\cite{KV92}, who differed from Ref.~\cite{BM97a} in
evaluating the gluon content in the nucleon, to give a much smaller scattering length of $-0.05$~fm.
In a Lattice QCD (LQCD) calculation using the approach of Refs.~\cite{IAH06,AHI09}, 
Kawanai and Sasaki~\cite{KS10b,KS11,sasaki-1}
obtained an attractive $J/\psi$-N potential of the Yukawa form $V_{J/\psi N,J/\psi N} = - \alpha  e^{-\mu r}/{r}$
with $\alpha =  0.1$ and $\mu = 0.6$~GeV, which gives a scattering length of $- 0.09$~fm.

\begin{figure}[b]
\centering
\includegraphics[width=0.6\columnwidth,angle=0]{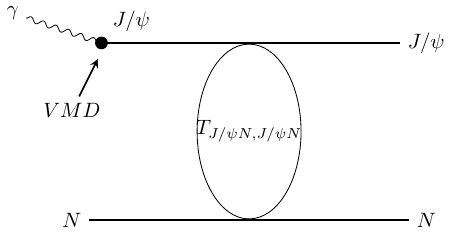}
\caption{The $\gamma +N\rightarrow  J/\psi +N$ reaction
 within the model based on  the Vector Meson Dominance (VMD) assumption. }  
\label{fig:vdm}
\end{figure}

\begin{figure}[t]
\centering
\includegraphics[width=0.6\columnwidth,angle=0]{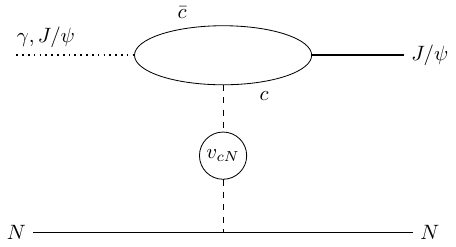}
\caption{Quark-antiquark loop mechanism of
$\gamma +N \rightarrow  J/\psi +N$  and 
$J/\psi +N \rightarrow  J/\psi +N$ due to a 
phenomenological charm quark(c)-nucleon(N) potential $v_{cN}$.}         
\label{fig:cqm-loop}
\end{figure}

To make progress, it is necessary  to have experimental information on the $J/\psi$-N scattering
to test the theoretical results described above and future LQCD calculations.
 One can employ  the traditional approach to determine the 
vector meson-nucleon (VN) interaction  by applying the Vector Meson Dominance (VMD) assumption.
This method is commonly used
to extract $J/\psi$-N cross  sections from the data of $J/\psi$  photo-production reactions.
In this approach, the incoming photon is converted into a vector meson $V$ which is then scattered
from the  nucleon, as illustrated in Fig.~\ref{fig:vdm}. 
However, the approach based on VMD is not valid for $J/\psi$  because  the
VMD coupling constant is determined by the $J/\psi \to \gamma \to  e^+e^-$ decay width
at $q^2=m^2_{J/\psi} \sim 9~\mbox{GeV}^2$ which is far from $q^2=0$ of the $\gamma + p \to J/\psi+p$ reaction. 
Furthermore, the use of VMD for $J/\psi$ is questionable as discussed in Refs.~\cite{du,XCYBCR21}.
In addition, 
the transition amplitude $t_{J/\psi N, J/\psi N}(k,q,W)$ for $J/\psi + N \rightarrow J/\psi +N $ 
near threshold
is far off-shell. For instance, at $W= (m_N+m_{J/\psi})+ 0.5 $~GeV, the incoming $\gamma N$
relative momentum is $q= 0.8$~GeV, which is much larger than the outgoing $J/\psi$-N relative momentum
$k=0.1$~GeV in the  center of mass (CM) system.
Thus,  the VMD  approach is  clearly not directly applicable for
describing $J/\psi+N \rightarrow J/\psi+N$ in the near threshold energy region.

In this paper,  we present a reaction
model to extract the $J/\psi$-N scattering amplitudes
from  the  data of
$\gamma +N \rightarrow J/\psi+N$ reactions, 
specifically from the experiments at Jefferson Laboratory
 (JLab)~\cite{GlueX-19,GlueX-23,jlab-hallc}.
In the meantime, we will obtain  phenomenological $J/\psi$-N
potentials, $V_{J/\psi N,J/\psi N}$, for investigating nuclear reactions involving $J/\psi$ meson.
 No VMD is assumed by taking the
$c\bar{c}$ structure of $J/\psi$ into account to define the model Hamiltonian.
For simplicity in this exploring work,  we will follow  the  Pomeron-exchange model of  Donnachie  and Landshoff (DL)~\cite{DL84} 
to neglect the quark sub-structure of the nucleon and
assume that the interactions between the charm-anticharm ($c\bar{c}$) quarks
in $J/\psi$ and the nucleon 
 can be defined by a phenomenological quark-N potential $v_{cN}$. 
It follows that  the $\gamma+ N
\rightarrow J/\psi+N$ transition amplitude, $B_{\gamma N , J/\psi N}$,
 and the $J/\psi+N\rightarrow J/\psi+N$ potential, 
$V_{J/\psi N}$, 
are defined by $c\bar{c}$-loop mechanisms, as illustrated in  Fig.~\ref{fig:cqm-loop}.
Following  the dynamical formulation~\cite{SL96,MSL06,JLMS07,KNLS13} within which
the unitarity condition requires that 
$J/\psi$-N final state interaction (FSI)  effects must be included,
 the  total amplitude of $\gamma + N \rightarrow J/\psi +N$
 then has the following form:
\begin{eqnarray}
T^{\rm D}_{\gamma N , J/\psi N}
&=&B_{\gamma N , J/\psi N}+ T^{({\rm fsi}) }_{\gamma N,J/\psi N},\nonumber  \\
&&\label{eq:totamp-00}
\end{eqnarray}
with 
\begin{eqnarray}
T^{({\rm fsi})}_{\gamma N,J/\psi N}=
B_{\gamma N , J/\psi N}\,\,G_{J/\psi N}\,\,
T_{J/\psi N , J/\psi N}, 
\label{eq:totamp-01}
\end{eqnarray}
where $G_{J/\psi N}$ is  the $J/\psi$-N propagator, and
$T_{J/\psi N , J/\psi N}$ is the $J/\psi$-N scattering amplitude  calculated from
the  $J/\psi$-N potential,  $V_{J/\psi N}$, by  solving
the following  Lippman-Schwinger Equation 
\begin{eqnarray}
T_{J/\psi N , J/\psi N}&=&V_{J/\psi N , J/\psi N}\nonumber \\
&& +V_{J/\psi N , J/\psi N}\,G_{J/\psi N}\,
T_{J/\psi N , J/\psi N}.
 \nonumber  \\
&&\label{eq:totamp-02}
\end{eqnarray}

To also describe data up to 300  GeV~\cite{GlueX-19,GlueX-23,jlab-hallc, ZEUS-95b,H1-00b},
 we add the Pomeron-exchange amplitude $T^{\rm Pom}_{\gamma N, J/\psi N}$ of 
Donnachie and Landshoff\cite{DL84} (DL), 
reviewed  in Ref.~\cite{lso23}, such that
the total  amplitude can be used to investigate nuclear reactions involving $J/\psi$ at all energies.
To be consistent, the Pomeron-exchange amplitude
should also be defined by the similar $c\bar{c}$-loop mechanism of Fig.~\ref{fig:cqm-loop}.
By using a hadron model~\cite{RW94,Roberts94} based on Dyson-Schwinger equation (DSE) of QCD, such a quark-loop
Pomeron-exchange model was explored in Refs.~\cite{PL96,PL97}.
It will be interesting to use the recent DSE models~\cite{MR97a,CLR10,CCRST13,QR20,YBCR22,AV00,SEVA11,EF11}
to improve the results of Refs.~\cite{PL96,PL97}.
Within the Hamiltonian formulation of this work, it  requires
a realistic Constituent Quark Model (CQM) to generate
the $J/\psi$  wavefunction and careful treatments 
of relativistic kinematic effects within 
Dirac's formulation of relativistic Quantum Mechanics~\cite{KP91}.
We therefore will not pursue this here.
Rather we focus on the near-threshold region, and  any effects
from Pomeron-exchange which is very weak in the  near threshold
region, can be considered as an estimate of the uncertainties
of the results presented in this  paper.

Since the Pomeron-exchange amplitude $T^{\rm Pom}_{\gamma N, J/\psi N}$  has been determined
in  Ref.~\cite{lso23}, our task is to develop a   model for
calculating  the amplitude $T^{\rm D}_{\gamma N, J/\psi N}$ defined 
by Eqs.~(\ref{eq:totamp-00})-(\ref{eq:totamp-02}).
As  defined  by the loop-mechanism illustrated in Fig.~\ref{fig:cqm-loop}, the $J/\psi$-N potential  $V_{J/\psi N}(r)$ is constructed  
by folding $v_{cN}(r)$ into the  $J/\psi$ wavefunction $\phi_{J/\psi}$.
 By using $\phi_{J/\psi}$
from the  CQM of Ref.~\cite{SEFH13}, the
amplitude  $T^{\rm D}_{\gamma N, J/\psi N}$  
is completely determined  by  $v_{cN}(r)$.
 To establish correspondence with the LQCD  calculations,
 the parametrization of $v_{cN}(r)$  is chosen such
 that the predicted $V_{J/\psi N}(r)$ at large distances  exhibits the same
Yukawa potential form  extracted from a   LQCD
calculation of \cite{KS10b,sasaki-1}. 

We  determine the  parameters of  $v_{cN}$  by fitting  the
 total cross section data from the 
JLab experiments~\cite{GlueX-19,GlueX-23}.
As will be  presented later,
 the resulting differential cross sections $d\sigma/dt$ are
in reasonably
good agreements with  the data~\cite{GlueX-19,GlueX-23,jlab-hallc} from JLab.
More importantly, it is shown that
the FSI effects dominate the cross section
in the very near-threshold region, allowing for sensitive
testing of the predicted $J/\psi$-N scattering amplitudes.  
Within the experimental uncertainties, this procedure
allows us to obtain several $J/\psi$-N potentials which all
fit the available JLab  data reasonably well.
They, however, predict rather different  cross sections near threshold and
the  resulting $J/\psi$-N scattering lengths.
More extensive  and precise data in the very near  threshold
region 
are needed for making further progress. 

By using the determined $c$-N potential $v_{cN}(r)$ and the wavefunctions
generated from the same CQM of  Ref.~\cite{SEFH13}, we can apply the
constructed dynamical model to  predict
the cross  sections  of photo-production of the other 
charmonium states.
The  results
for the  production of $\eta_c(1S)$  and $\psi(2S)$ mesons are presented
for future experimental tests at JLab and the future Electron-ion colliders (EIC).

Here we note that the JLab data of  $J/\psi$ photo-production 
 had also been investigated by
using models based on two- and  three-gluons exchange mechanisms~\cite{BCHL01},
Generalized  Parton  Distribution  (GPD) of the nucleon~\cite{GJL21}, and
the Holographic QCD~\cite{MZ19}. All of these approaches have rather different
assumptions in treating the quark sub-structure of $J/\psi$. They are distinctively
different from our approach which  accounts for the $c\bar{c}$-loop mechanisms in
calculating both the $J/\psi$ photo-production amplitudes 
and $J/\psi$-N  final-state interactions. 
Thus their objective is not to extract
$J/\psi$-N interactions  at low  energies, 
as we are  trying to achieve in  this work.

Without using VMD,  
the JLab  data had also been investigated~\cite{du} by using the  effective
Lagrangian approach.  Their  objective was to 
demonstrate  that  with appropriate parameters 
the cusp  structure of JLab data at $W\sim 4.2-4.3$ GeV can be
explained by the  box-diagram
mechanisms $\gamma N \rightarrow \bar{D}^*\Lambda_c\rightarrow  J/\psi N$ due to
the exchanges  of $\bar{D}^*$ and $\Lambda_c$ mesons. This approach can in principle
be  extended to extract $J/\psi$-N  interaction from $J/\psi$  photo-production data, 
but  has not been pursued.

In Section II, we present our formulation. The results  are presented in Section III.
In Section IV, we provide a summary and discuss
possible future improvements.

\section{Formulation}
 
We  follow Ref.~\cite{marvin-gold-ken} to
use the normalization  $\la{\bf k}|{\bf k}'\ra= \delta({\bf k}-{\bf k}')$
for plane wave  state $|{\bf k}\ra$ and  $\la\phi_{\alpha}|\phi_\beta\ra=\delta_{\alpha,\beta}$
for bound  state $|\phi_\alpha\ra$. The $J/\psi$ meson will be
denoted as $V$ in the rest of the paper. 

In  the center of mass (CM) frame, 
the  differential cross  section of vector meson ($V$) photo-production reaction,
$\gamma({\bf q}, \lambda_\gamma)+ N({-\bf q},m_{s})\rightarrow
V({\bf k},\lambda_V) + N(-{\bf k},m_{s}')$, is calculated from~\cite{{lso23}}

\begin{eqnarray}
 \frac{d\sigma_{VN,\gamma N}}{d\Omega}
&=&\frac{(2\pi)^4}{|{\bf q}|^2}\frac{|{\bf k}|\, E_V({\bf k})E_N({\bf k})}{W}\frac{|{\bf q}|^2E_N({\bf q})}{W}
\nonumber \\ &&\hskip -1.8cm \mbox{} \times 
\frac{1}{4} \sum_{\lambda_V,m'_s}\sum_{\lambda_\gamma,m_s}
\left| \braket{\mathbf{k},\lambda_V m'_s|T_{VN,\gamma N}(W)|\mathbf{q},\lambda_\gamma  m_s} \right|^2,\;\;\;
\label{eq:crst-gnjn}
\end{eqnarray}
where  
 $m_s$ ($m'_s$) denotes the $z$-component of the initial (final) state nucleon spin, and $\lambda_V$ and $\lambda_\gamma$
are the helicities of vector meson $V$ and photon $\gamma$, respectively.
The magnitudes $q=|\mathbf{q}|$ and  $k=|\mathbf{k}|$ are defined by the invariant mass
$W=q+E_N(q)=E_V(k)+E_N( k)$. 

The reaction amplitude $T_{VN,\gamma N}(W)$
can be decomposed into the sum of 
the dynamical scattering amplitude $T_{VN,\gamma N}^{\rm D}(W)$ 
and the Pomeron-exchange amplitude $T_{VN,\gamma N}^{\rm Pom}(W)$ as~\cite{lso23}
\begin{eqnarray}
T_{VN,\gamma N}(W)= T^{\rm D}_{VN,\gamma N}(W) + T^{\rm Pom}_{VN,\gamma N}(W).
\end{eqnarray}
We shall describe each amplitude in the following subsections.

\subsection{Dynamical model for $T^{\rm D}_{VN,\gamma N}(W)$}

Following the dynamical approach of Refs.~\cite{SL96,MSL06,JLMS07,KNLS13}, 
the amplitude $T^{\rm D}_{VN,\gamma N}(W)$  is calculated  from
using  the following Hamiltonian
\begin{eqnarray}
H=H_0 +\Gamma_{\gamma,c\bar{c}} + v_{ c\bar{c}}+v_{cN},
\label{eq:h}
\end{eqnarray} 
where $H_0$ is the free Hamiltonian,
$v_{cN}$ is a phenomenological quark-nucleon potential to be
determined, $v_{c\bar{c}}$ is the  $c{\bar c}$ potential of CQM, and 
 $\Gamma_{\gamma,c\bar{c}}$ is 
the electromagnetic coupling of 
 $\gamma\rightarrow {c\bar{c}} $ defined  by
\begin{eqnarray}
\braket{{\bf q}|\Gamma_{\gamma,c\bar{c}}|{\bf k}_1,{\bf k}_2}
&=&\frac{1}{\sqrt{2|{\bf q}|}}
\frac{1}{\sqrt{2E_c({\bf k}_1)}}\frac{1}{\sqrt{2E_c({\bf k}_2)}}\nonumber\\
&&\hskip -0.5cm \mbox{} \times \frac{e_c}{(2\pi)^{3/2}} [\bar{u}({\bf k}_1)\gamma^\mu\epsilon_\mu(q)v({\bf k}_2)].
\end{eqnarray}
Here $e_c$ is the charge of the  charmed quark, $\epsilon_\mu(q)$ is
the photon polarization vector,
 $\bar{u}({\bf k}_1)$
and $v({\bf k}_2)$ are the Dirac spinors with the normalization
$\bar{u}({\bf k})u({\bf k})=\bar{v}({\bf k})v({\bf k})=1$.

 Using the potential $v_{ c\bar{c}} $
in the Hamiltonian, the wavefunction $\ket{\phi_V}$ of the $J/\psi$
is obtained  by solving the bound-state equation
within a CQM developed by  Segovia et al.~\cite{SEFH13}, expressed as:
\begin{eqnarray}
(H_0+v_{c\bar{c}}) \ket{\phi_{V}} = E_V \ket{\phi_{V}}.
\label{eq:wf-v}
\end{eqnarray}
Here, we assume a simple $s$-wave wavefunction
defined  in  momentum-space as
\begin{eqnarray}
 \phi^{J_Vm_V}_{V,{\bf p}_V}({\bf k} m_{s_c}, {\bf k'} m'_{s_c})
&=&\braket{J_Vm_V|\frac{1}{2}\frac{1}{2}m_{s_c}m'_{s_{\bar{c}}}}
 \phi({\bf  \bar{k}})\nonumber \\
&& \times  \delta(  {\bf p}_V- {\bf k}-{\bf k'})\,,
\label{eq:phi-v} 
\end{eqnarray}
where ${\bf p}_V$ is the momentum of $J/\psi$, 
${\bf k} ({\bf k^{\prime}}$) is the momentum of $c ({\bar c})$, and
 ${\bf \bar k}= ({\bf k}-{\bf k'})/2$.
 The total angular momentum and its magnetic quantum number of  $J/\psi$ are denoted by
 $J_V$ and $m_V$, respectively, and  $m_{s_c} (m'_{s_{\bar{c}}})$ is the magnetic quantum number of $c (\bar{c})$ spin angular momentum.

With the Hamiltonian given by Eq.~(\ref{eq:h})  and neglecting
the quark-quark  scattering, 
the scattering amplitude $T(W)$ for $\gamma+ N \rightarrow V+N$ process
is defined  by the following Lippmann-Schwinger equation,
\begin{eqnarray}
T(W)= H'+T(W)\frac{1}{W-H_0+i\epsilon} H' .
\label{eq:lseq-0}
\end{eqnarray}
where
\begin{eqnarray}
H'=\Gamma_{\gamma,c\bar{c}} + v_{cN}.
\label{eq:h-int}
\end{eqnarray}
Inserting the intermediate states 
of $\ket{VN}$,$\ket{c\Bar{c}N}$ and keeping only
 the first order in electromagnetic coupling $e$ in Eq.~(\ref{eq:lseq-0}), we obtain
\begin{eqnarray}
T^{\rm D}_{VN,\gamma N}(W) = B_{VN,\gamma N}(W) + T^{\rm (fsi)}_{VN,\gamma N}(W) ,
\label{eq:t-cqm-0}
\end{eqnarray}
where $B_{VN,\gamma N}(W)$ is the
 Born term of $J/\psi$ photoproduction 
and the  FSI term $T^{\rm (fsi)}_{VN,\gamma N}(W) $
 is required by the unitary condition and is defined by
\begin{eqnarray}
T^{\rm (fsi)}_{VN,\gamma N}(W)&=&T_{VN,VN}(W)\frac{1}{W-H_0+i\epsilon}\nonumber \\
&&\times B_{VN,\gamma N}(W) .
\label{eq:fsi-term}
\end{eqnarray}

In this work,  we assume that the $VN$  potential can be constructed
by the Folding model~\cite{feshbach} using the quark-N interaction $v_{c N}$  and 
 the wavefunction $\phi_V$ generated from Eq.~(\ref{eq:wf-v}) 
\begin{eqnarray}
V_{VN,VN}= \braket{\phi_V, N|\sum_{c}v_{cN}|\phi_V,N}.
\label{eq:v-vnvn}
\end{eqnarray}
The wavefunction and $v_{cN}$ potential are also used 
to construct the $J/\psi$ photo-production process with the following form
\begin{eqnarray}
B_{VN,\gamma N}(W) &=&  
\bra{ \phi_V,N } \left[\sum_{c} v_{cN} \frac{\ket{c\bar{c}} \bra{c\bar{c}}}{E_{c\bar{c}}-H_0}\, 
\Gamma_{\gamma,c\bar{c}} \right] \, \ket{\gamma,N}  ,\nonumber\\
\label{eq:photo-b}
\end{eqnarray}
where 
 $E_{c\bar{c}}$ is the energy available to the propagation of $c\bar{c}$.

In the following, we give explicit expressions of
the matrix  elements of  $V_{VN,VN}$, $B_{VN,\gamma N}(W)$, and 
$T^{(\rm fsi)}_{VN,\gamma N}(W)$.

\subsubsection{Matrix elements of $V_{VN,VN}$}
\label{sub_vnnvn}

\begin{figure}[b]
\centering
\includegraphics[width=0.6\columnwidth,angle=0]{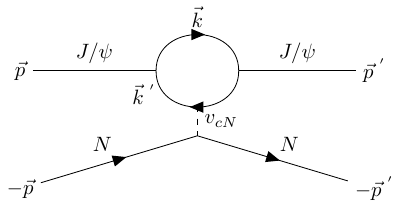}
\caption{$J/\psi$-N potential defined by the quark-nucleon potential
$v_{cN}$, with the momentum variables in
 Eq.~(\ref{eq:v-vnvn-0}).}
\label{fig:vnvn}
\end{figure}

To evaluate Eq.~(\ref{eq:v-vnvn}), we assume for simplicity that quark-N interaction, $v_{cN}$, is independent of spin variables, 
\begin{eqnarray}
&& \bra{{\bf k}\,m_{s_c}\,, {\bf p}\,m_{s_N}} v_{cN} \ket{{\bf  k}'\,m'_{s_c}\,, {\bf p}'\,m'_{s_N}}
\nonumber\\
&&\hskip 0.2cm =\delta_{m_{s_c},m'_{s_c}}\delta_{m_{s_N},m'_{s_N}} 
\delta({\bf k+p-k'-p'}) \braket{{\bf q}|v_{cN}|{\bf q'}},
 \nonumber \\
\label{eq:vc-pot}
\end{eqnarray}
where the relative momenta of quark and nucleon are  defined by
\begin{eqnarray}
{\bf q}&=& \frac{m_N{\bf k}-m_c{\bf p}}{m_N+m_c}, \label{eq:kin1}  \\
{\bf q}'&=& \frac{m_N{\bf k}'-m_c{\bf p}'}{m_N+m_c} . \label{eq:kin2} 
\end{eqnarray}
Here $m_c$ and $m_N$ are the masses of the quark $c$ and the nucleon, 
 respectively.

With the  $J/\psi$ wavefunction given 
in Eq.~(\ref{eq:phi-v}) and the spin independent quark-N potential defined by Eq. (\ref{eq:vc-pot}), 
we  can evaluate
 the matrix element of the  potential $V_{VN,VN}$ given by Eq.~(\ref{eq:v-vnvn}).
In the CM frame, where ${\bf p}_V=-{\bf p}$ and ${\bf p}'_V=-{\bf p}'$ 
as  illustrated in Fig.~\ref{fig:vnvn}, we then have
\begin{eqnarray}
&& \braket{ {\bf p}_V m_{V}, {\bf  p} m_{s}|V_{VN,VN}|{\bf p}'_V m'_{V}, {\bf p}' m'_{s}}
\nonumber\\
&&\hskip 0.2cm = \delta_{m_{V},m'_{V}}\delta_{m_{s},m'_{s}} 
\delta({\bf p}_V+{\bf p}-{\bf p}'_V-{\bf p}')
\braket{{\bf p}|V_{VN}|{\bf p}' },
\nonumber\\
 \label{eq:vjpsin} 
\end{eqnarray}
where
\begin{eqnarray}
\braket{ {\bf p}|V_{VN}|{\bf p}'} &=& 2 \int  d{\bf k} \,  \textstyle \phi^* \left({\bf k}-\frac{{\bf p}}{2} \right)
\nonumber \\ && \hskip -0.5cm\mbox{} \times\Braket{{\bf p}-\frac{ m_N}{m_N+m_c}{\bf k}|v_{cN}|{\bf p}'-\frac{ m_N}{m_N+m_c}{\bf k}}
\nonumber \\ && \hskip -0.5cm\mbox{} \times
\textstyle \phi \left({\bf k}-\frac{{\bf p}'}{2}\right) .
\label{eq:v-vnvn-0}
\end{eqnarray}
 Here the factor 2 arises from the  
summation of  the  contributions from the two quarks within the $J/\psi$ meson and 
we have used the definitions of Eqs.~\eqref{eq:kin1} and~\eqref{eq:kin2}.
 
For a potential $v_{cN}(r)$ depending  only on  the relative distance $r$ between $c$ and $N$,  we have
\begin{eqnarray}
 \braket{{\bf q}|v_{cN}|{\bf q'}} &=&
v_{cN}({\bf q}-{\bf q}')\nonumber\\ 
&=&\frac{1}{(2\pi)^3}\int\, d{\bf r}\,
e^{i{\bf ({\bf q}-{\bf q}')}\cdot {\bf r}} v_{cN}(r).
\label{eq:vcn-t}
\end{eqnarray}  
The matrix element of $v_{cN}$ in Eq.~\eqref{eq:v-vnvn-0} can then
be written as
\be
\Braket{ {\bf p}-\frac{ m_N}{m_N+m_c}{\bf k}|v_{cN}|{\bf p}'-\frac{ m_N}{m_N+m_c}{\bf k} }
= v_{cN}({\bf p}-{\bf p}'). 
\label{eq:yukawa}
\ee
For later calculations, we note here that for  a  Yukawa form
$v_{cN}(r)=\alpha \frac{e^{-\mu r}}{r}$,
Eq.~(\ref{eq:vcn-t}) leads  to
\begin{eqnarray}
v_{cN}({\bf p}-{\bf p}') 
&=&\alpha\frac{1}{(2\pi)^2}
\frac{1}{({\bf p}-{\bf p}')^2+\mu^2}.
\label{eq:vcn-y-t}
\end{eqnarray}

Using Eq.~(\ref{eq:yukawa}), Eq.~(\ref{eq:v-vnvn-0}) can now be expressed in the following factorized form:
\begin{eqnarray}
\braket{ {\bf p}|V_{VN}|{\bf p}' } &=& F_V({\bf t}) \left[ 2v_{cN}({\bf t}) \right] ,
\label{eq:v-vn-f}
\end{eqnarray}
where ${\bf t}={\bf p}-{\bf  p}'$, and
\begin{eqnarray}
F_V({\bf t}) &=& \int  d{\bf k} \, \textstyle
\phi^*\left({\bf k}- \frac{{\bf p}}{2}\right) \phi \left({\bf k}-\frac{{\bf p}'}{2}\right) \nonumber \\
&=&\int  d{\bf k} \,\textstyle
\phi^*\left({\bf k}-\frac{{\bf t}}{2}\right) \phi\left({\bf k}\right)
\label{eq:v-ff}
\end{eqnarray}
is the form factor of the vector meson $V$ and $\phi({\bf k})$ is the wavefunction of $J/\psi$  in momentum space.

\subsubsection{Matrix element  of $B_{VN,\gamma  N}(W)$}

\begin{figure}[h]
\centering
\includegraphics[width=0.6\columnwidth,angle=0]{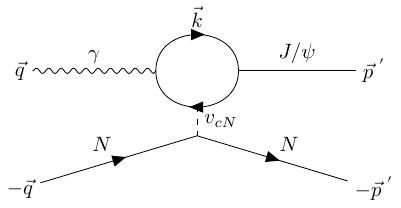}
\caption{$J/\psi$ photo-production on the nucleon target 
with the momentum variables indicated in Eq.~(\ref{eq:tmx-b-2}).}
\label{fig:gnvn}
\end{figure}

By using the $J/\psi$ wavefunction  and  Eq.~(\ref{eq:vc-pot}) for quark-N potential, the matrix element of photo-production of Eq.~(\ref{eq:photo-b}) can be calculated.
With the variables in the CM system, as illustrated in Fig.~\ref{fig:gnvn}, we obtain 
\begin{eqnarray}
&&\Braket{ {\bf p}'m_V m'_s|B_{VN,\gamma N}(W)|{\bf q}\lambda m_{s}}
\nonumber\\
&&  =
\sum_{m_c,m_{\bar{c}}}\frac{1}{(2\pi)^3}\frac{e_c}{\sqrt{2|{\bf q}|}}
\int d{\bf k} \textstyle
\braket{J_Vm_V|\frac{1}{2}\frac{1}{2}m_c m_{\bar{c}} } \phi\left({\bf k}-\frac{1}{2}{\bf p}' \right)
\nonumber \\ 
&& \hskip 0.4cm \mbox{} \times  \delta_{m_{s},m'_s} 
\Braket{{\bf p}'-\frac{ m_N}{m_N+m_c}{\bf k}|v_{cN}|{\bf q}-\frac{ m_N}{m_N+m_c}{\bf k} }
\nonumber \\ 
&& \hskip 0.4cm \mbox{} \times
\frac{1}{W-E_N({\bf q})-E_c({\bf q-k})-E_c({\bf k})+i\epsilon} 
\nonumber\\
&& \hskip 0.4cm \mbox{} \times
\bar{u}_{m_c}({\bf k})[{\bf \epsilon_\lambda}\cdot {\bf \gamma}] v_{m_{\bar{c}}}({\bf q-k}).
\label{eq:t-gnvn-0}
\end{eqnarray}
If one chooses the  Yukawa form for $v_{cN}(r)$, one obtains the following factorized form:
\begin{eqnarray}
&& \Braket{ {\bf p}' m_V m'_s|B_{VN,\gamma N}(W)|{\bf q}\lambda m_{s} } 
\nonumber\\
&& \hskip 0.2cm 
= C_{\lambda,m_V} \delta_{m_s,m_s'} B({\bf p}',{\bf q},W)
\left[ 2 v_{cN}({\bf q}-{\bf p}') \right] ,
\label{eq:mtx-b-1}
\end{eqnarray}
where
\be
C_{\lambda,m_V}=\sum_{m_c,m_{\bar{c}}} \textstyle
\braket{J_Vm_V|\frac{1}{2}\frac{1}{2}m_c m_{\bar{c}} }
\braket{m_{\bar{c}}|{\bf \sigma}\cdot {\bf \epsilon_\lambda}|m_c},
\ee
and
\begin{eqnarray}
 B({\bf p}',{\bf q},W)
&&=\frac{1}{(2\pi)^3}\frac{e_c}{\sqrt{2|{\bf q}|}}
\int d{\bf k} \, \phi \left({\bf k}-{\textstyle\frac{1}{2}}{\bf p}' \right) 
\nonumber\\
&&\times
\frac{1}{W-E_N({\bf q})-E_c({\bf q-k})-E_c({\bf k})+i\epsilon} 
\nonumber  \\ 
&& \mbox{} \times
\sqrt{\frac{E_c({\bf k})+m_c}{2E_c({\bf k})}}\sqrt{\frac{E_c({\bf q- k})+m_c}{2E_c({\bf q- k})}}
\nonumber\\
&& \mbox{} \times
\left(1-\frac{{\bf k}\cdot{\bf (q-k)}}{[E_c({\bf k})+m_c][E_c({\bf q- k})+m_c]} \right).
\nonumber\\
\label{eq:tmx-b-2}
\end{eqnarray}    

\subsubsection{Final State Interactions}

\begin{figure}
\centering
\includegraphics[width=0.6\columnwidth,angle=0]{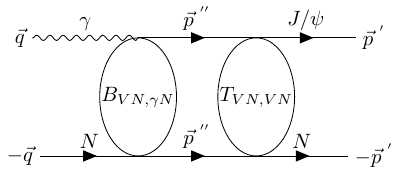}
\caption{J/$\psi$ photo-production on the nucleon  with final state interaction
given in Eq.~(\ref{eq:tfsi-mx}).}
\label{fig:fsi}
\end{figure}

\begin{figure}
\centering
\includegraphics[width=1.0\columnwidth,angle=0]{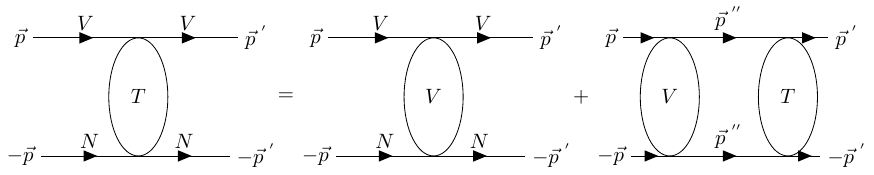}
\caption{The VN scattering equation defined by Eq.~(\ref{eq:lseq-00}).}
\label{fig:fsi-vn}
\end{figure}

Including the final state interaction, as illustrated 
in Fig.~\ref{fig:fsi}, the matrix element of the total amplitude in Eq.~(\ref{eq:t-cqm-0}) is
\begin{eqnarray}
&& \Braket{ {\bf p}' m_V m'_s | T_{VN,\gamma N}(W) | {\bf q} \lambda m_s}
\nonumber\\
&&=
\Braket{{\bf p}'m_Vm'_s|B_{VN,\gamma N}(W)|{\bf q}\lambda m_s}
\nonumber\\
&& \hspace{0.2cm}
+ \braket{{\bf p}'m_Vm'_s|T^{\rm (fsi)}_{VN,\gamma N}|{\bf q}(W)\lambda m_s},
\label{eq:pom-cqm-t0}
\end{eqnarray}
with
\begin{eqnarray}
&& \Braket{{\bf p}'m_Vm'_s|T^{\rm (fsi)}_{VN,\gamma N}(W)|{\bf q}\lambda m_s}
\nonumber\\
&&\hskip 0.2cm =
\sum_{m^{\prime\prime}_V,m^{\prime\prime}_s} \int d{\bf  p}^{\prime\prime} 
\Braket{{\bf p}\, m_Vm'_s|T_{VN,VN}(W)|{\bf  p}^{\prime\prime}m^{\prime\prime}_V,m^{\prime\prime}_s} 
\nonumber \\ 
&&\hskip 1.2cm \mbox{} \times
\frac{1}{W-E_N(p^{\prime\prime})-E_V(p^{\prime\prime})+i\epsilon}
\nonumber\\
&&\hskip 1.2cm \mbox{} \times
\Braket{{\bf p}^{\prime\prime}m^{\prime\prime}_Vm^{\prime\prime}_s|B_{VN,\gamma N}(W)|{\bf q}\lambda, m_s} . 
\label{eq:tfsi-mx}
\end{eqnarray}

With the spin independent quark-N potential defined by Eq. (\ref{eq:vc-pot}),
 the $V+N \rightarrow V+N$  scattering amplitude in the above equation can be  written as 
\begin{eqnarray}
&& \Braket{ {\bf p}\, m_V m_{s}|T_{VN,VN}(W)|{\bf p}' \, m'_V m'_{s} }
\nonumber\\
&&\hskip 0.2cm = \delta_{m_V,m'_V}\delta_{m_s,m'_s} 
\Braket{ {\bf p}' |T_{VN}(W)|{\bf p} },
\label{eq:t-vnvn-0}
\end{eqnarray}
where $\braket{ {\bf p}'|T_{VN}(W)|{\bf p} }$ is defined  by
the following  Lippmann-Schwinger Equation, as  illustrated in Fig.~\ref{fig:fsi-vn}:

\begin{eqnarray}
\braket{{\bf p}'|T_{VN}(W)|{\bf p}} &=& \braket{ {\bf p}'|V_{VN}|{\bf p} } 
\nonumber \\ && 
+ \int d{\bf p^{\prime\prime}} 
\frac{\braket{{\bf p}'|V_{VN}|{\bf p}^{\prime\prime}}\braket{{\bf p}^{\prime\prime}|T_{VN}(W)|{\bf p}}}
{W- E_N(p^{\prime\prime})-E_V(p^{\prime\prime})+i\epsilon}.
\nonumber\\
\label{eq:lseq-00}
\end{eqnarray}
Here $\braket{ {\bf p}'|V_{VN}|{\bf p} }$ has been defined by Eq.~(\ref{eq:v-vn-f}).

We solve Eq.~(\ref{eq:lseq-00}) in the partial-wave representation
by using  the following expansions
\be\label{eq:v-pw}
\braket{ {\bf p}'|V_{VN}|{\bf p}}=\sum_{L}\frac{2L+1}{4\pi}
V_L(p',p)P_L(x), 
\ee
 and
\be\label{eq:t-pw}
\braket{ {\bf p}'|T_{VN}(W)|{\bf p}}=\sum_{L}\frac{2L+1}{4\pi}
T_L(p',p,W)P_L(x), 
\ee
where $x={\bf \hat{p}\cdot\hat{p}'}$  and $P_L(x)$ is the 
Legendre function of the first kind.
With Eq.~(\ref{eq:v-pw}) 
 together with $\braket{ {\bf p}'|V_{VN}(W)|{\bf p}}$ defined 
in Eq.~(\ref{eq:v-vn-f}), 
the  partial-wave matrix element  of potential can be calculated  by 
\be
V_L(p',p) = (2\pi)\int_{-1}^{+1} dx P_L(x) 
\braket{ {\bf p}'|V_{VN}|{\bf p}},
\label{eq:pw-v}
\ee
If  we set $F_V({\bf t})=1$ in Eq.~(\ref{eq:v-vn-f})
and use Eq.~(\ref{eq:vcn-y-t}) for
 a Yukawa form  of the $c$-N potential
$v_{cN}(r)=\alpha\frac{e^{-\mu r}}{r}$, one finds
\begin{eqnarray}
V_L(p',p)=\frac{2}{\pi}\left(\frac{\alpha}{2pp'}\right)Q_L(Z),
\end{eqnarray}
where $Z=\frac{p^2+p^{\prime 2}+\mu^2}{2pp'}$
and $Q_L(Z)$ is the Legendre  function  of the second  kind.

By using Eqs.~(\ref{eq:v-pw}) and (\ref{eq:t-pw}),
Eq.~(\ref{eq:lseq-00}) then leads to
\begin{eqnarray}
&&T_L(p',p,W)=V_L(p',p)\nonumber\\
&&\hspace{2.2cm}+\int dp^{\prime\prime}p^{\prime\prime 2} 
\biggl[V_L(p',p^{\prime\prime})
\nonumber\\
&&\hspace{3cm}\times\;
\frac{1}{W-E_N(p^{\prime\prime})+E_V(p^{\prime\prime})+i\epsilon}\nonumber  \\
&&\hspace{3cm} \times\; T_L(p^{\prime\prime},p,W) \biggr].
 \label{eq:lseq-01}
\end{eqnarray} 
We solve Eq.~(\ref{eq:lseq-01}) 
by using the standard numerical method described in Ref.~\cite{Hftel-frank}.
The scattering phase shifts $\delta_L$ are  calculated from the resulting $T_L(p^{\prime\prime},p,W)$
as follows:
\begin{eqnarray}
e^{i\delta_L}\sin\delta_L=-\pi \frac{p_0E_N(p_0)E_V(p_0)}{E_N(p_0)+E_V(p_0)} T_L(p_0,p_0,W),
\end{eqnarray}
where 
$p_0$ represents the on-shell momentum
and the invariant mass $W=E_N(p_0)+E_V(p_0)$.
We will also calculate  the scattering length $a$, which  is defined 
for the $L=0$ partial-wave at $p_0\rightarrow 0$  as:  
\begin{eqnarray}
p_0\,\cot\delta_{0}=\frac{-1}{a}\,.
\end{eqnarray}

\subsection{Pomeron-exchange amplitude  $T^{\rm Pom}_{VN,\gamma N}(W)$}
\begin{figure}[b]
\centering
\includegraphics[width=0.6\columnwidth,angle=0]{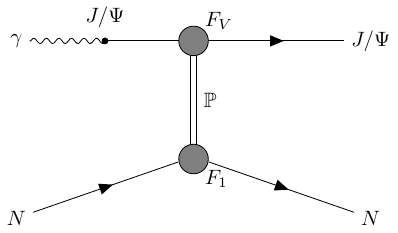}
\caption{The Pomeron-exchange model of Donnachie and Landshoff\cite{DL84}
for the  $\gamma +  N \rightarrow J/\psi+N$ reaction.}
\label{fig:pom_1}
\end{figure}

Following  the approach of Donnachie and Landshoff~\cite{DL84,DL92,DL95,DL98}, 
the Pomeron-exchange amplitude  is constructed within Regge Phenomenology and is of the following 
\begin{eqnarray}
&&\braket{ \mathbf{k},m_V m'_s | T^{\rm Pom}_{VN,\gamma N}(W) | \mathbf{q},\lambda_\gamma m_s}
\nonumber\\
&&=\frac{1}{(2\pi)^3}\sqrt{\frac{ m_Nm_N }{4 E_{V}(\mathbf{k}) E_N(\mathbf{p}')
|\mathbf{q}|E_N(\mathbf{p}) }}\nonumber \\
&& \mbox{} \times 
[\bar{u}(p',m'_s)\epsilon^*_\mu(k,\lambda_{V})\mathcal{M}^{\mu\nu}_\mathbb{P}(k,p',q,p)
\epsilon_\nu(q,\lambda_\gamma) u(p,m_s)].
\nonumber\\
\end{eqnarray}
In this approach, the incoming photon is 
 converted to a  vector meson which  is then 
scattered from the nucleon by the  Pomeron-exchange mechanism,
as illustrated in Fig.~\ref{fig:pom_1}.
The  amplitude $\mathcal{M}^{\mu\nu}_\mathbb{P}(k,p',q,p)$ is given by
\begin{equation}
\mathcal{M}^{\mu\nu}_\mathbb{P}(k,p',q,p) = G_\mathbb{P}(s,t)
\mathcal{T}^{\mu\nu}_\mathbb{P}(k,p',q,p),
\label{eq:MP}
\end{equation}
and
\begin{eqnarray}
\mathcal{T}^{\mu\nu}_\mathbb{P}(k,p',q,p) &=& i \, 2 \frac{e \, m_V^2}{f_V}
 [2\beta_{q_{V}}F_V(t)][3\beta_{u/d} F_1(t)] 
\nonumber \\ && \mbox{}  \{ \slashed{q} g^{\mu\nu} - q^\mu \gamma^\nu \}  \, ,
\label{eq:pom-a}
\end{eqnarray}
where $m_V$ is the mass of the vector meson, and 
$f_V= 5.3$, $15.2$, $13.4$, $11.2$, $40.53$  
for $V=\rho, \omega, \phi, J/\psi, \Upsilon$ are traditionally 
determined by the widths of the $V \to \gamma \to  e^+e^-$ 
decays. The parameters $\beta_{q_{V}}$ ($\beta_{u/d}$) define the coupling of the Pomeron with the quark $q_{V}$ ($u$ or $d$) in the vector meson 
$V$ (the nucleon $N$).
In Eq.~(\ref{eq:pom-a}), a form factor for the Pomeron-vector meson vertex is also introduced with
\begin{eqnarray}
F_V(t)=\frac{1}{m_V^2-t} \left( \frac{2\mu_0^2}{2\mu_0^2 + m_V^2 - t} \right) ,
\label{eq:f1v}
\end{eqnarray}
where $t=(q-k)^2=(p-p')^2$. 
By using the Pomeron-photon analogy, the form factor for the Pomeron-nucleon vertex is defined by the isoscalar electromagnetic form 
factor of the nucleon as  follows

\begin{equation}
F_1(t) = \frac{4m_N^2 - 2.8 t}{(4m_N^2 - t)(1-t/0.71)^2}.
\label{eq:f1}
\end{equation}
Here $t$ is in the unit of GeV$^2$, and $m_N$ is the proton mass.

\begin{figure}[t]
\centering
\includegraphics[width=0.9\columnwidth,angle=0]{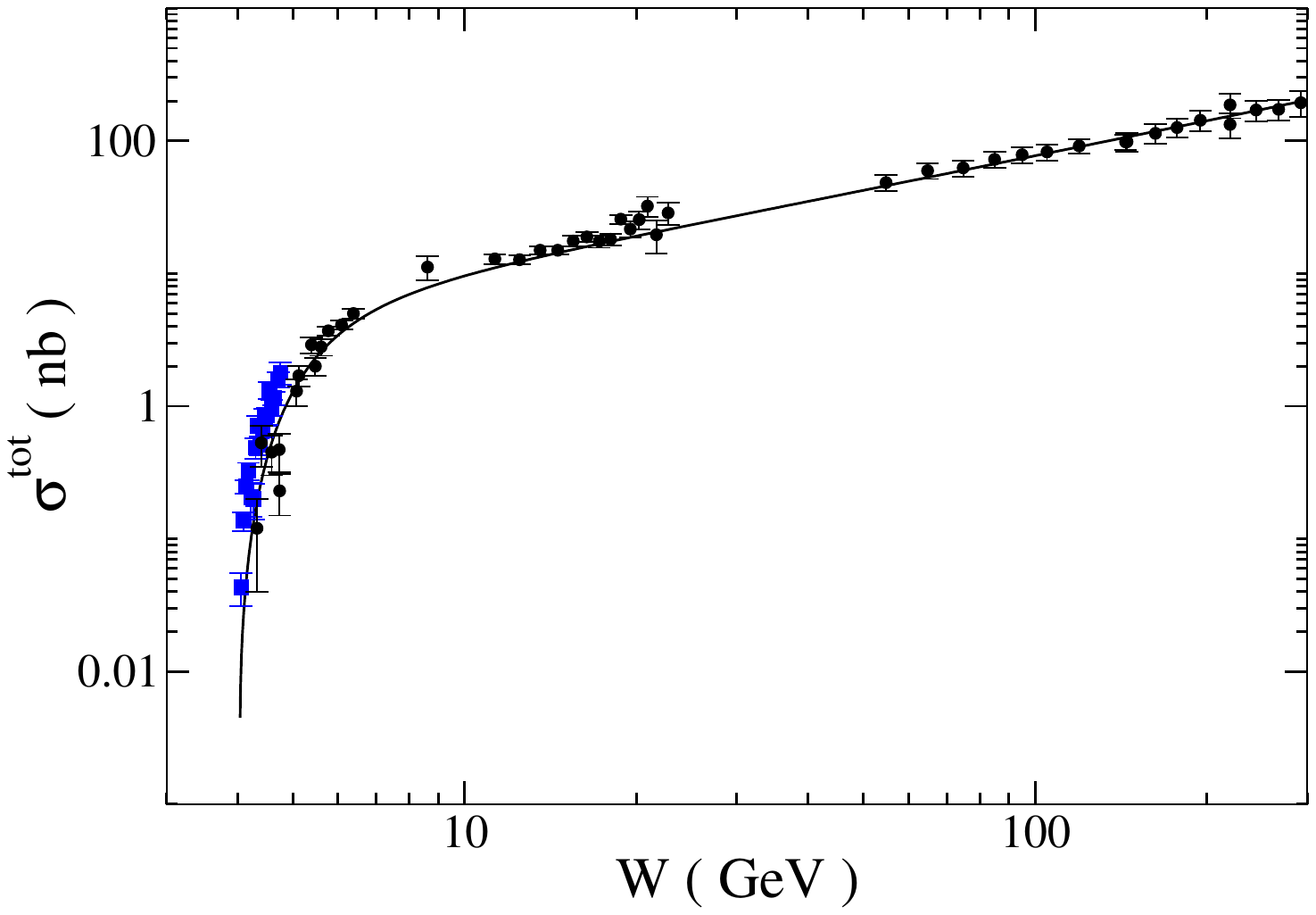} 
\includegraphics[width=0.9\columnwidth,angle=0]{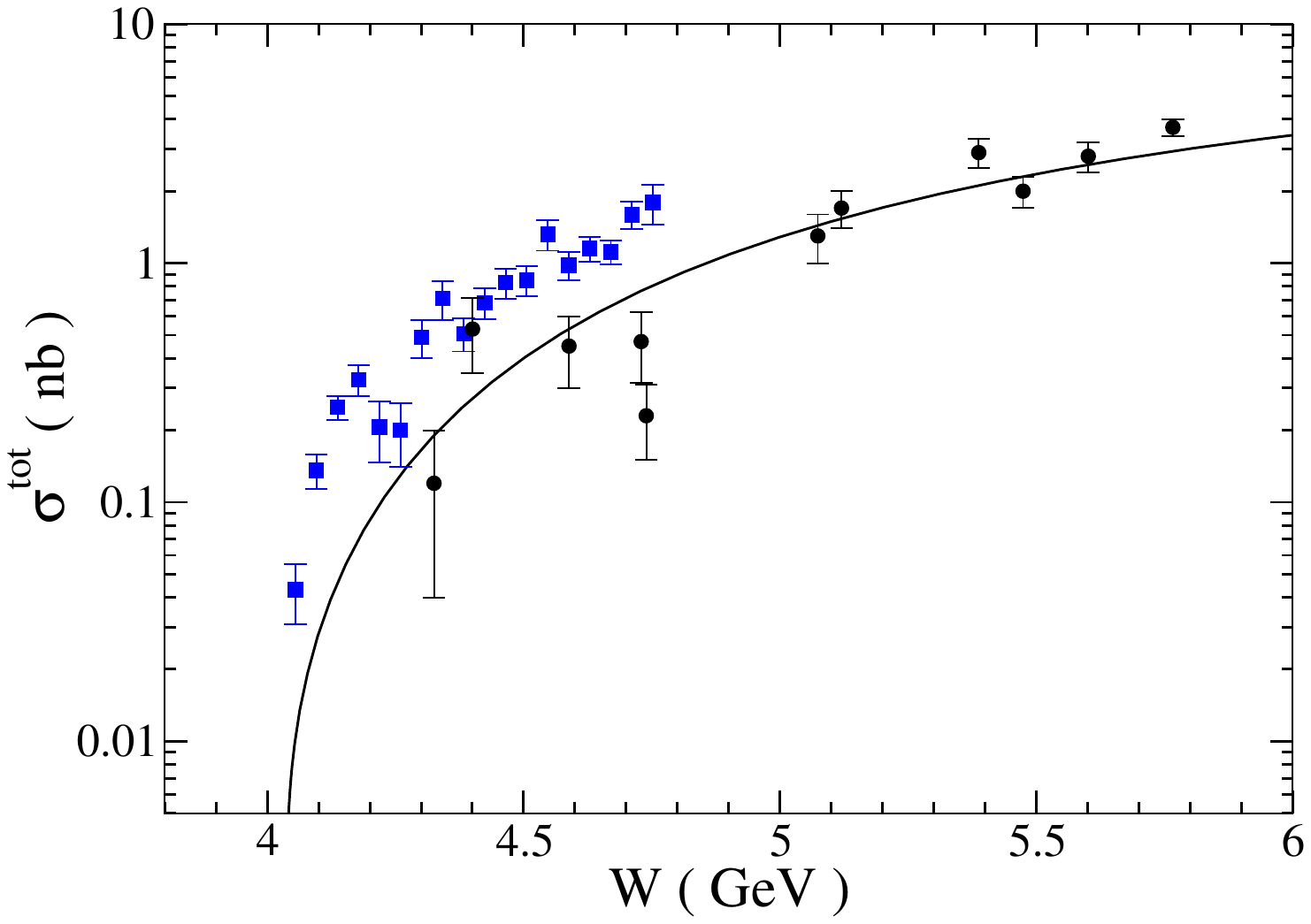}
\caption{Total cross sections from Pomeron-exchange amplitude for two different energy ranges:
$4\leq W\leq 300$ GeV (top)  and for $4\leq W\leq 6$ GeV (bottom). 
Data are taken from~\cite{ZEUS-95b,H1-00b} (black circles) 
and ~\cite{GlueX-23} (blue squares)respectively.}
\label{fig:totcrst-pom} 
\end{figure}

The propagator $G_\mathbb{P}$ of the Pomeron in Eq.~(\ref{eq:MP})
 follows the Regge phenomenology form:
\begin{equation}
G_\mathbb{P} = \left(\frac{s}{s_0}\right)^{\alpha_P(t)-1}
\exp\left\{ - \frac{i\pi}{2} \left[ \alpha_P(t)-1 \right]
\right\} \,,
\label{eq:regge-g}
\end{equation}
where $s=(q+p)^2=W^2$, $ \alpha_P (t) = \alpha_0 + \alpha'_P t$, and $s_0=1/\alpha'_P$.
 We use the value  of $s_0=0.25$ GeV 
from the works~\cite{DL84,DL92,DL95,DL98} of Donnachie and Landshoff. 

$T^{\rm Pom}_{VN,\gamma N}(W)$ has been  determined  in Ref.~\cite{WL13,Lee20,lso23} by fitting  the data
of total cross  sections up to 300 GeV. 
 The resulting  parameters for $\rho^0$, $\omega$, 
$\phi$ photo-production~\cite{OL02} have been determined as follows:
$\mu_0=  1.1$ GeV$^2$, $\beta_{u/d}=2.07$ GeV$^{-1}$, $\beta_{s}=1.38$ GeV$^{-1}$, $\alpha_0=1.08$ for 
$\rho$ and $\omega$, $\alpha_0=1.12$ for $\phi$. 
For the heavy quark systems, we find that 
 using the same $\mu_0^2$, $\beta_{u/d}$, and $\alpha'_P$ values, the 
photo-production  data for $J/\psi$ and $\Upsilon$ can be fitted by setting $\beta_c = 0.32$ GeV$^{-1}$ and $\beta_b = 0.45$ GeV$^{-1}$, along with a larger 
$\alpha_0=1.25$. 

 Figure~\ref{fig:totcrst-pom} depicts the total cross section for the $J/\psi$ photo-production on the nucleon 
obtained solely from the contribution of the Pomeron-exchange amplitude,
and compares it with  the  data  from  
Refs.~\cite{GlueX-19,GlueX-23}\footnote{
To make clearer comparisons, this analysis includes only the latest total cross section data with higher statistics from Ref.~\cite{GlueX-23}.
}  
and \cite{ZEUS-95b,H1-00b}.
One can see from Fig.~\ref{fig:totcrst-pom} that while the Pomeron-exchange mechanism effectively describes 
the  data (black circles)~\cite{ZEUS-95b,H1-00b} at very high energies, 
it falls short in accurately describing the 
the data (blue squares)~\cite{GlueX-19,GlueX-23} from the GlueX experiment at JLab.

\section{Results}
\subsection{Determination of quark-Nucleon potential $v_{cN}$}
We are guided by the Yukawa form  of
$J/\psi$-N potential extracted from the LQCD  calculation
of Ref.~\cite{KS10b} to determine the quark-N potential  $v_{cN}$.
We observe that the factorized form of $V_{VN,VN}$ in
Eq.~(\ref{eq:v-vn-f}) suggests that if $v_{c N}(r)$ is also in the Yukawa form,
the  resulting  $V_{VN,VN}$  can approach the  $J/\psi$-N potential
of  Ref.~\cite{KS10b} at large distance $r$.
We therefore consider the following parameterization
\begin{eqnarray}
v_{cN}(r) =\alpha\left( \frac{^{-\mu r}}{r}-c_s\frac{e^{-\mu_1 r}}{r} \right).
\label{eq:vcn-r}
\end{eqnarray}
We first consider a model (1Y) with $c_s=0$, meaning that the resulting $J/\psi$-N potential takes on the Yukawa form extracted from LQCD.
The  calculation then only  has  two free parameters, $\alpha$ and $\mu$. 
We find that the  total  cross  section data~\cite{GlueX-19,GlueX-23} from JLab for the small $W$ region 
can  be  best fitted by choosing  
$\alpha=-0.067$ and  $\mu=0.3$ GeV if the  $J/\psi$  wavefunction is taken from the CQM model of Ref.~\cite{SEFH13}.

\begin{figure}[t]
\centering
 \includegraphics[width=0.9\columnwidth,angle=0]{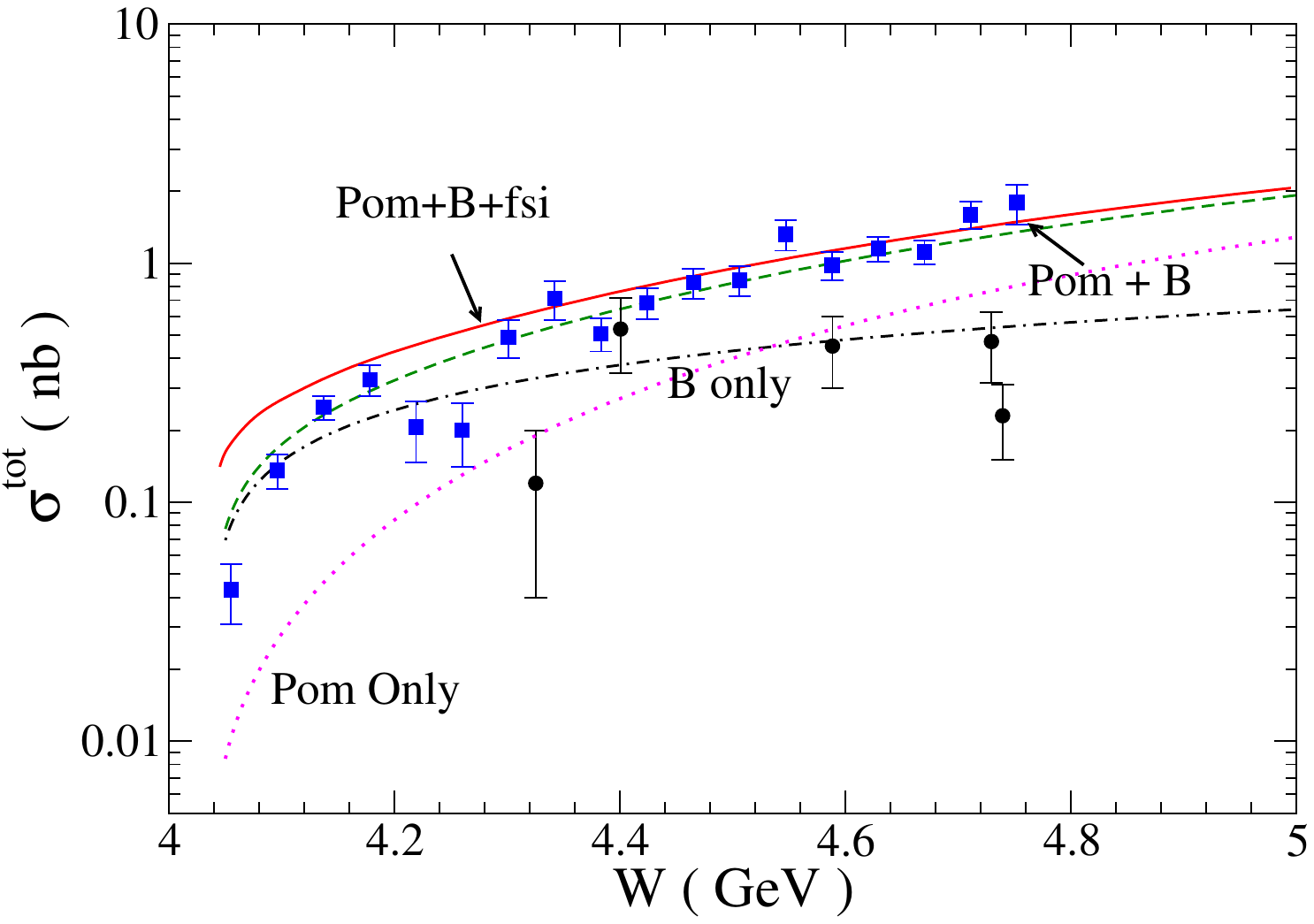}
\caption{Total cross sections  for $4\leq W\leq 5$ GeV from the 1Y model. 
The same  data points are used as in Fig.~\ref{fig:totcrst-pom}.}
\label{fig:totcrst-1}
\end{figure}
We show in Fig.~\ref{fig:totcrst-1} the total cross sections 
 for the $J/\psi$ photo-production in the small $4\leq W\leq 5$ GeV region 
obtained from the 1Y model. 
The dotted, dot-dashed, and dashed lines represent the results obtained from the Pomeron-exchange, Born term, and the sum of Pomeron-exchange and Born term
contributions, respectively. The solid line represents the full result, including the final state interaction in addition to the Pomeron-exchange and Born terms.
As seen in Fig.~\ref{fig:totcrst-1}, the Pomeron-exchange mechanism 
alone is insufficient 
to fit the JLab data~\cite{GlueX-19,GlueX-23}, particularly for low-energy regimes. 
However, the contribution from the $c$-N interaction $v_{cN}$ is essential for fitting the data in this 1Y model. 
In particular, 
the cross sections in the very near threshold 
region are largely  determined by the FSI term. 
 These results demonstrate that $J/\psi$-N  interactions can be extracted rather clearly from the  
 $J/\psi$ photo-production data within this model, which does not use  the VMD assumption. 
 More importantly, the calculations properly account for off-shell effects and satisfy the unitarity condition, a feature that is not considered in most, 
 if not all, previous approaches discussed in Ref.~\cite{lso23}.
 
\begin{figure}[t]
\centering
\includegraphics[width=0.9\columnwidth,angle=0]{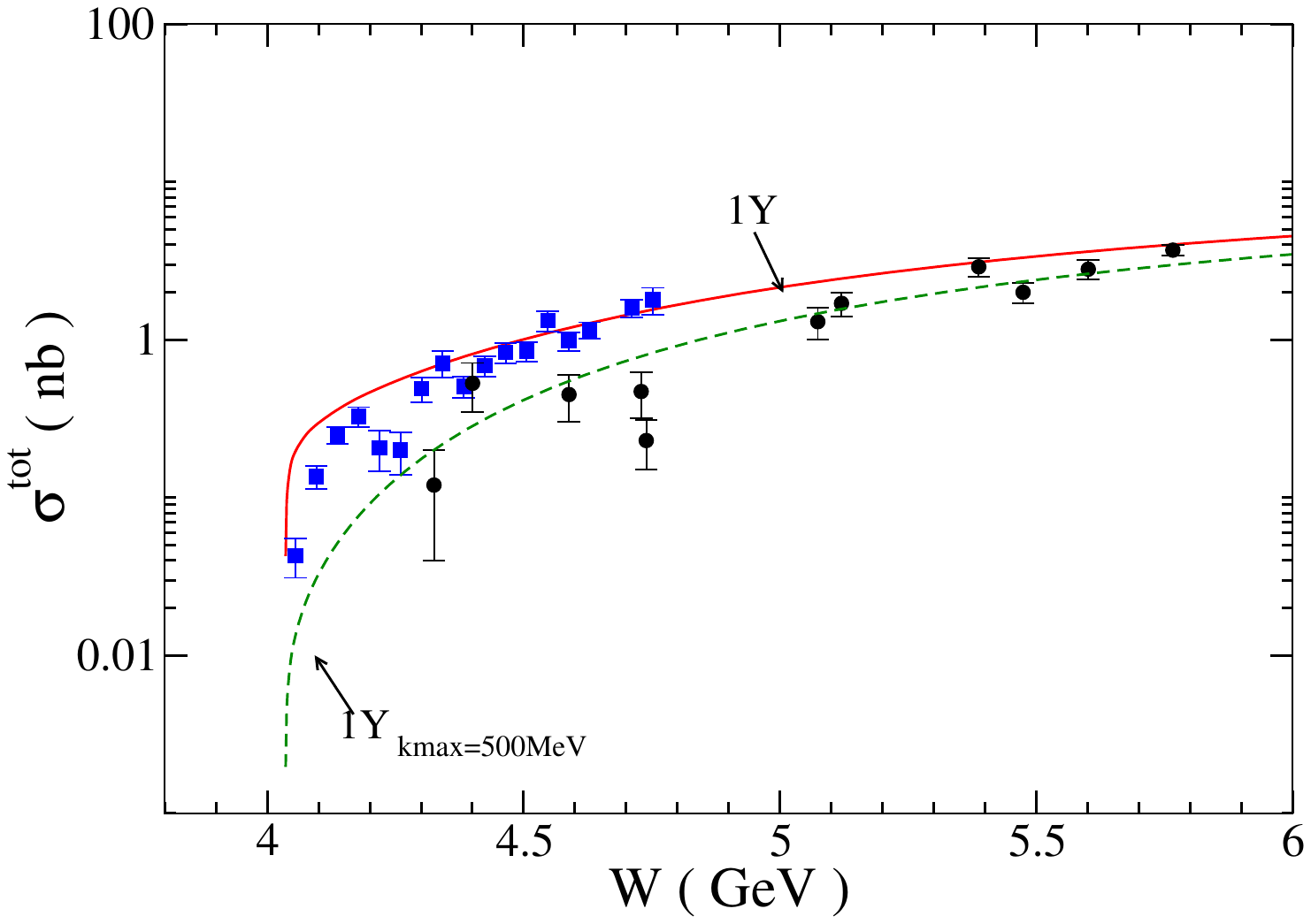}
\includegraphics[width=0.9\columnwidth,angle=0]{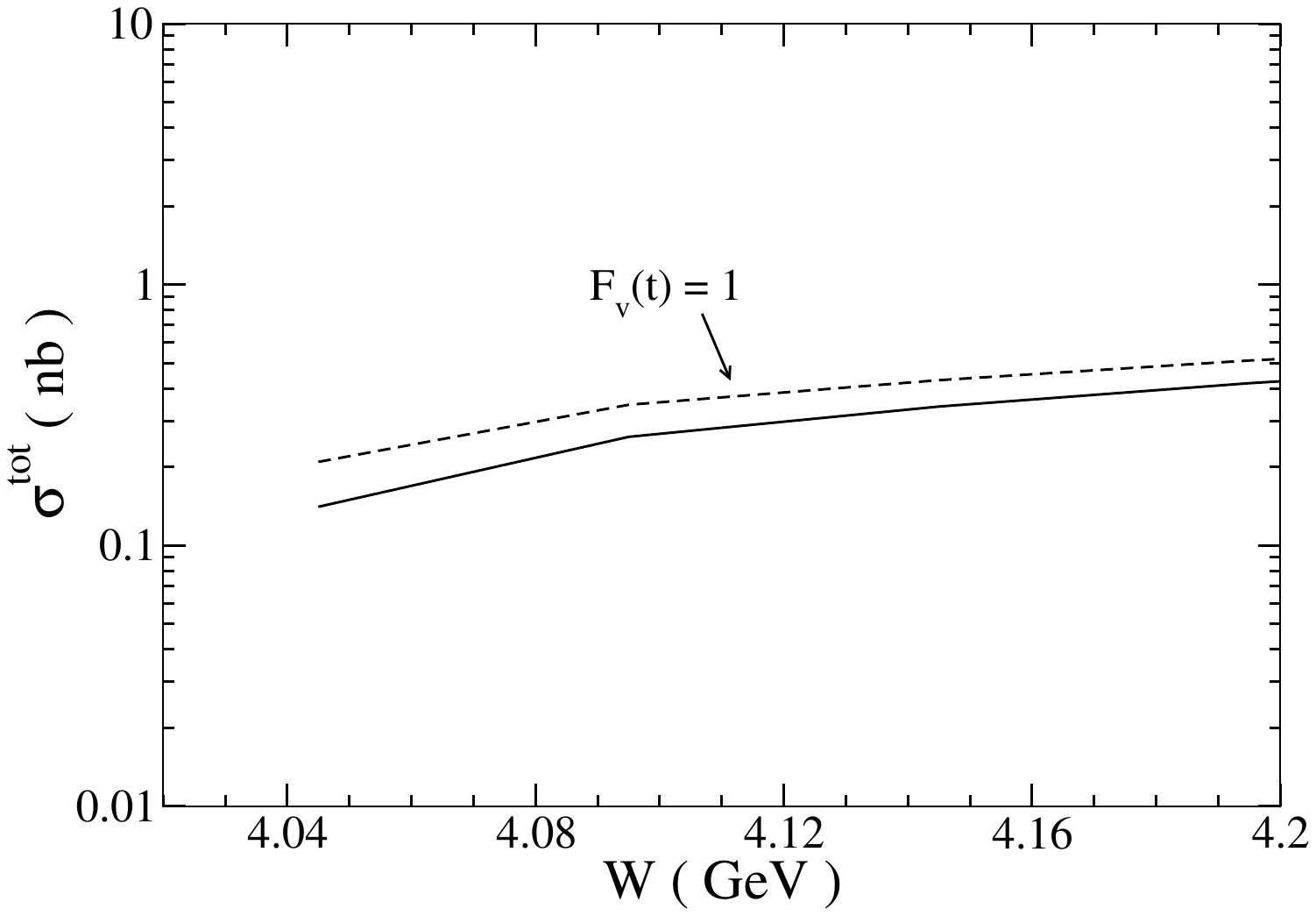}
\caption{The effect of the momentum cut-off $k_{\rm max}=500$ MeV (top),
 and the effect of the form factor $F_V(t)$  (bottom), on the total cross section.
The same  data points are used as in Fig.~\ref{fig:totcrst-pom}.}
\label{fig:cross-mch}
\end{figure}

To see the importance of using a realistic $J/\psi$ wavefunction,
we compare the  cross section from the full  calculation
and that from using  a wavefunction  that employs a 
 momentum cut-off in evaluating  $B_{VN,\gamma  N}$ amplitude. 
Specifically, we replace the integral
$B= \int_0^\infty  \phi(k) (\cdots) dk$ with $\int_0^{k_{\rm max}} \phi(k) (\cdots) dk$ 
 to evaluate the cut-off effect.

In the top panel of Fig.~\ref{fig:cross-mch}, we illustrate the impact 
of the momentum cut-off $k_{\rm max}$  on the total cross section.
The solid line represents the result obtained without the cut-off (i.e. $k_{\rm max}\to\infty$), 
while the dashed line corresponds to  
the result obtained with the cut-off value of $k_{\rm max}=500$ MeV.
Clearly, the high momentum tail of the $J/\psi$ wavefunction 
is crucial for fitting the data,  
particularly the JLab data~\cite{GlueX-19,GlueX-23} in the low-energy region.
We also note that using the Gaussian wavefunction determined by $J/\psi\rightarrow e^+e^-$ of Ref.~\cite{lso23} results in much larger cross sections, 
and fitting to  the JLab data requires a much smaller value of $\alpha$.

As one can see from Eq.~(\ref{eq:v-vn-f}),  the $J/\psi$-N potential is determined by 
the form factor $F_V(t)$, which is obtained from the convolution of the initial and final state $J/\psi$ wavefunctions.
In the bottom panel of Fig.~\ref{fig:cross-mch}, we illustrate the 
impact of $F_V(t)$ on the total cross section by comparing 
the results from including
the momentum-dependent form factor $F_V(t)$ (solid line) and
setting  $F_V(t)=1$ (dashed line). 
It is evident that the effect of $F_V(t)$ on the total cross section is significant, underscoring the importance of using a realistic $J/\psi$ wavefunction 
in determining the $J/\psi$-N interaction  and the FSI effects.

\subsection{The 2Y model}

\begin{figure}
\centering
\includegraphics[width=0.9\columnwidth,angle=0]{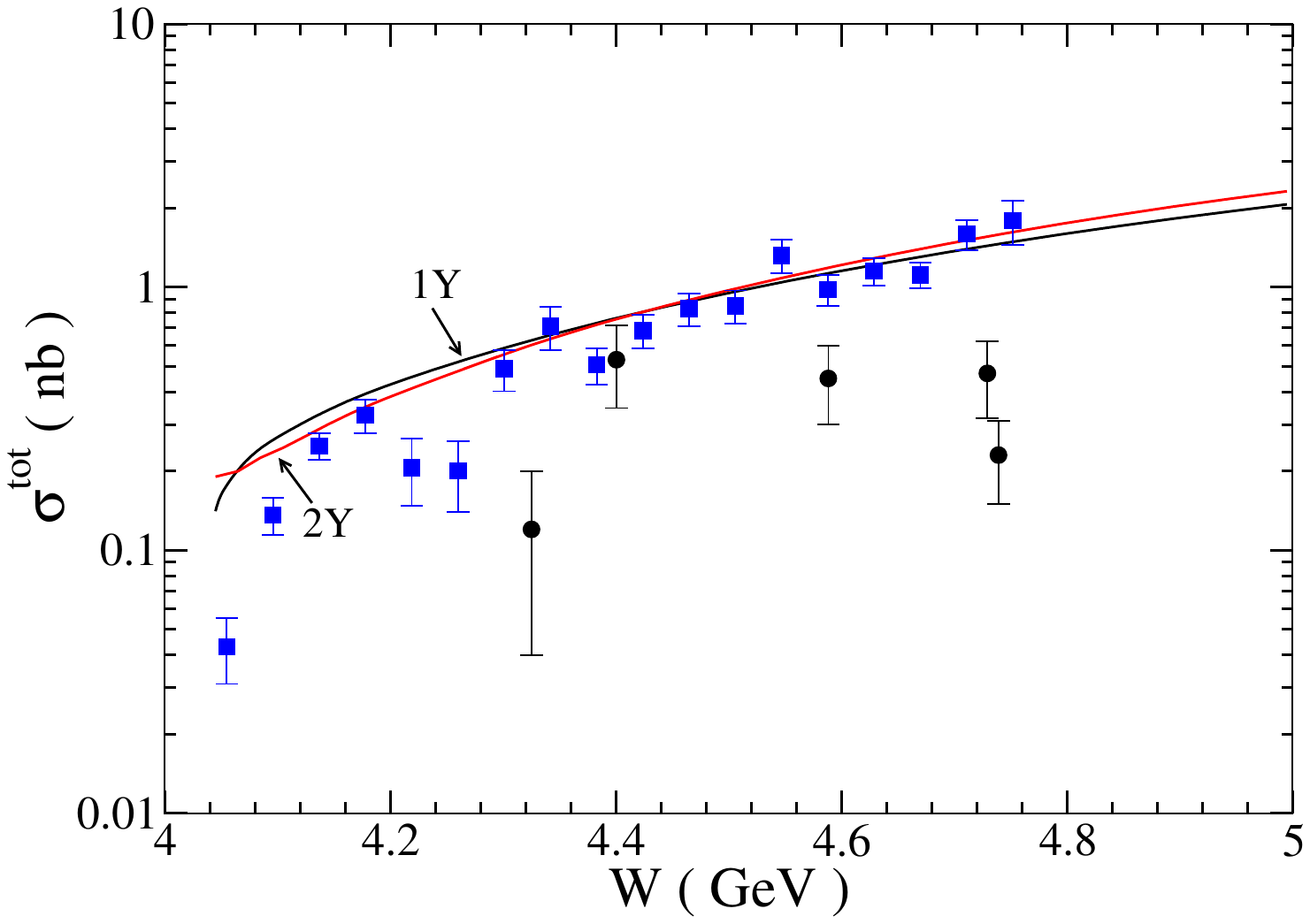}
\includegraphics[width=0.9\columnwidth,angle=0]{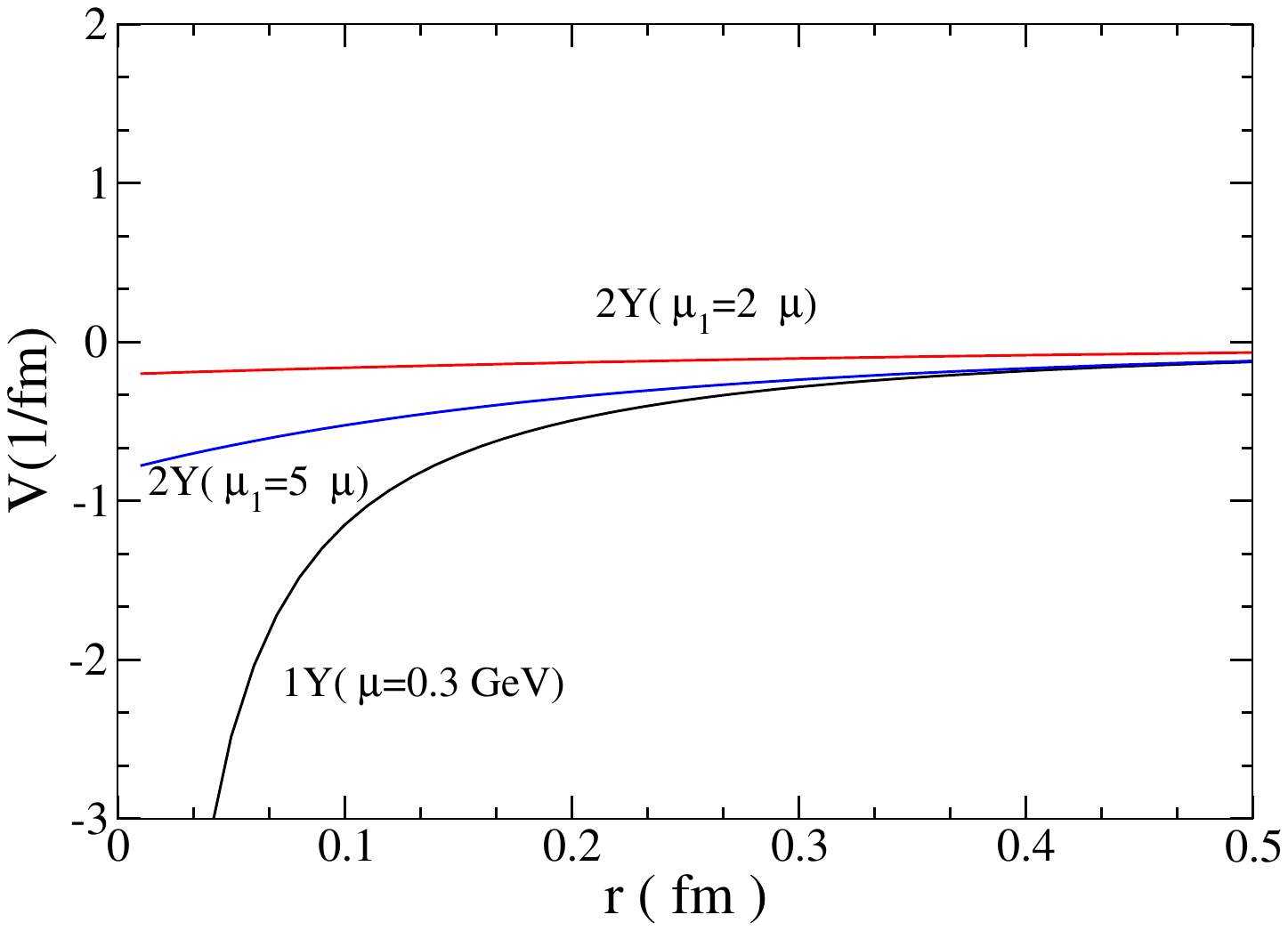}
\caption{ Top panel: Total cross sections from 1Y model,
  and 2Y model. Bottom panel: Short-range behavior 
of $v_{cN}(r)$ for the 1Y and 2Y models.The same
  data points are used as in Fig.~\ref{fig:totcrst-pom}.}
\label{fig:totcrst-paper-5}
\end{figure}

\begin{figure*}[t]
\centering
\includegraphics[width=0.68\columnwidth,angle=0]{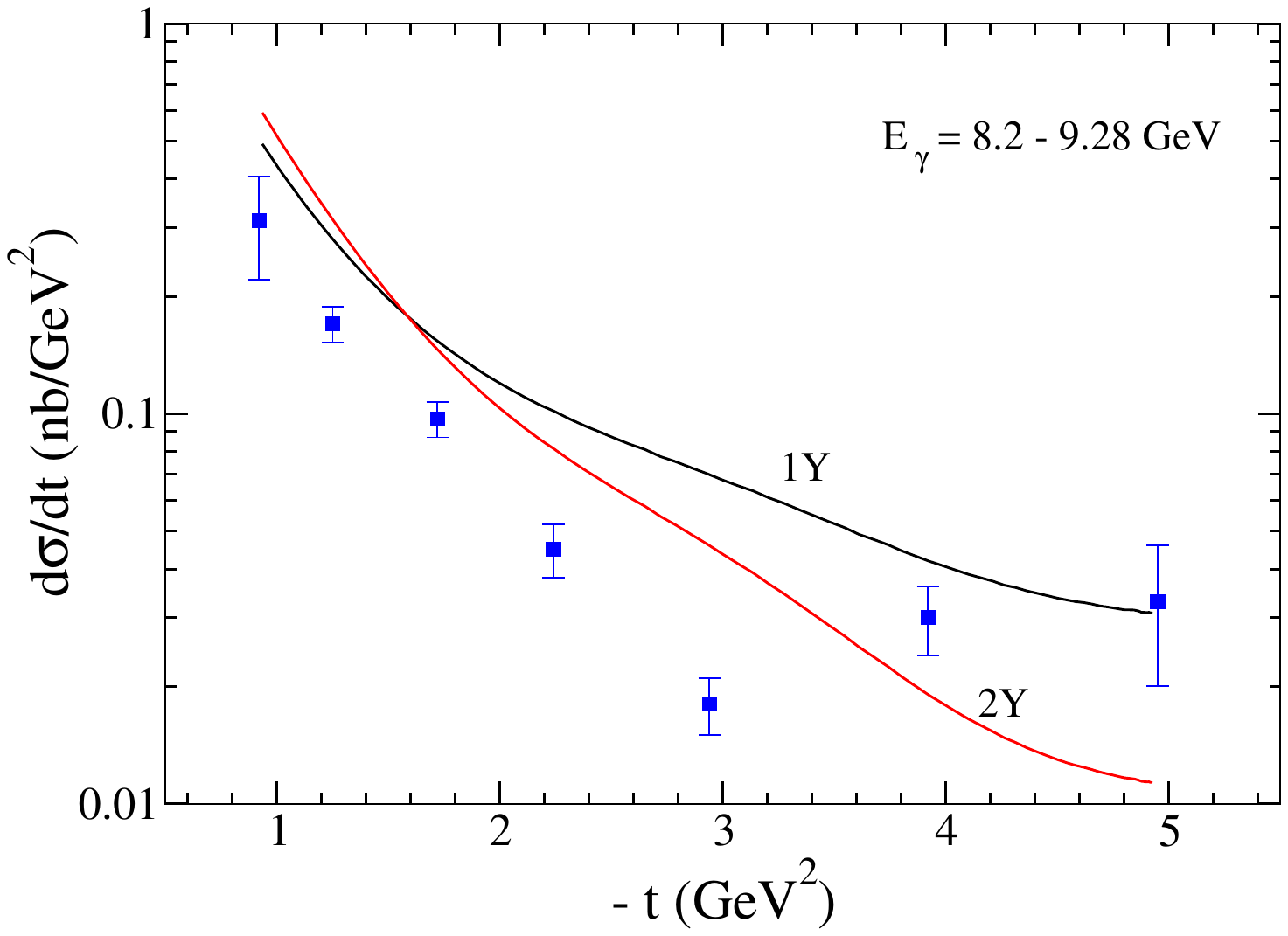}
\includegraphics[width=0.68\columnwidth,angle=0]{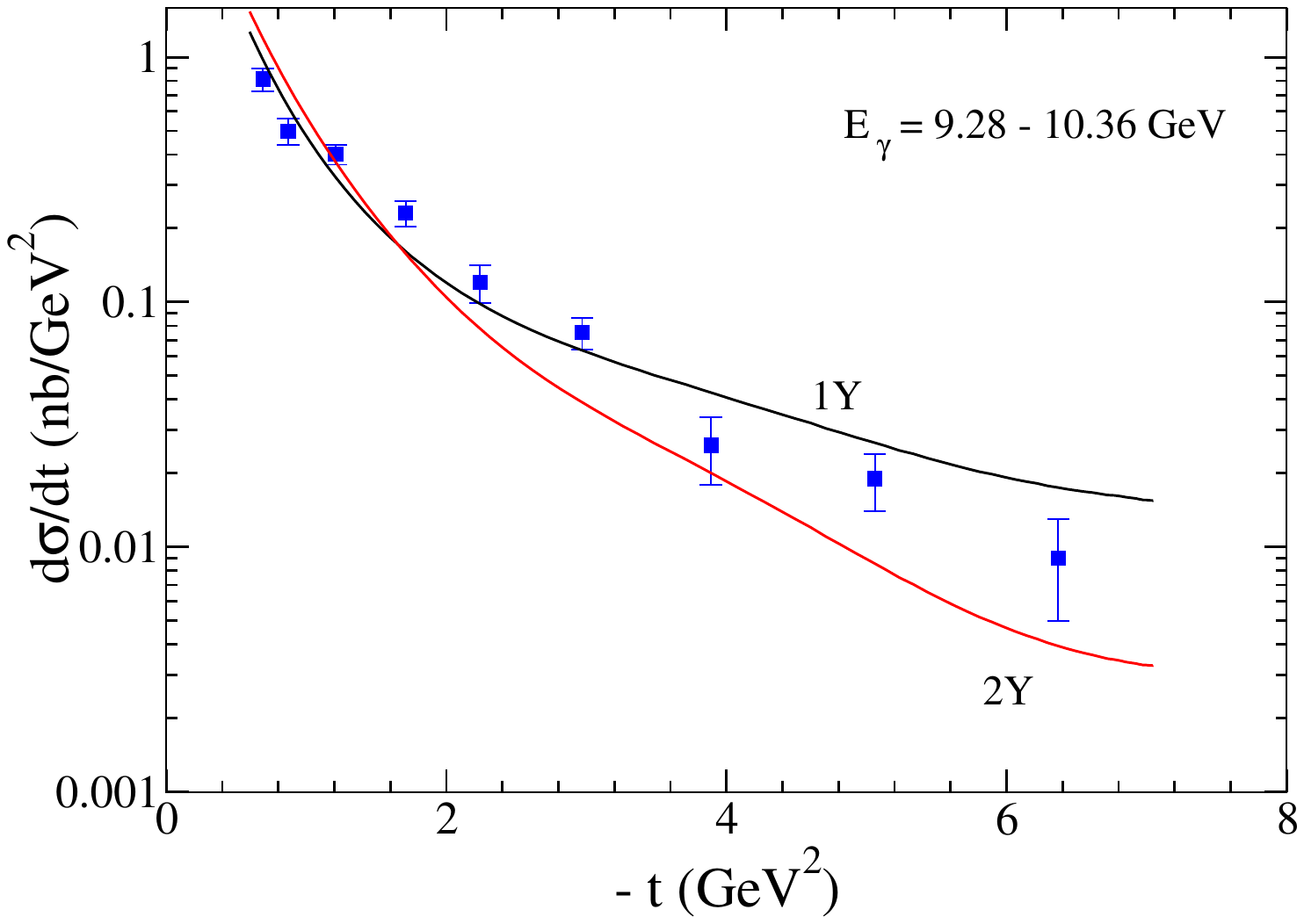}
\includegraphics[width=0.68\columnwidth,angle=0]{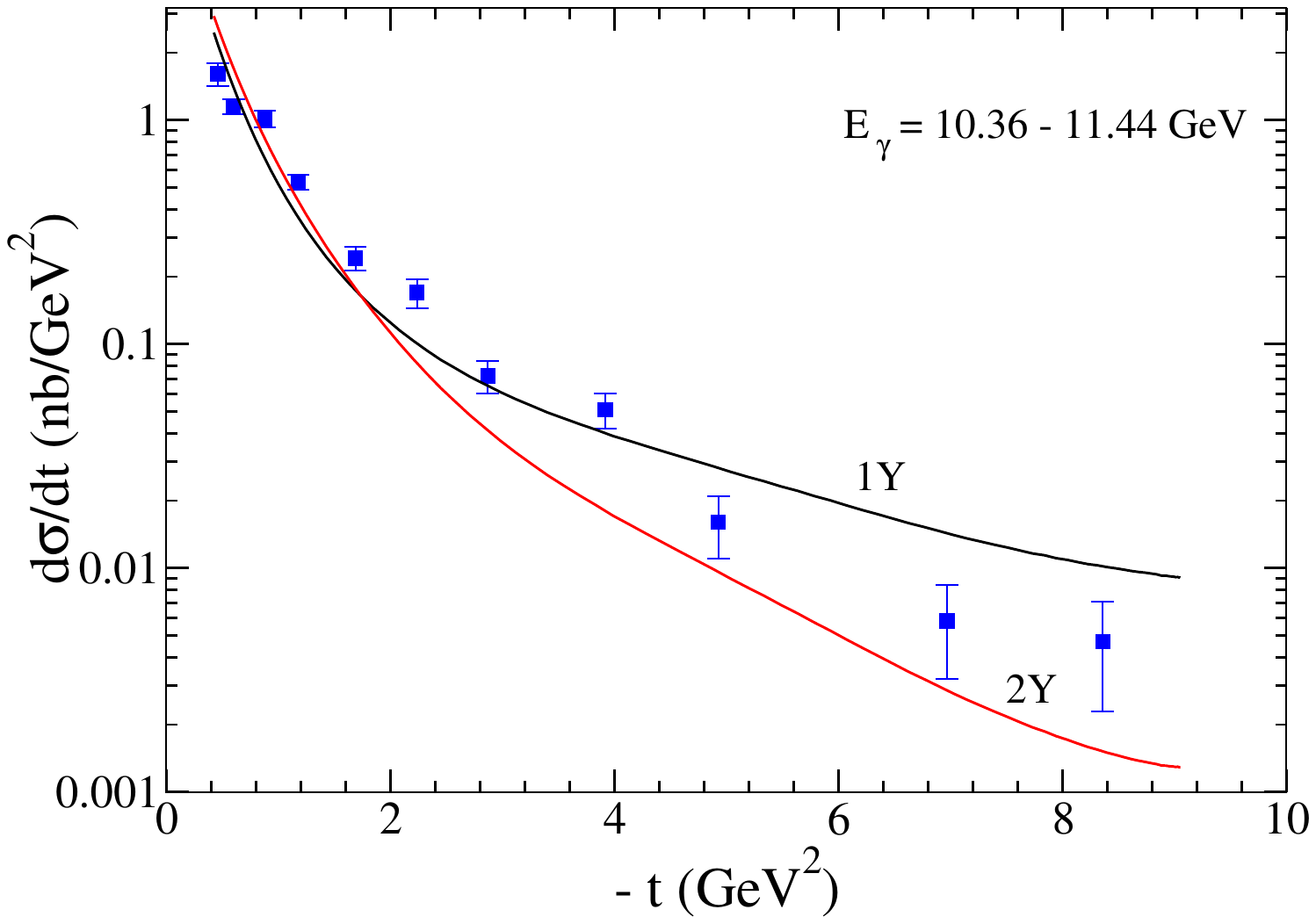}
\includegraphics[width=0.68\columnwidth,angle=0]{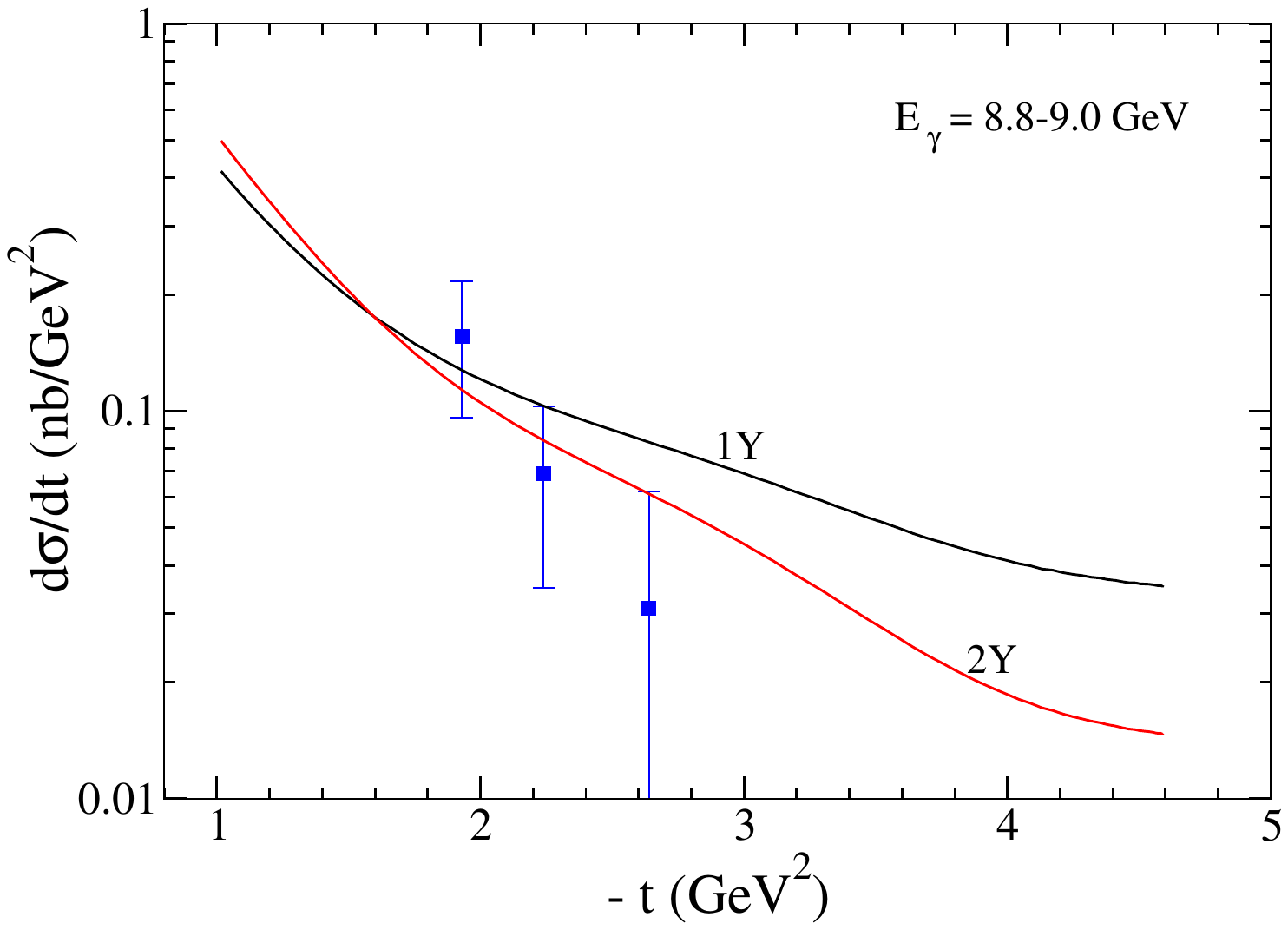}
\includegraphics[width=0.68\columnwidth,angle=0]{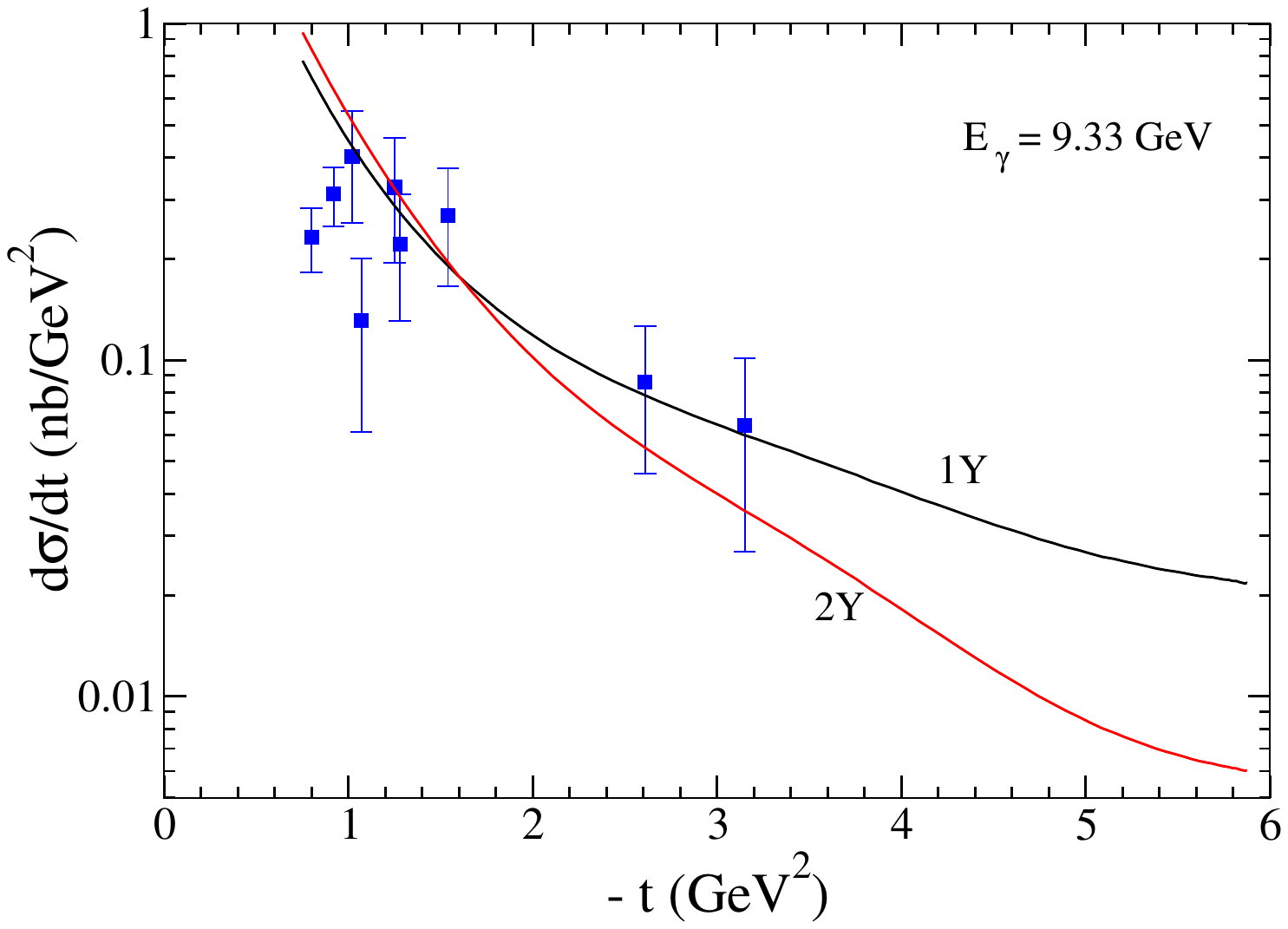}
\includegraphics[width=0.68\columnwidth,angle=0]{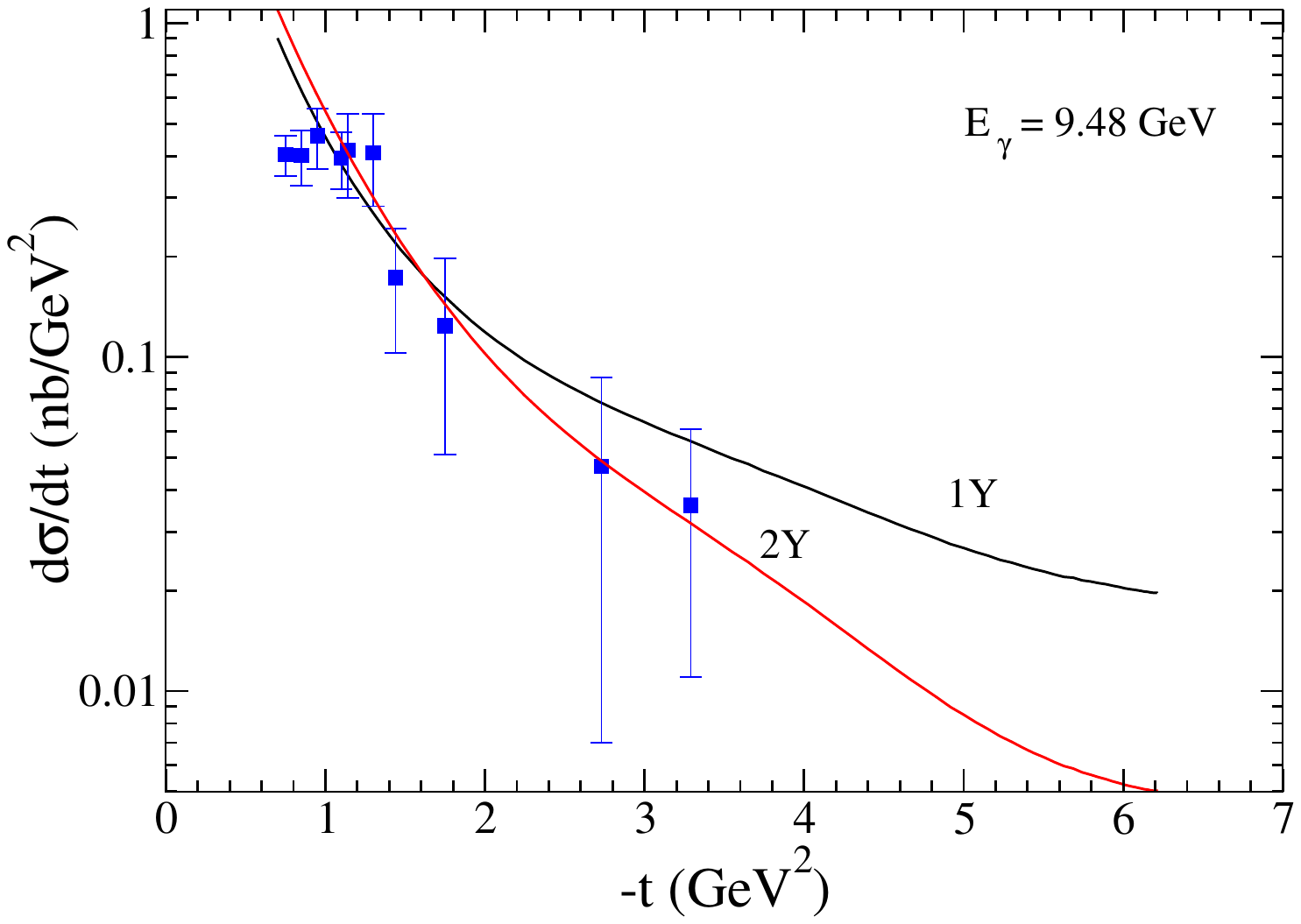}
\includegraphics[width=0.68\columnwidth,angle=0]{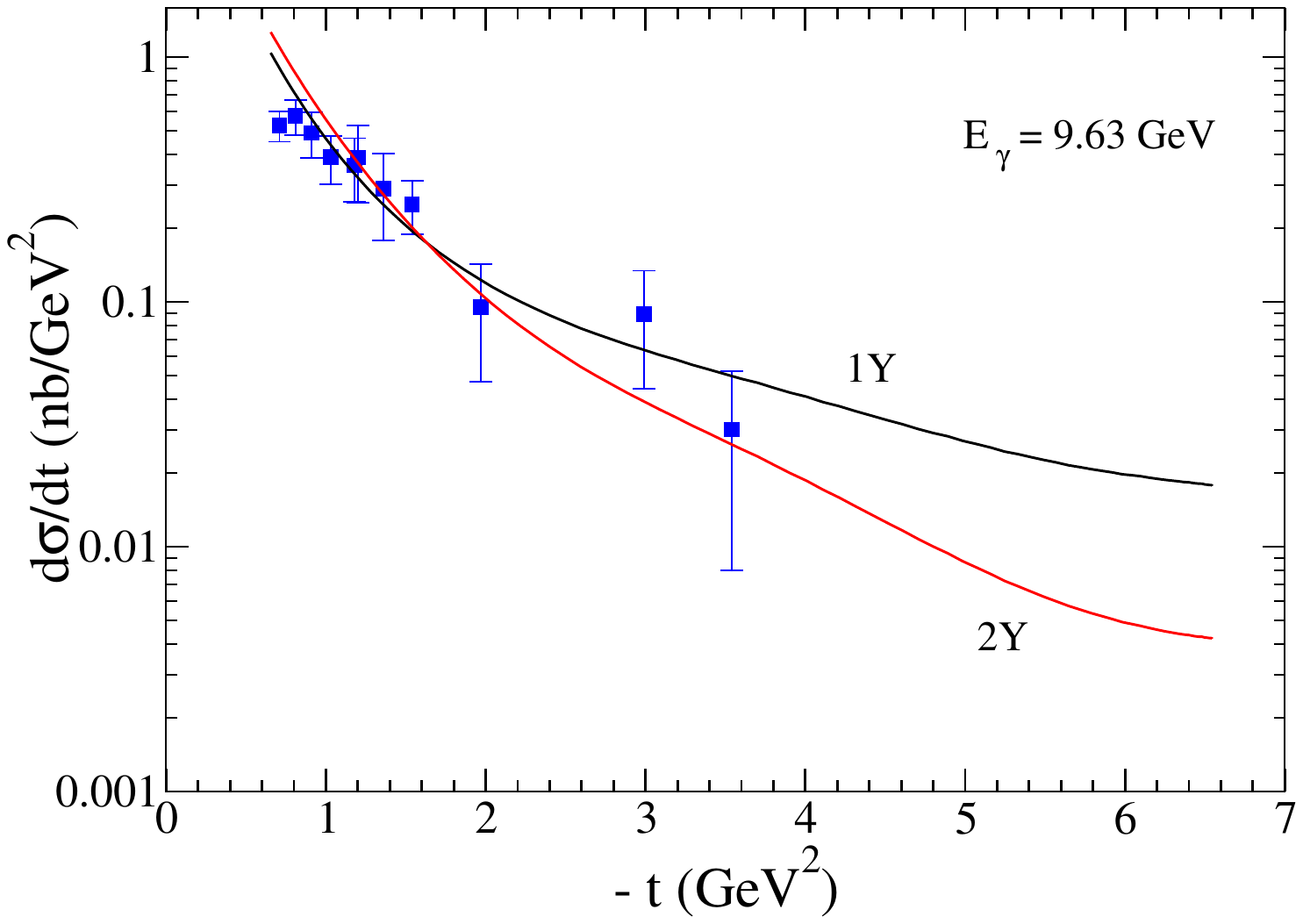}
\includegraphics[width=0.68\columnwidth,angle=0]{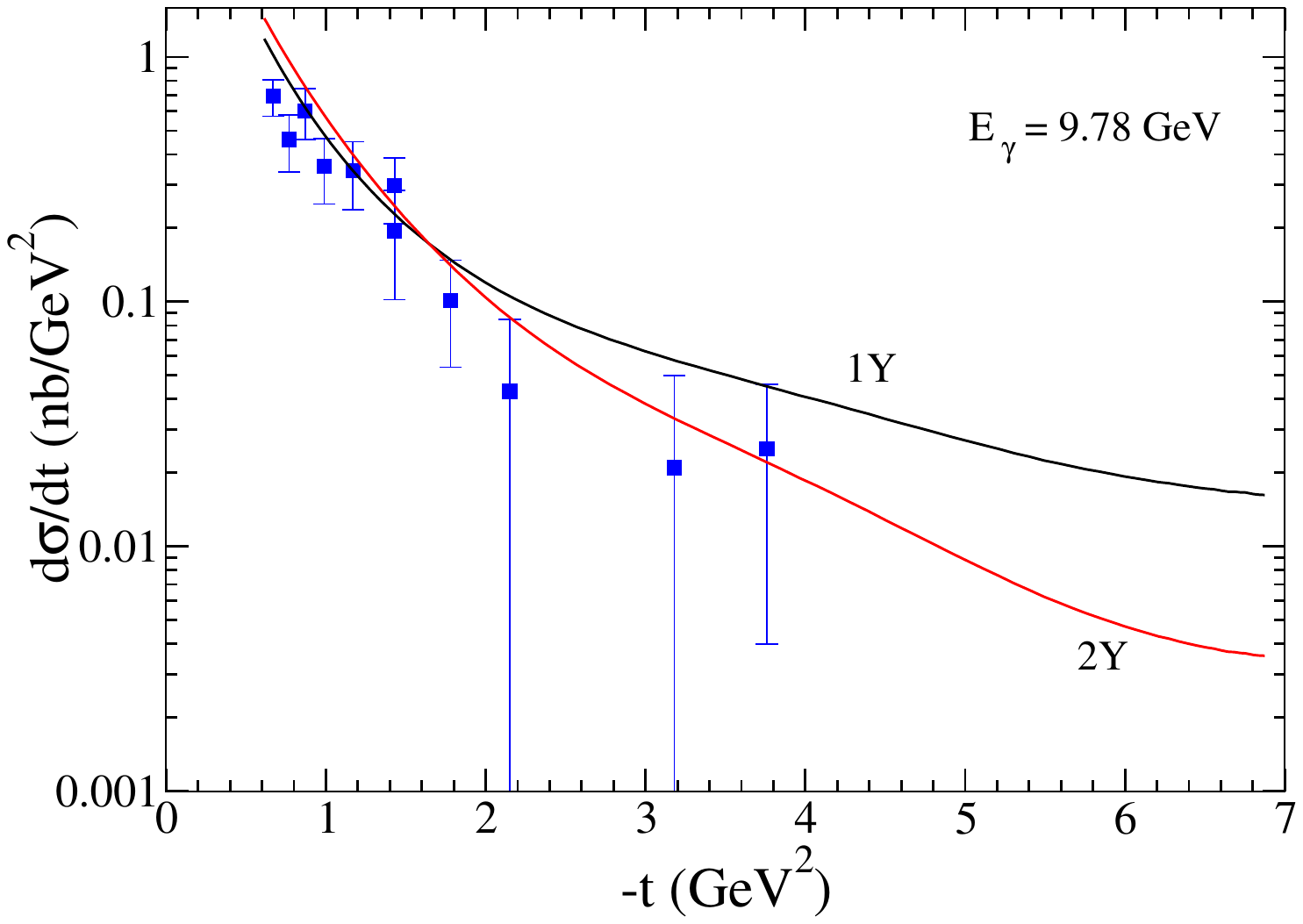}
\includegraphics[width=0.68\columnwidth,angle=0]{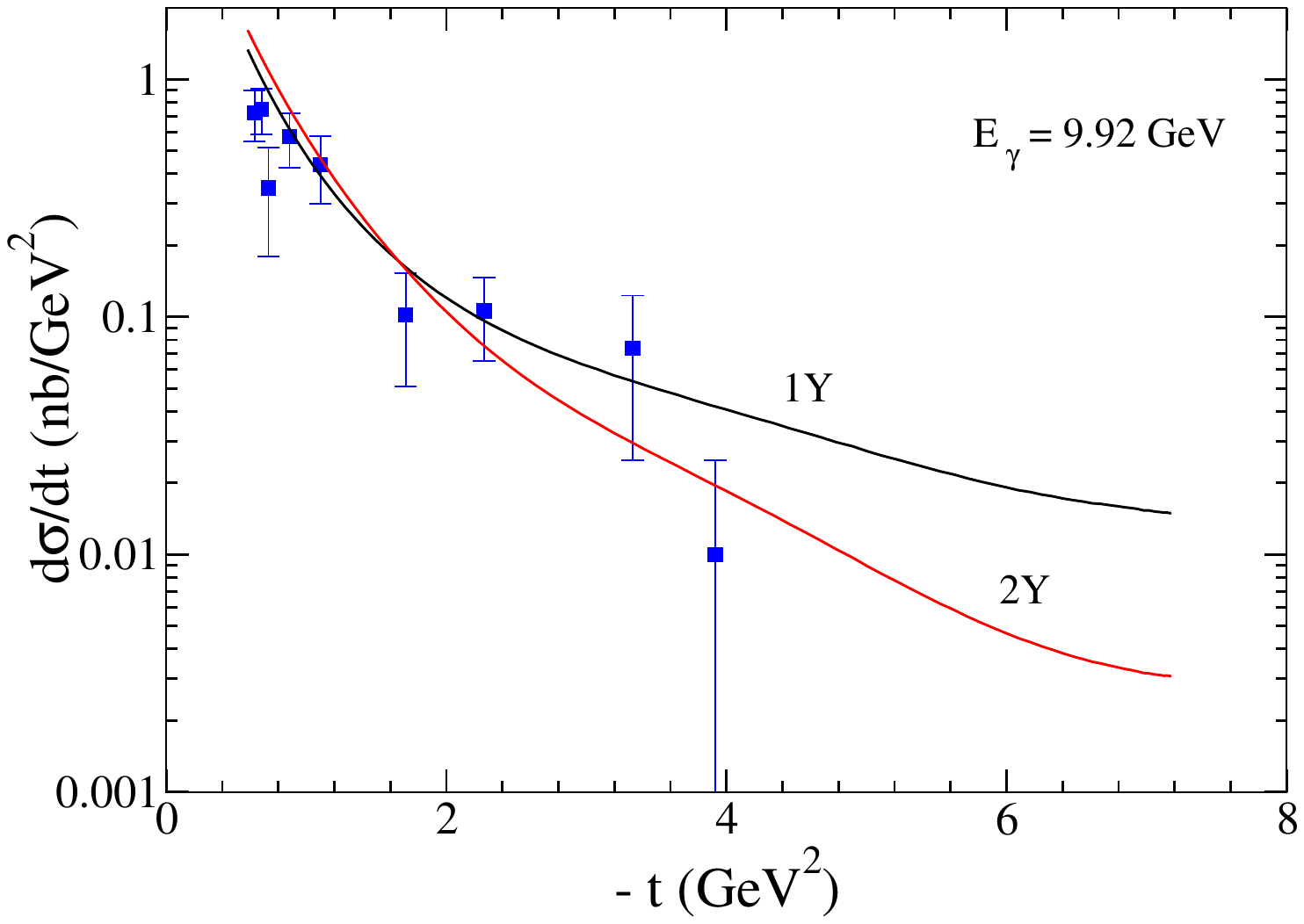}
\includegraphics[width=0.68\columnwidth,angle=0]{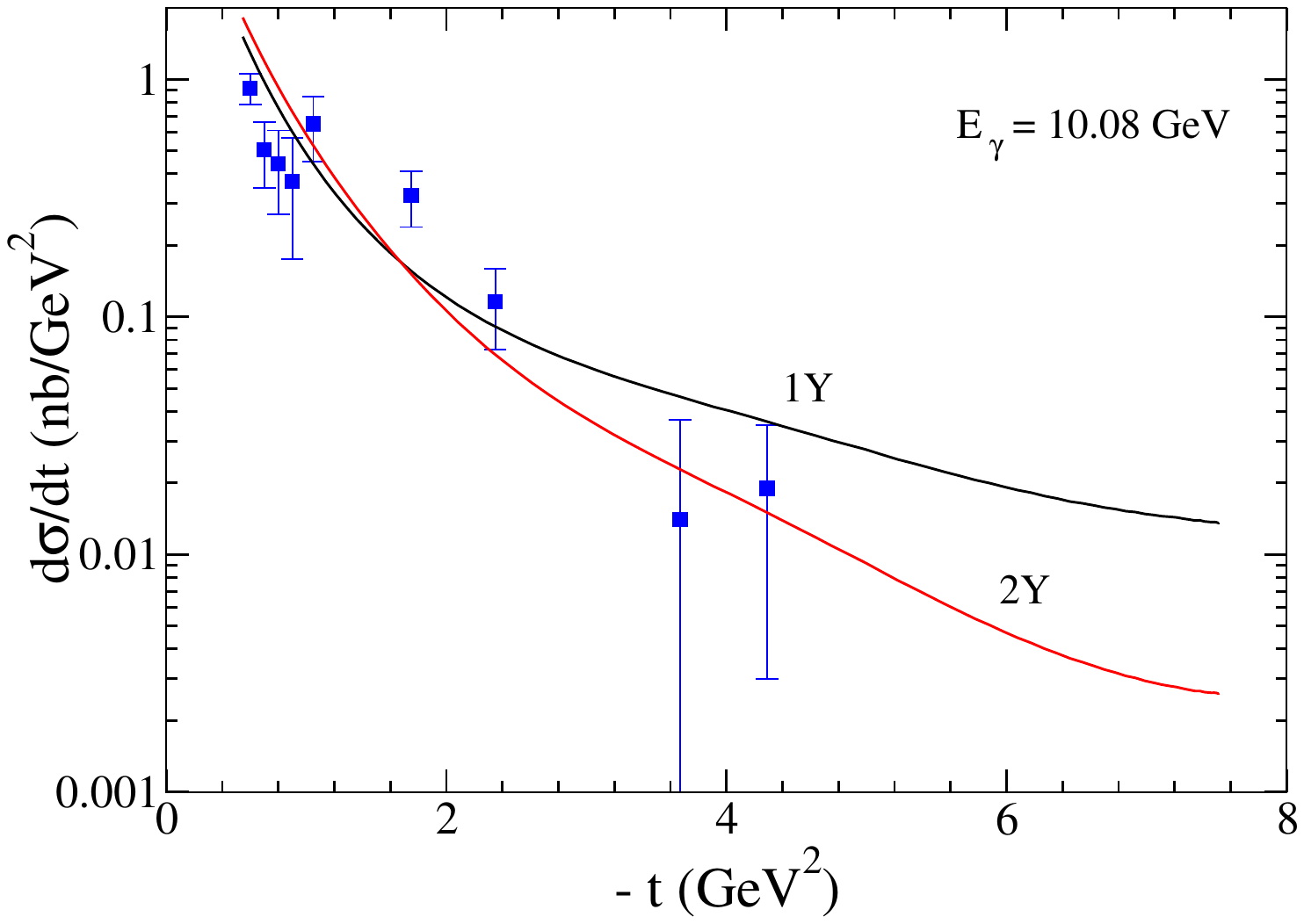}
\includegraphics[width=0.68\columnwidth,angle=0]{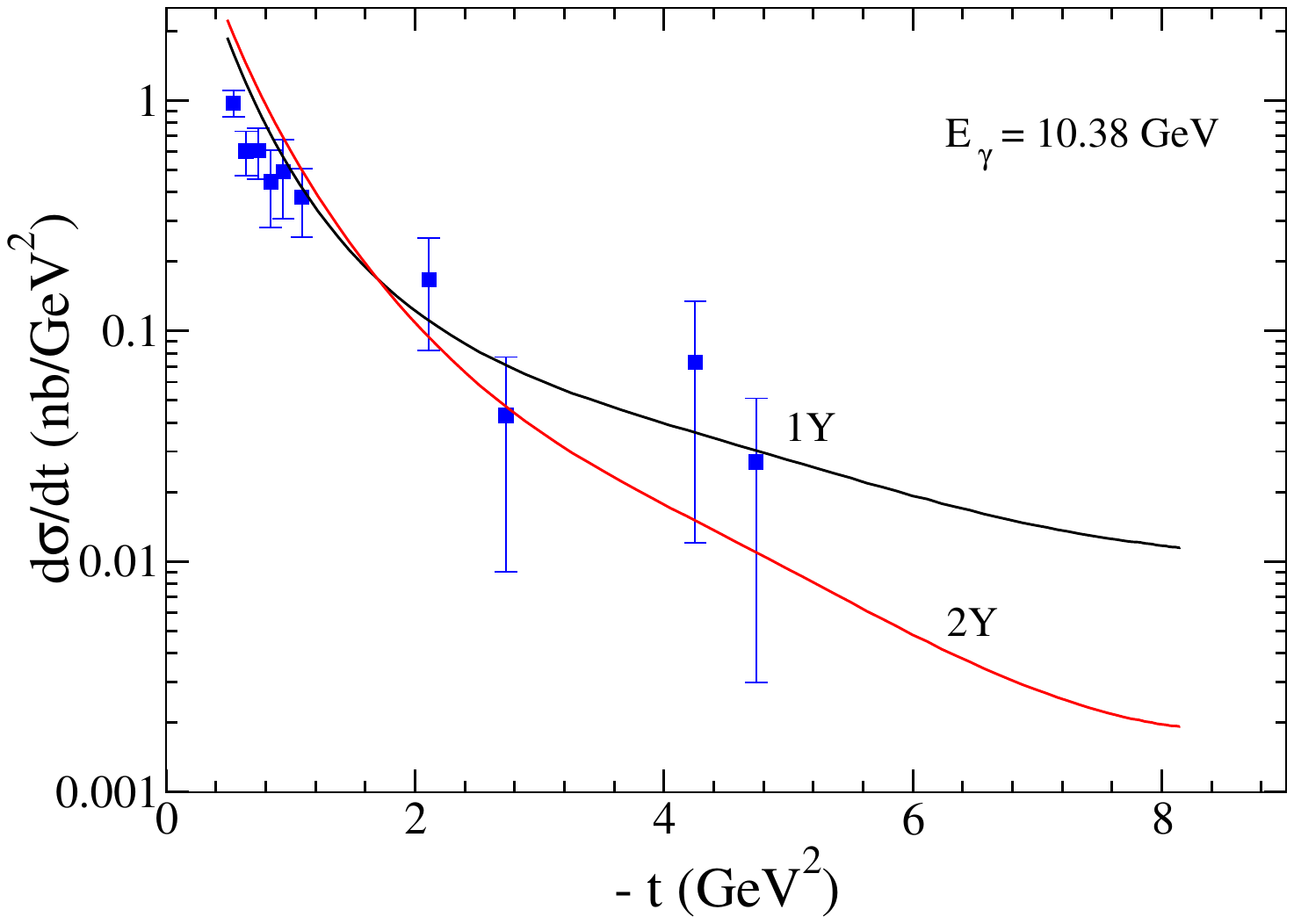}
\includegraphics[width=0.68\columnwidth,angle=0]{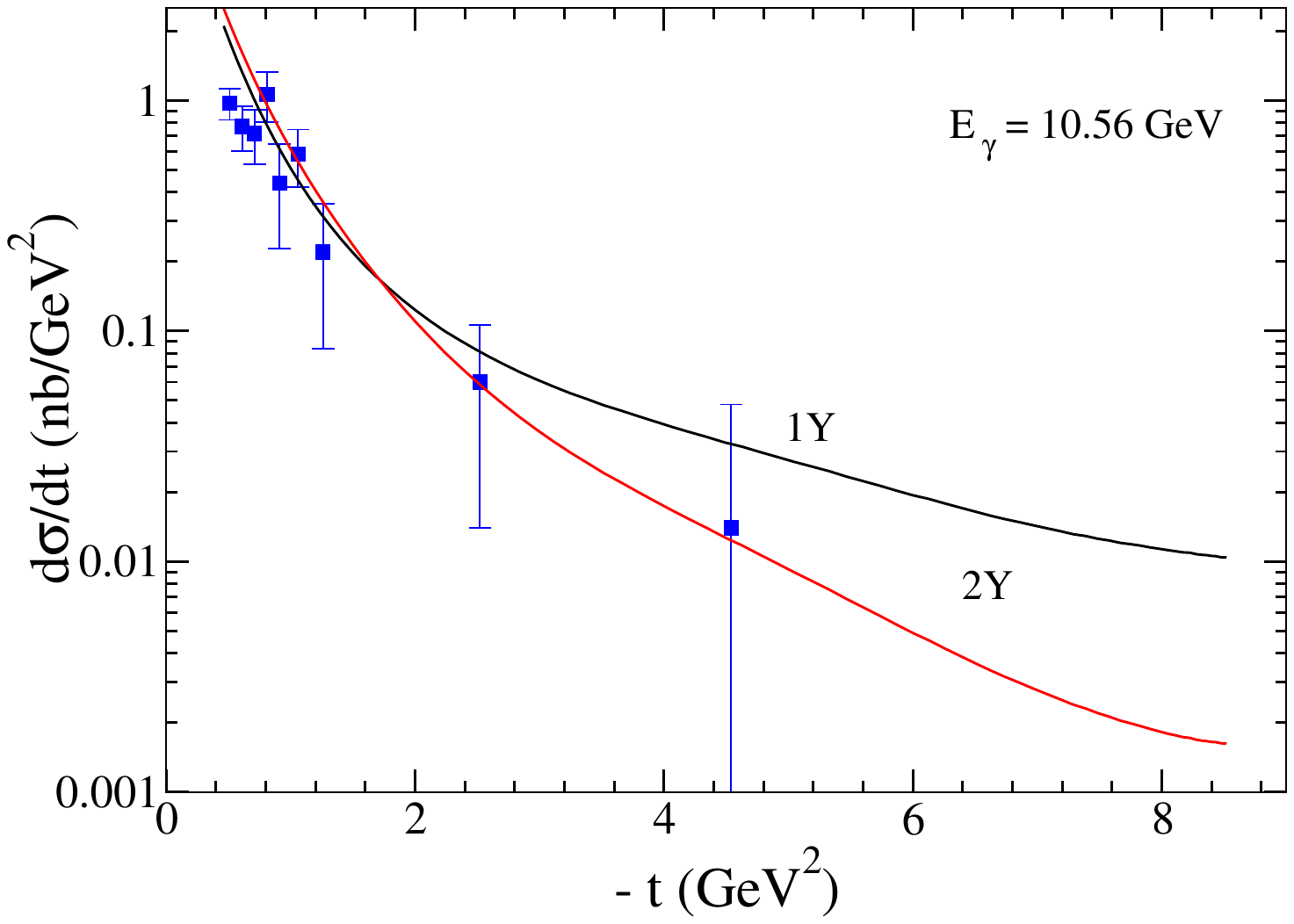}
\caption{Differential  cross sections from the 1Y model  and 2Y model.
The data in the top three panels are 
 taken from the  GlueX  Collaboration~\cite{GlueX-23}. The other data  are
taken  from an experiment at JLab's Hall-C~\cite{jlab-hallc}.}. \label{fig:dsdt-low1Y2Y}
\end{figure*}
\begin{figure}[t]
\centering
\includegraphics[width=0.9\columnwidth,angle=0]{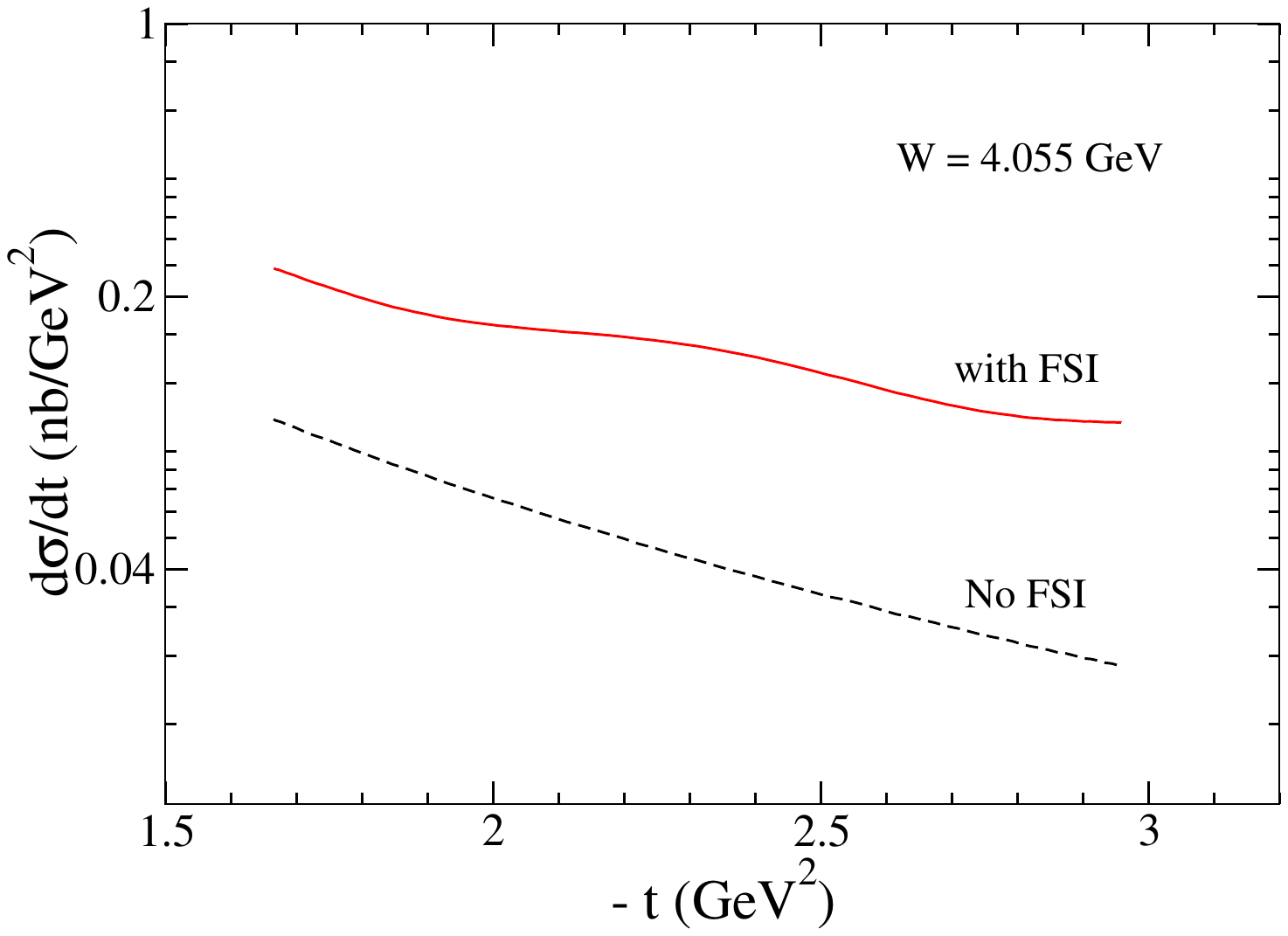}
\caption{Prediction of the differential cross sections near threshold at $W= 4.055$ GeV by using 2Y model.}
\label{fig:dsdt-thres0}
\end{figure}
 We now consider  the model ($2Y$) by incorporating the 
second term with  $c_s=1$ in addition to the first term in Eq.~\eqref{eq:vcn-r}.
The cut-off of the short-range  part is  set as $\mu_1=N\mu$.
We observe that for large values of $N > 50$, the 2Y model closely approximates the 1Y model, 
and the JLab data can be fitted with any $N < 20$ by adjusting the coupling constant $\alpha$.
 The result of the total cross section  from the 2Y model  with $N=5$  and $\alpha=-0.145$  is  comparable to that of  the 1Y model
with $\alpha=-0.067$, as shown in 
 the top panel of Fig.~\ref{fig:totcrst-paper-5}.  
The two models fit the JLab 
data~\cite{GlueX-23,jlab-hallc} equally well, but have very large difference at very near threshold, which
is dominated completely by FSI. The origin of  it is that  they have very different short range 
 behaviors of the potentials  $v_{cN}(r)$, 
as shown in the bottom panel.\footnote{To compare the potential for the 1Y model (black line),
we evaluate for the 2Y model with $N=2$ (red line) and $N=5$ (blue line), using the same $\alpha=-0.145$ for both $N$ values.
This ensures clear comparisons of their $r$-dependence at short distances.}
We thus expect 
 that their predictions on the  differential cross sections must
be  large at large $t$.
This is shown in Fig.~\ref{fig:dsdt-low1Y2Y}, where the differential cross sections as 
a function of $t$ from the 1Y model with $\alpha=-0.067$ and the 2Y 
model with $\alpha=-0.145$ and $N=5$ are compared. 
Both models reasonably describe the JLab data from the GlueX Collaboration~\cite{GlueX-23} and Hall-C~\cite{jlab-hallc}.
We observe notable differences between the predictions of the 1Y and 2Y models
 at large values of $t$. More precise data are clearly
needed for making
 further progress, while the 2Y model appears to be better.
For future experiments at very near threshold, we present in Fig.~\ref{fig:dsdt-thres0}
the differential cross sections at $W=4.055$ GeV. 
 We compare the predictions of the 2Y model 
 with (solid line) and without (dashed line) including the FSI.

\subsection{LQCD constraints}
\begin{figure}
\begin{center}
\includegraphics[clip,width=0.4\textwidth,angle=0]{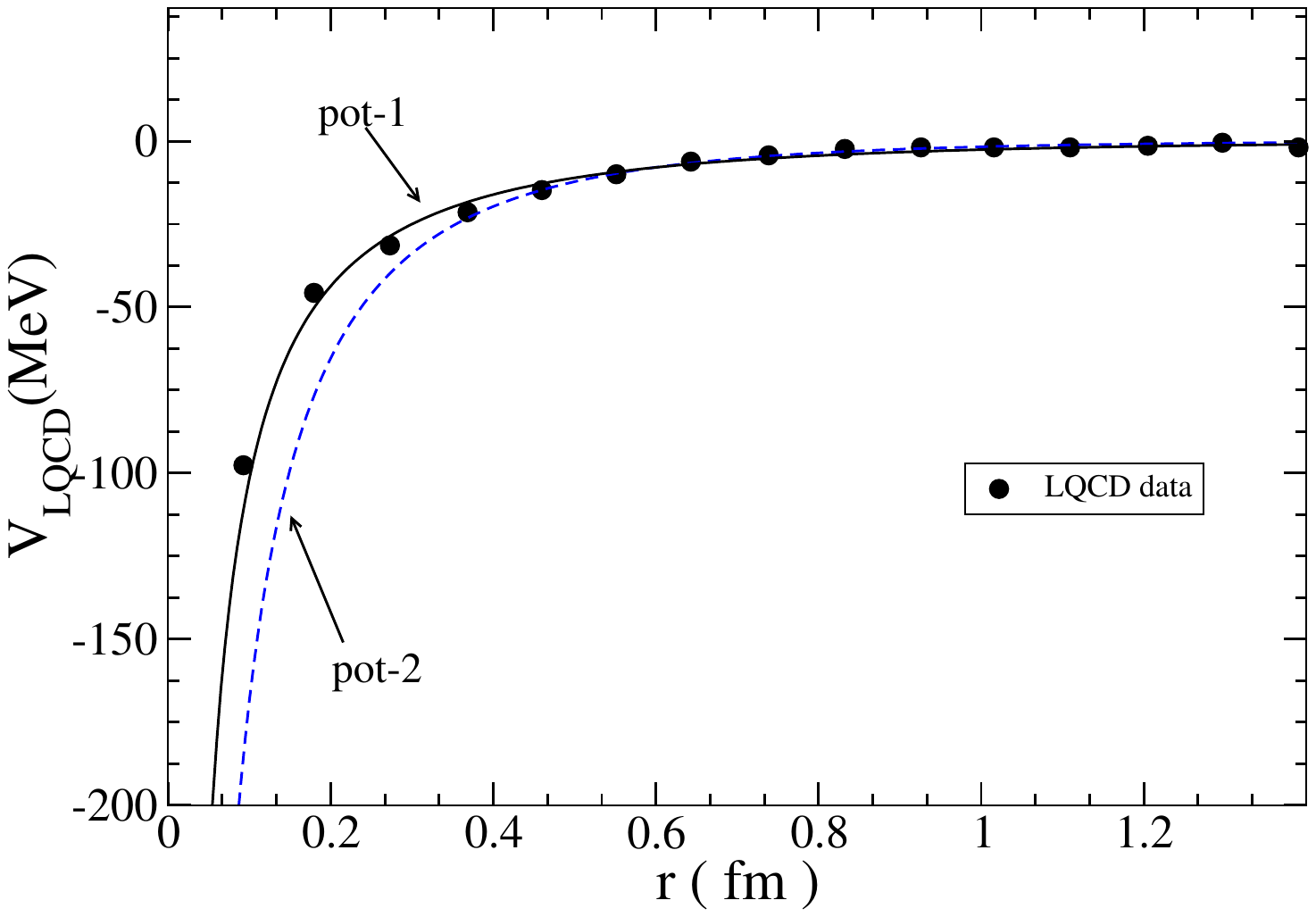}
\includegraphics[clip,width=0.4\textwidth,angle=0]{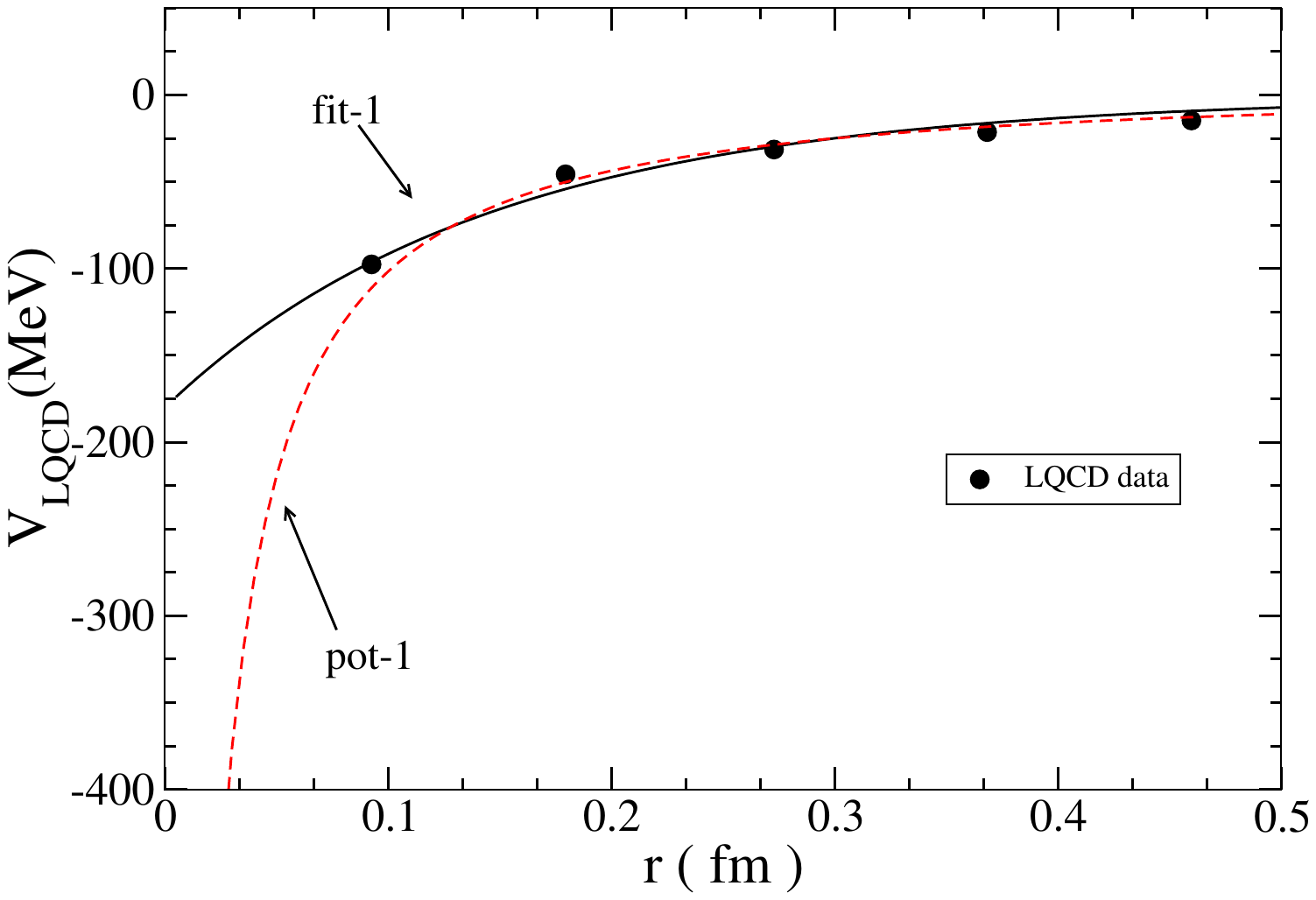}
\caption{Two different forms of  $J/\psi$-N potential $V^{\rm LQCD}(r)$: $\al_L\frac{e^{-\mu_L r}}{r}$ (top panel)
and $\alpha_{L}(\frac{e^{-\mu_{L} r}}{r}-\frac{e^{-N_L\mu_{L} r}}{r})$ (bottom panel),
obtained from fitting the LQCD Data~\cite{KS10b,sasaki-1}.}
\label{fig:v-lqcd}
\end{center}
\end{figure}
\begin{figure}
\centering
\includegraphics[clip,width=0.9\columnwidth,angle=0]{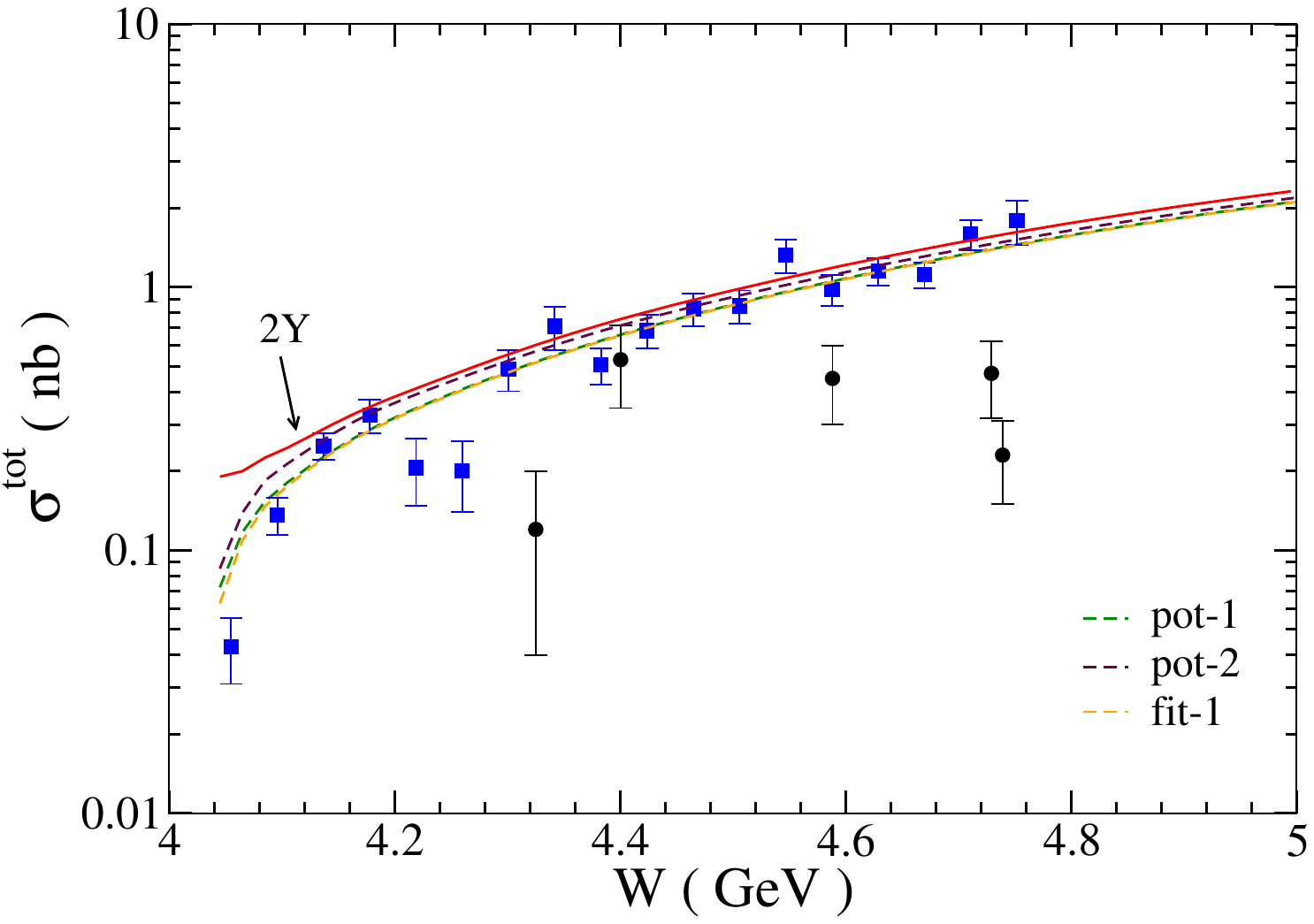}
\includegraphics[clip,width=0.9\columnwidth,angle=0]{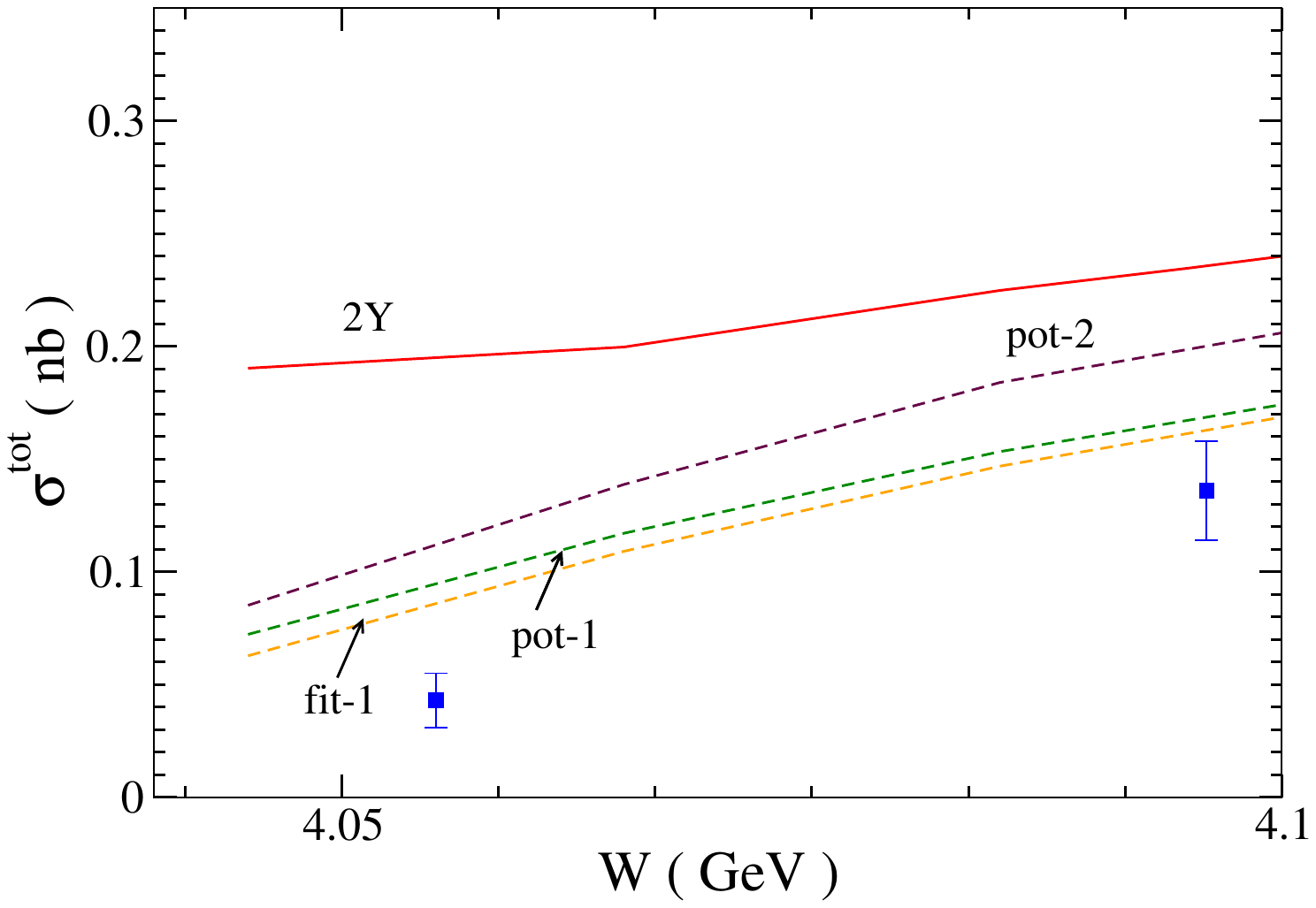}
\caption{ Total cross-sections of $J/\psi$ calculated 
from the three models pot-1, pot-2, and fit-1 constrained by the LQCD data.
Top panel: for $4\leq W\leq 5$ GeV. 
Bottom panel: for the near threshold region. 
The same  data points are used as in Fig.~\ref{fig:totcrst-pom}.}
\label{fig:totcrst-lqcd-tsh}
\end{figure}

We  have found that the  $J/\psi$-N potentials resulting  from fitting the JLab  data in the previous section 
give the  $J/\psi$-N scattering lengths, $-a\sim 0.3 - 0.7$ fm, which are rather  different from those extracted  
from the LQCD data of Ref.~\cite{KS10b}.
Thus, it raises the question of our parametrization of  the quark-N potential $v_{cN}(r)$. At low energies, the $J/\psi$-N system
has a small relative momentum, and the multi-gluon exchange involving two quarks in  $J/\psi$ may play important roles. Therefore,
we need to extend  our parametrization of $v_{cN}$ to include two-body operator.
However, we will not explore this possibility in this exploratory work because the model would then have too many 
parameters, which cannot be determined by the still very limited data in the near threshold
region. Instead, we next consider  the models  constructed
by imposing LQCD constraints on the calculations of the FSI. 
This is done by   using  the LQCD data of Refs.~\cite{KS10b,sasaki-1}
to determine the parameters of $v_{cN}(r)$  in calculating the matrix elements of the
 $J/\psi$-N potential $V_{VN,VN}$ defined by  Eqs.~(\ref{eq:vjpsin})-(\ref{eq:yukawa}), 
while the  Born term  amplitude $B_{\gamma N, VN}$  from the 2Y model described in  the previous
section is kept in the  calculations. This allows us to investigate the extent  to which
the extracted  $J/\psi$-N scattering amplitudes can be  related to the  LQCD data of Refs.~\cite{KS10b,sasaki-1}.

The  LQCD  data  of Ref.~\cite{sasaki-1}  for the $J/\psi$-N potential
have large uncertainties at small distances $r$. 
Therefore, we fit  the LQCD data for the region of large $r$ using
the  following form of $J/\psi$-N potential:
$V^{{\rm LQCD}}(r)=\alpha_{L}\frac{e^{-\mu_L r}}{r}$ with two different sets of parameters $(\alpha_{L}, \mu_{L})$,
namely $(\alpha_{L}, \mu_{L})=(-0.06, 0.3)$ and $(-0.11, 0.5)$, which we denote as ``pot-1" and ``pot-2", respectively.

The results of $V^{{\rm LQCD}}(r)=\alpha_{L}\frac{e^{-\mu_L r}}{r}$
obtained from the ``pot-1" and ``pot-2" are shown
in the upper panel of Fig.~\ref{fig:v-lqcd}. As one can see from the figure,
their short-range parts are rather different, while their long-range parts (i.e. $r\geq 0.6$ fm) are almost similar.
Following the derivations of Eqs.~(\ref{eq:v-vnvn-0})-(\ref{eq:v-ff}), one can see that
the matrix element of $V^{\rm LQCD}(r)$ is of the form of Eq.~(\ref{eq:v-vn-f}) with $F_V(t)=1$ and
$v_{cN}(r)$ calculated from the  1Y model,  i.e.,
$v_{cN}(r)=\alpha\frac{e^{-\mu r}}{r}$  with
 $\alpha=1/2\alpha_{L}$ and  $\mu=\mu_{L}$.
Thus, the parameters of  $v_{cN}(r)$ for this FSI calculation  are 
$(\al\equiv\al_{FSI}=\al_{\rm L}/2,\mu\equiv\mu_L)=(-0.03,0.3)$ 
fixed from ``pot-1" and $(-0.055,0.5)$ fixed from ``pot-2".

In the lower panel of Fig.~\ref{fig:v-lqcd}, we also fit the  LQCD data using another form of $J/\psi$-N potential,
$V^{\rm LQCD}(r)=\alpha_{L}(\frac{e^{-\mu_{L} r}}{r}-\frac{e^{-N_L\mu_{L} r}}{r})$
with $(\alpha_L,\mu_L, N_L)=(-0.2,0.3,2)$, which we denote as ``fit-1" and compare it with the result of ``pot-1".
The result of ``fit-1" has a rather
different short range  behavior compared  with that of  ``pot-1". 
The FSI calculation  with this potential can then be done  by
 using the 2Y model  of
$v_{cN}(r)=\alpha(\frac{e^{-\mu r}}{r}-\frac{e^{-N\mu r}}{r})$ to evaluate  
Eq.~(\ref{eq:v-vn-f}) with $F_V(t)=1$ and  
 $( \al\equiv\al_{FSI}=\al_{\rm L}/2, \mu\equiv\mu_L,N\equiv N)=(-0.1,0.3,2)$.

With  the  three $J/\Psi$-N potentials, i.e. (pot-1, pot-2, fit-2), we then have three models.
They have
the same Born term $B_{\gamma N, VN}(E)$ from the 2Y model of the previous section,
 but their FSI amplitude
$T^{({\rm fsi})}_{\gamma N, V N}$ defined in  Eq.~(\ref{eq:tfsi-mx})
are different because the
$V N \rightarrow V  N$ amplitude $T_{V N,V N}$ depends on the chosen $J/\psi$-N potential.
The resulting total cross sections are presented 
in Fig.~\ref{fig:totcrst-lqcd-tsh} for $4\leq W\leq 5$ (GeV) region (top) and for the
near threshold region (bottom), respectively. For comparison, we also show
 the result obtained from the  2Y model of the previous section.
As depicted in the top panel of Fig.~\ref{fig:totcrst-lqcd-tsh},
 no large difference is observed between the models 
with LQCD constraints and the  2Y model in the higher  energy region $W\geq 4.2$ GeV region.
However, as illustrated in the bottom panel, the 2Y model
predicts  a  much larger cross section in the very near threshold region $W \leq 4.1$ GeV.
These   results suggest  that imposing the  LQCD constraints significantly 
changes the threshold behavior and the predicted cross sections
are  closer to the JLab data~\cite{GlueX-19,GlueX-23} than the 2Y model.

As illustrated in Fig.~\ref{fig:dsdt-thres0}, we see that the cross sections near the threshold
shown  in the lower panel of Fig.~\ref{fig:totcrst-lqcd-tsh}
are mainly due to the  FSI  term $T^{({\rm fsi})}_{\gamma N, VN}$ given by Eq.~(\ref{eq:tfsi-mx}), which
depends on both  the Born amplitude $B_{\gamma N, VN}$ and the  
$VN\rightarrow VN$ amplitude $T_{VN,VN}$. 
Thus, the  large difference between 2Y and three models with the LQCD constraints could be reduced  if
each model's Born  term is  refined to fit  the data at higher energies as well as the 2Y model.  
To examine this  model dependence of our predictions, 
we next construct three  models by modifying  the parameter 
$\alpha\rightarrow  \alpha_B$ of the quark-nucleon potential
$v_{cN}(r)$ used in the calculation of the Born term $B_{\gamma N, VN}$ to fit the total cross 
sections from the  2Y model for $W\geq 4.1$ GeV.
 We denote the resulting models 
 as (a), (b), and (c) for
the models pot-1, pot-2, and fit-1, respectively. 
The determined parameter $\alpha_B$ along with the other parameters of
the three  models  are listed in Table~\ref{tab:tab-1}. 
Note that the scattering length $a$ is calculated solely from the
$VN\rightarrow VN$ amplitude $T_{VN,VN}(W)$.
Our predictions for $a$ obtained from models (a), (b), and (c) are
($-0.15, -0.233, -0.057$) fm, respectively, while we obtain $a=-0.67$ fm from the 2Y model.
These different scattering lengths are related to the differences in the total cross sections between the four models.
It will be interesting to have data to distinguish these predictions and test LQCD. 
This would allow for a comparison between the predicted scattering lengths and the actual experimental results.

\begin{figure}
\centering
\includegraphics[width=0.9\columnwidth,angle=0]{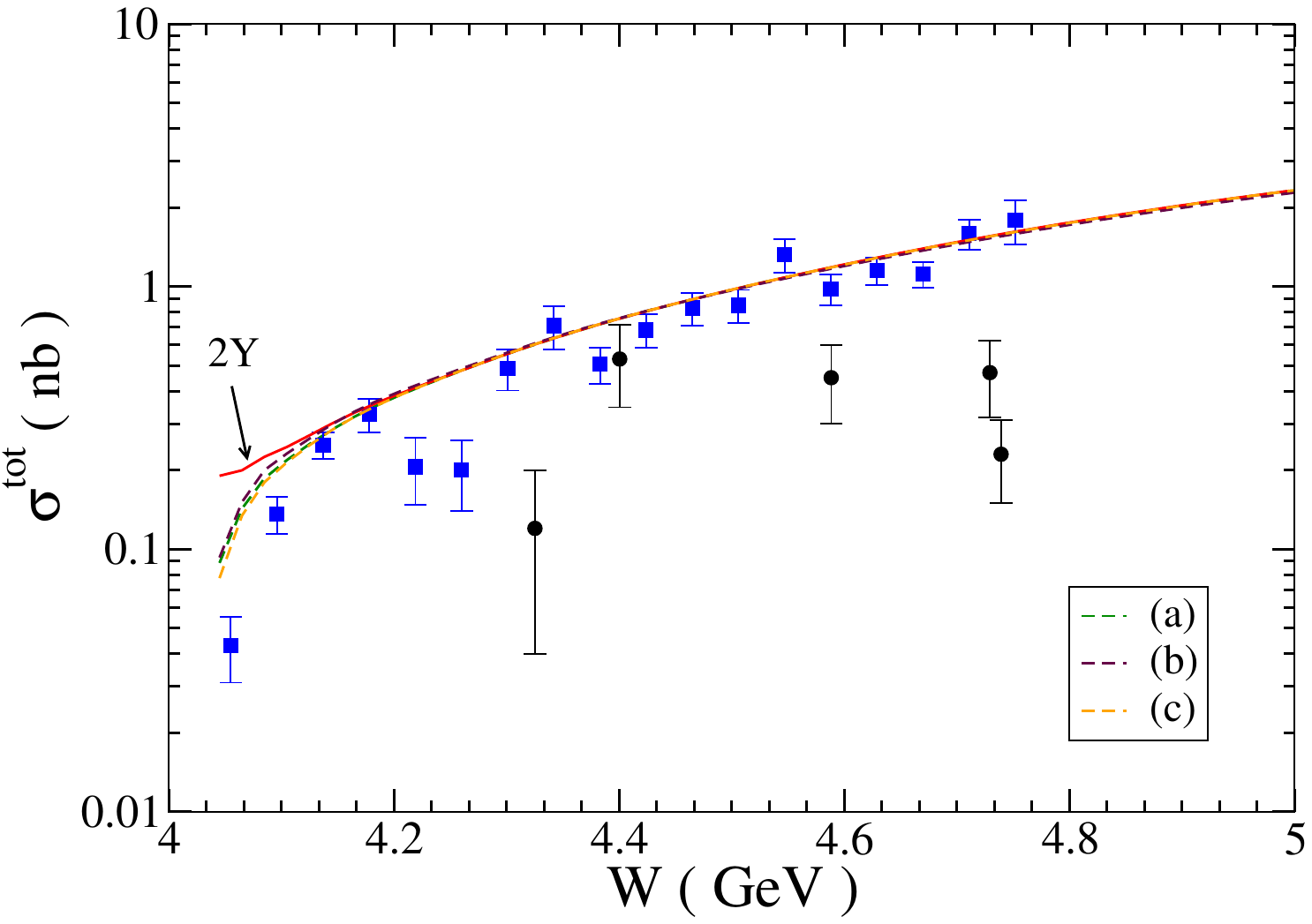}
\includegraphics[width=0.9\columnwidth,angle=0]{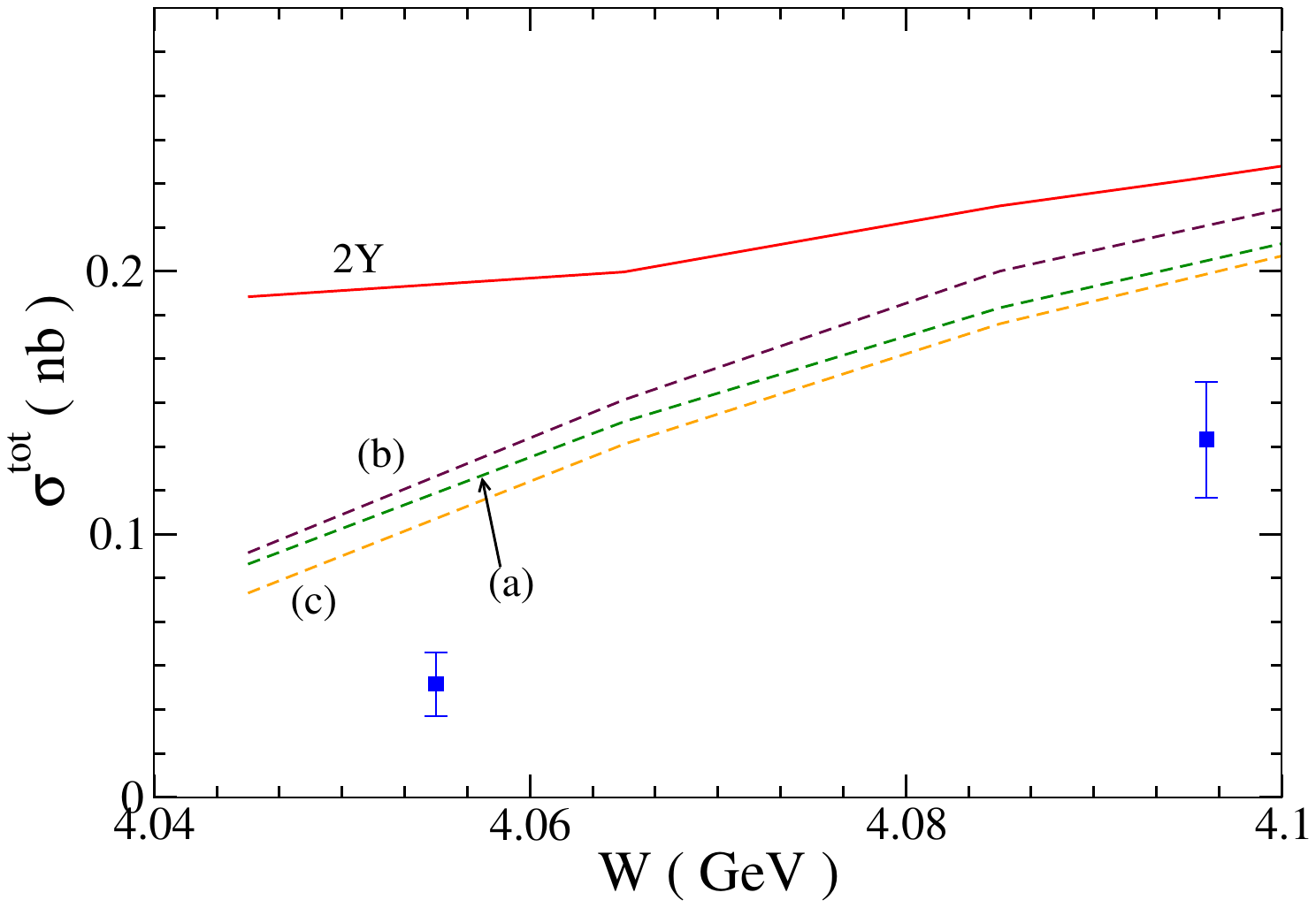}
\caption{ Total cross-sections  calculated 
  from the three models (a), (b), and (c)  constrained by the LQCD data. 
Top panel: for $4\leq W\leq 5$ GeV. 
Bottom panel: for the near threshold region. 
The same  data points are used as in Fig.~\ref{fig:totcrst-pom}.}
\label{fig:totcrst-lqcd-tsh-a}
\end{figure}

Using the model parameters listed in Table ~\ref{tab:tab-1}, we obtain 
the  total cross sections shown  in Fig.~\ref{fig:totcrst-lqcd-tsh-a}. 
They  are almost indistinguishable at high $W$, while their differences with the 2Y
model at near  threshold energy are large and comparable to that shown in 
Fig.~\ref{fig:totcrst-lqcd-tsh}. Thus, our predictions near threshold
are rather robust for future experimental tests.

\begin{figure*}[t]
\centering
\includegraphics[width=0.68\columnwidth,angle=0]{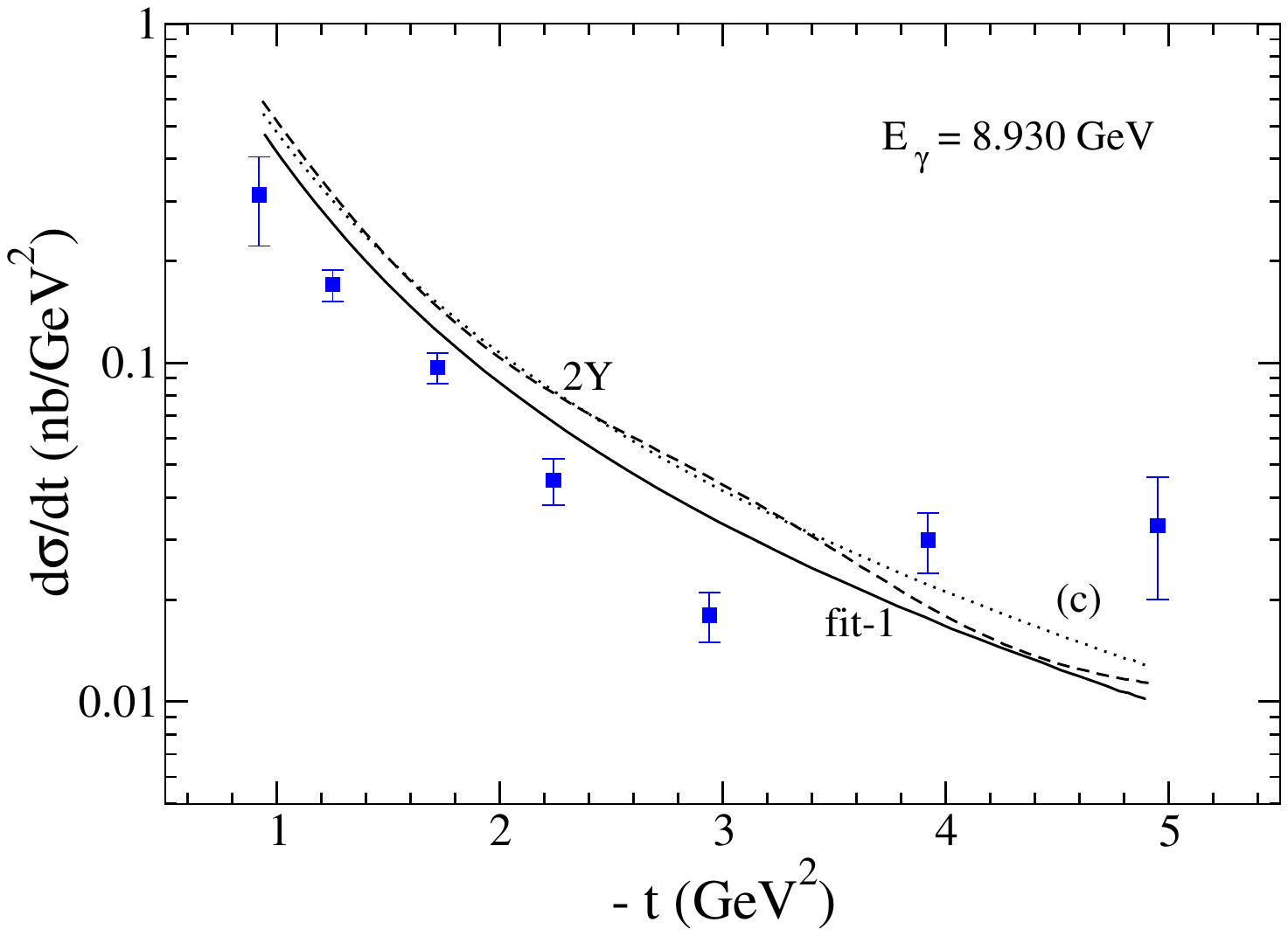}
\includegraphics[width=0.68\columnwidth,angle=0]{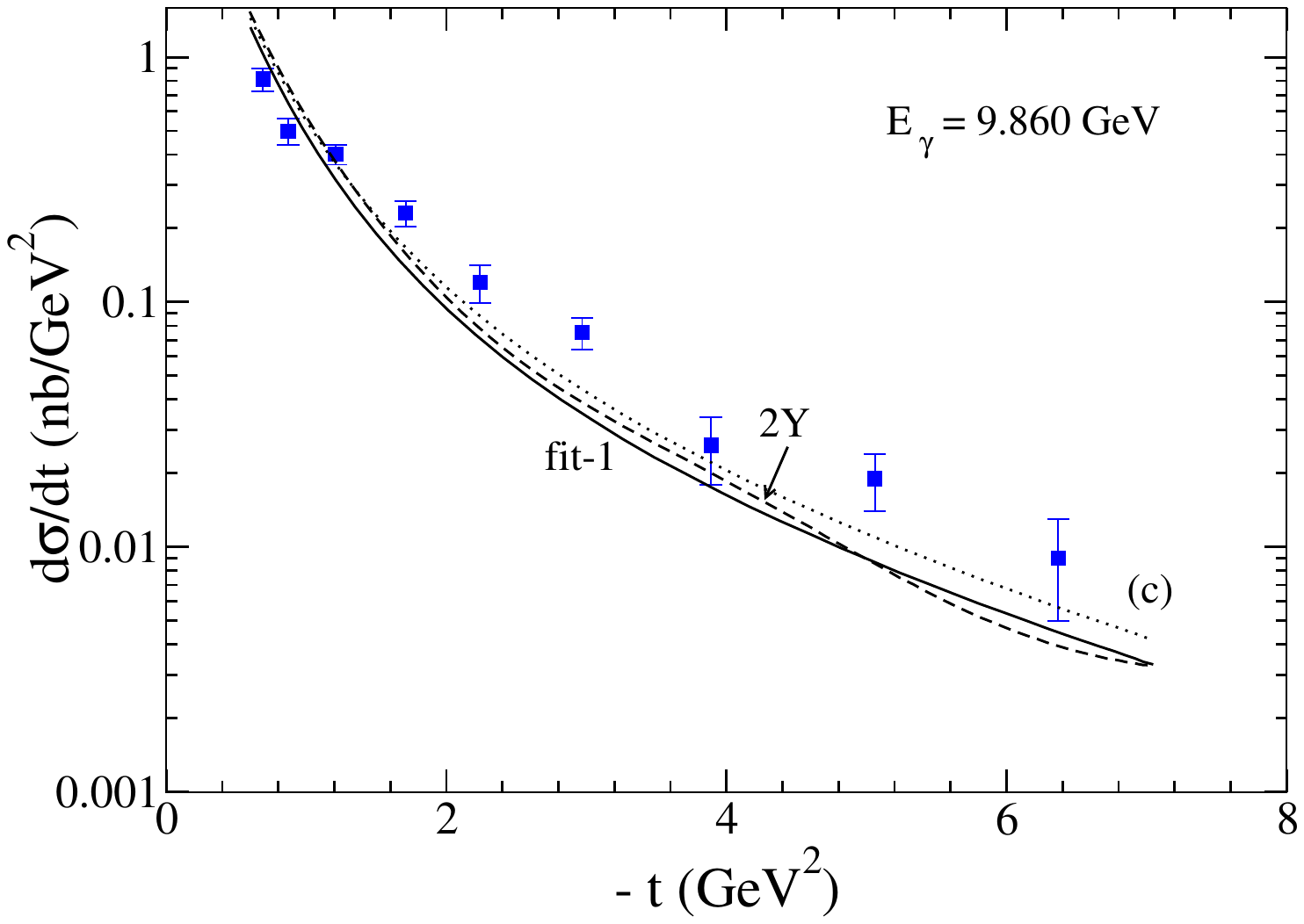}
\includegraphics[width=0.68\columnwidth,angle=0]{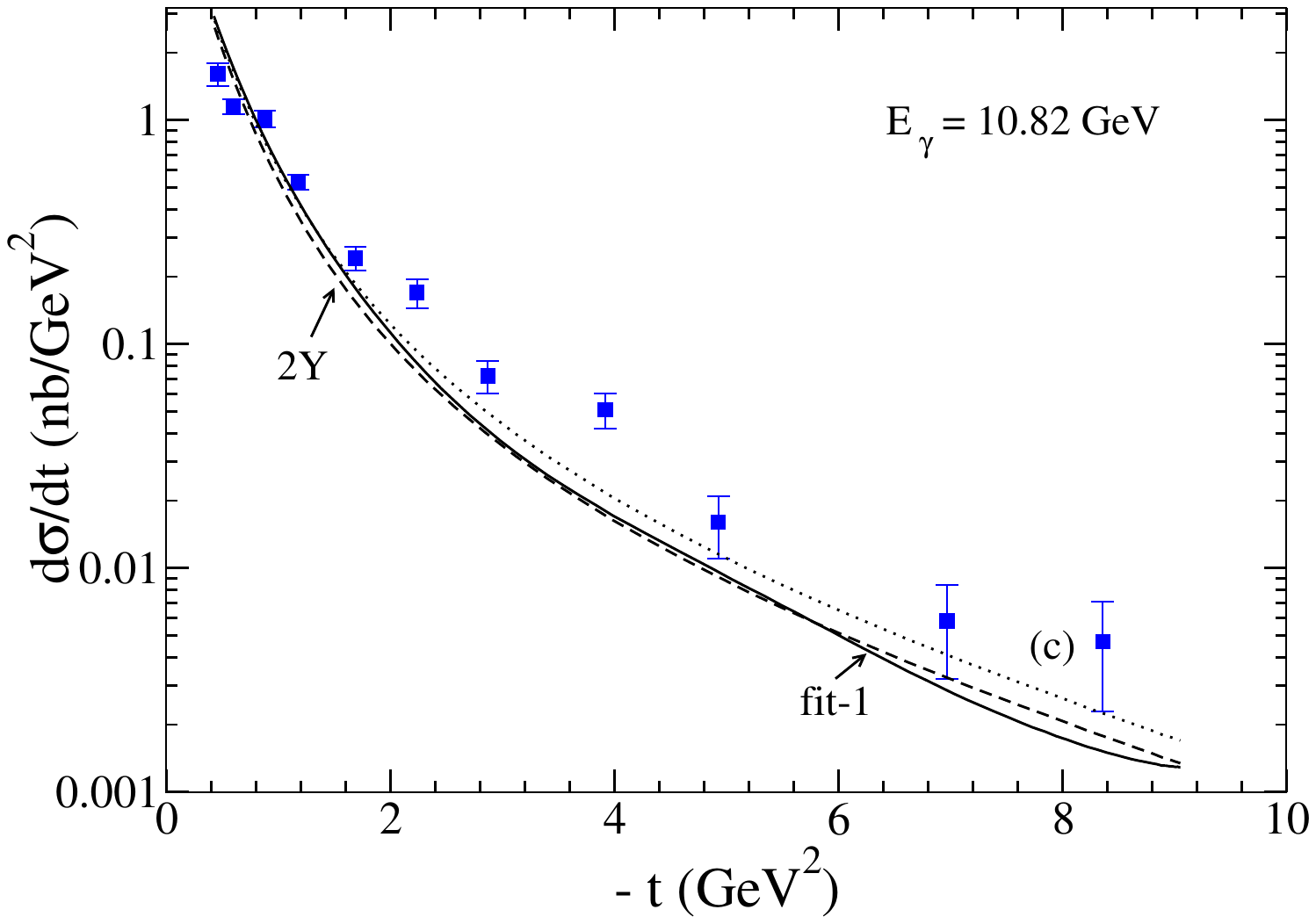}
\includegraphics[width=0.68\columnwidth,angle=0]{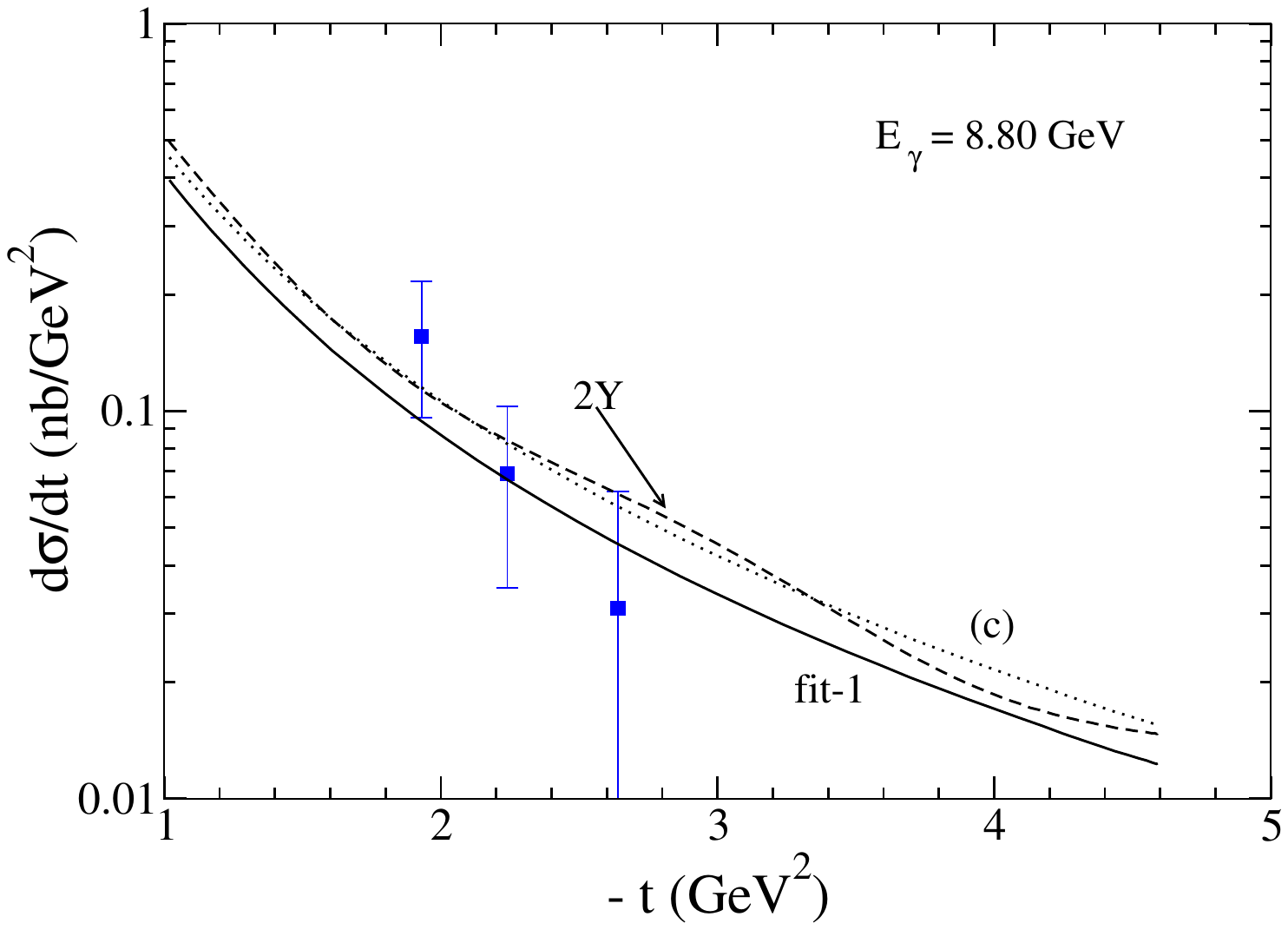}
\includegraphics[width=0.68\columnwidth,angle=0]{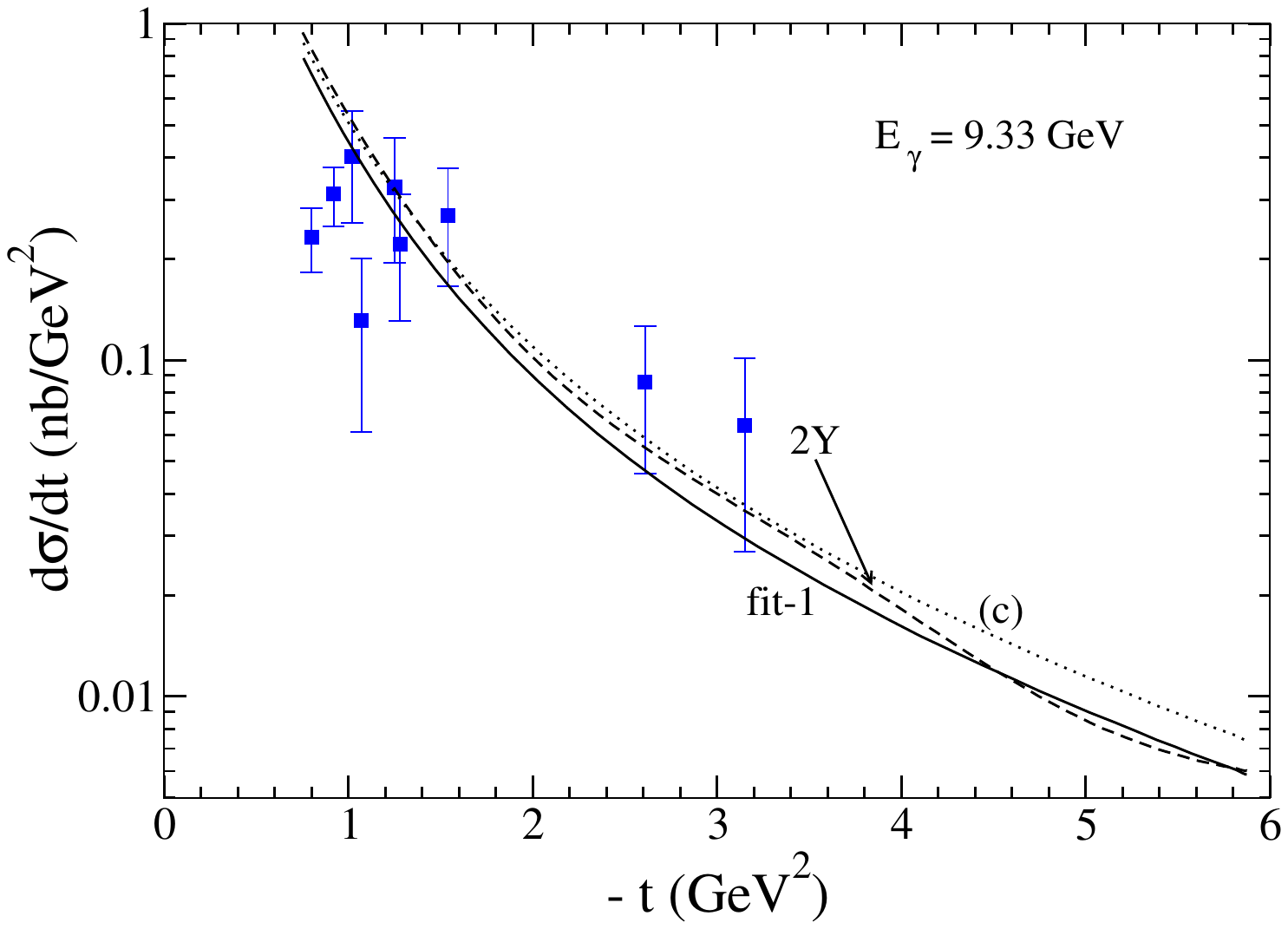}
\includegraphics[width=0.68\columnwidth,angle=0]{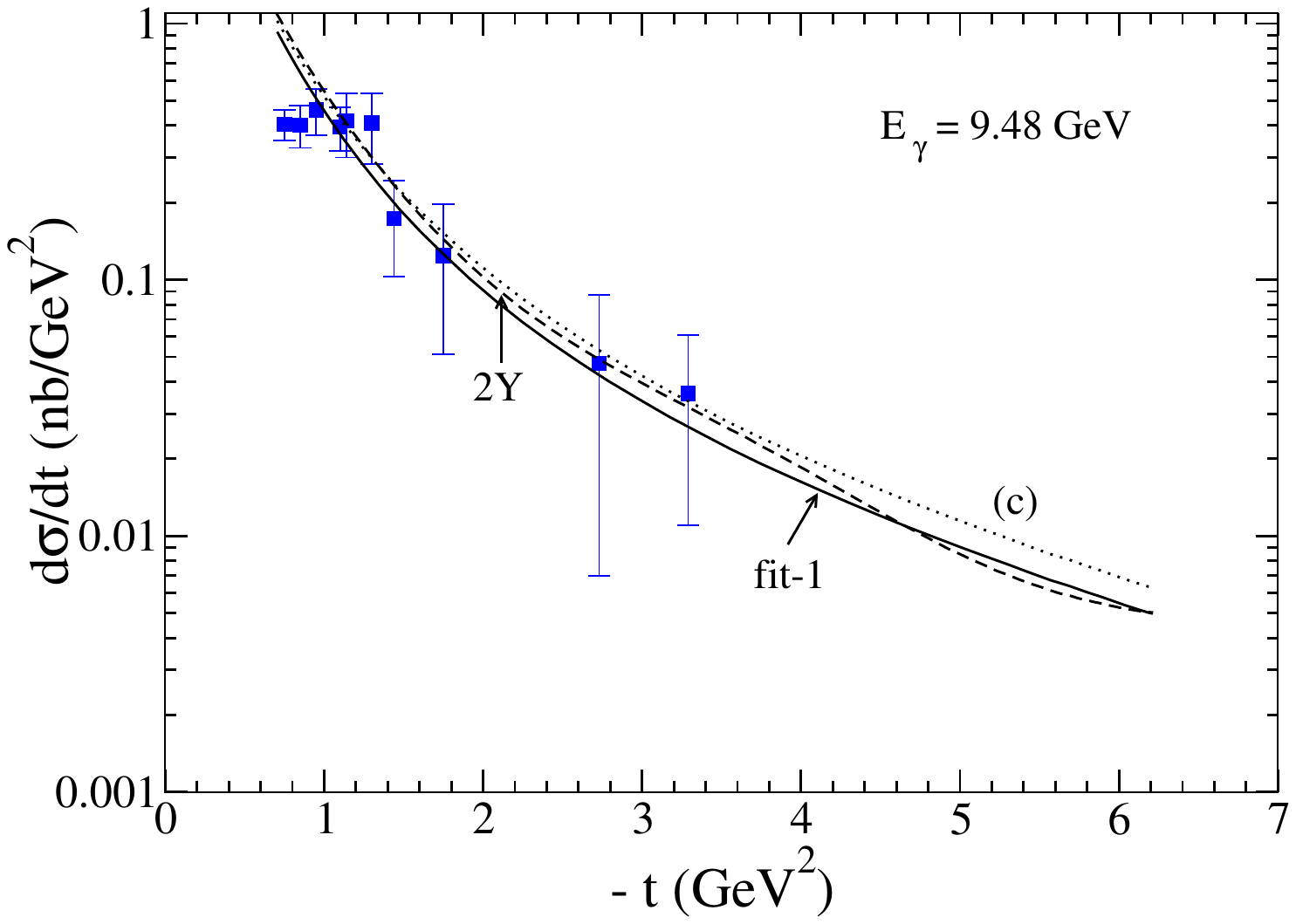}
\includegraphics[width=0.68\columnwidth,angle=0]{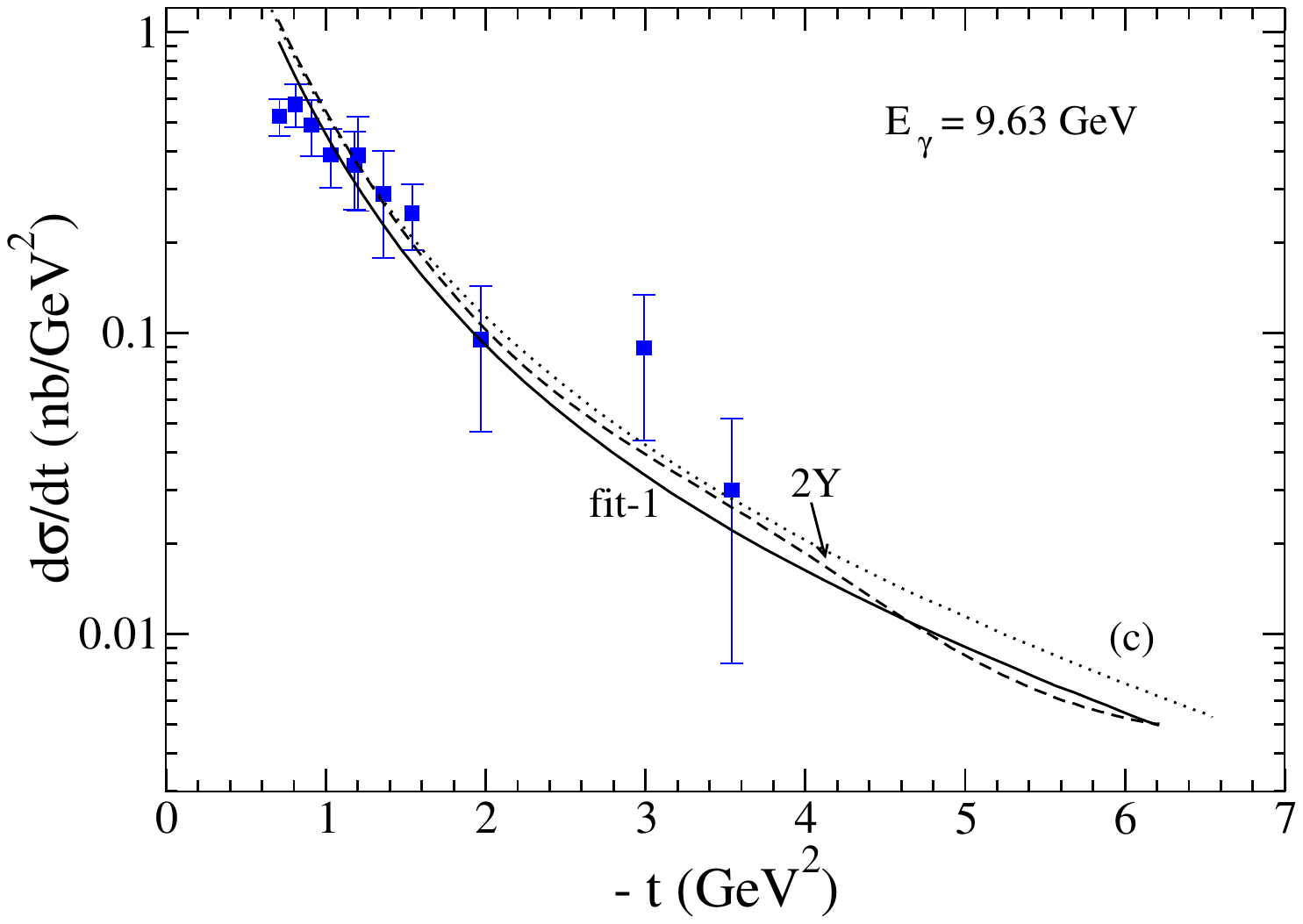}
\includegraphics[width=0.68\columnwidth,angle=0]{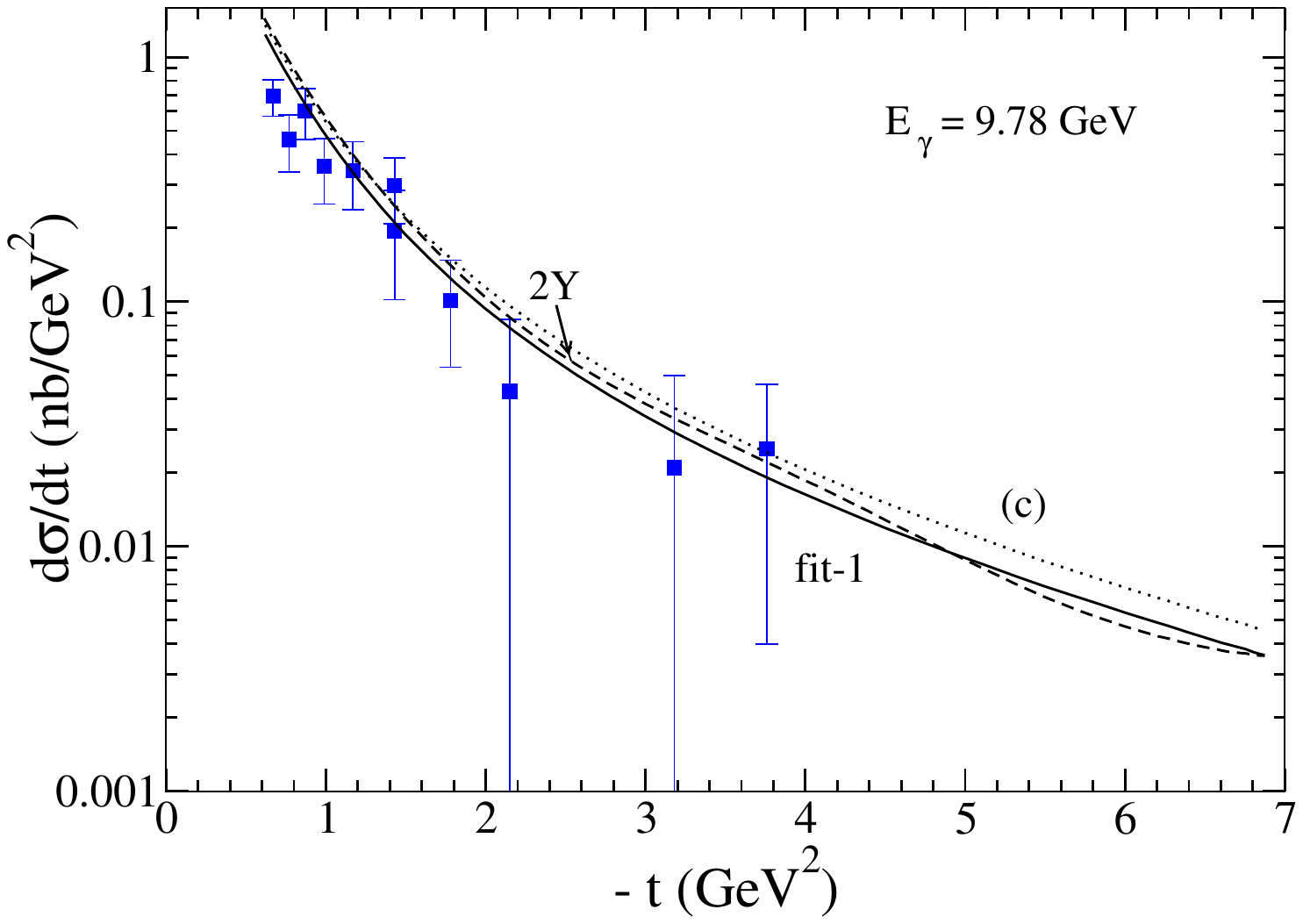}
\includegraphics[width=0.68\columnwidth,angle=0]{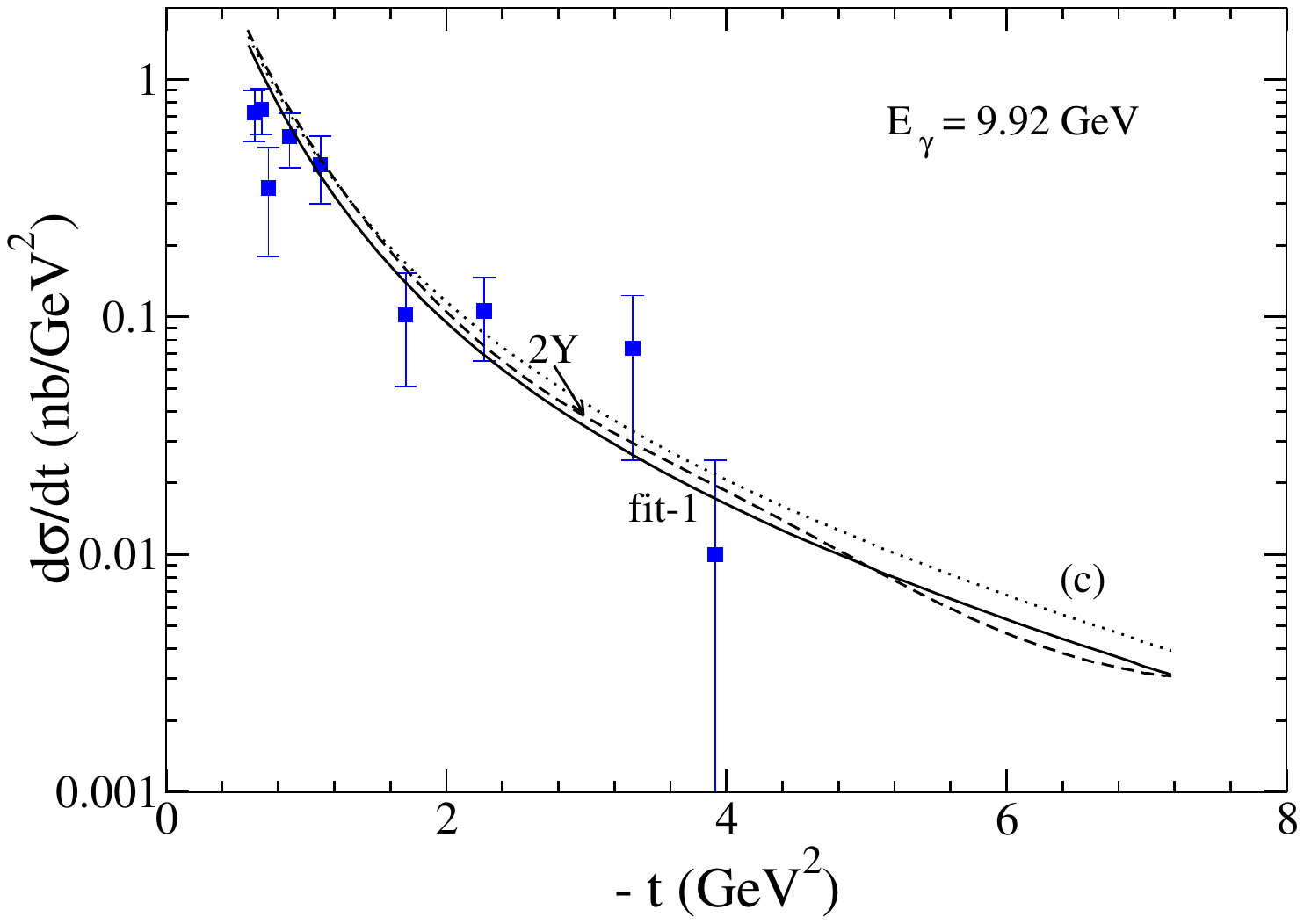}
\includegraphics[width=0.68\columnwidth,angle=0]{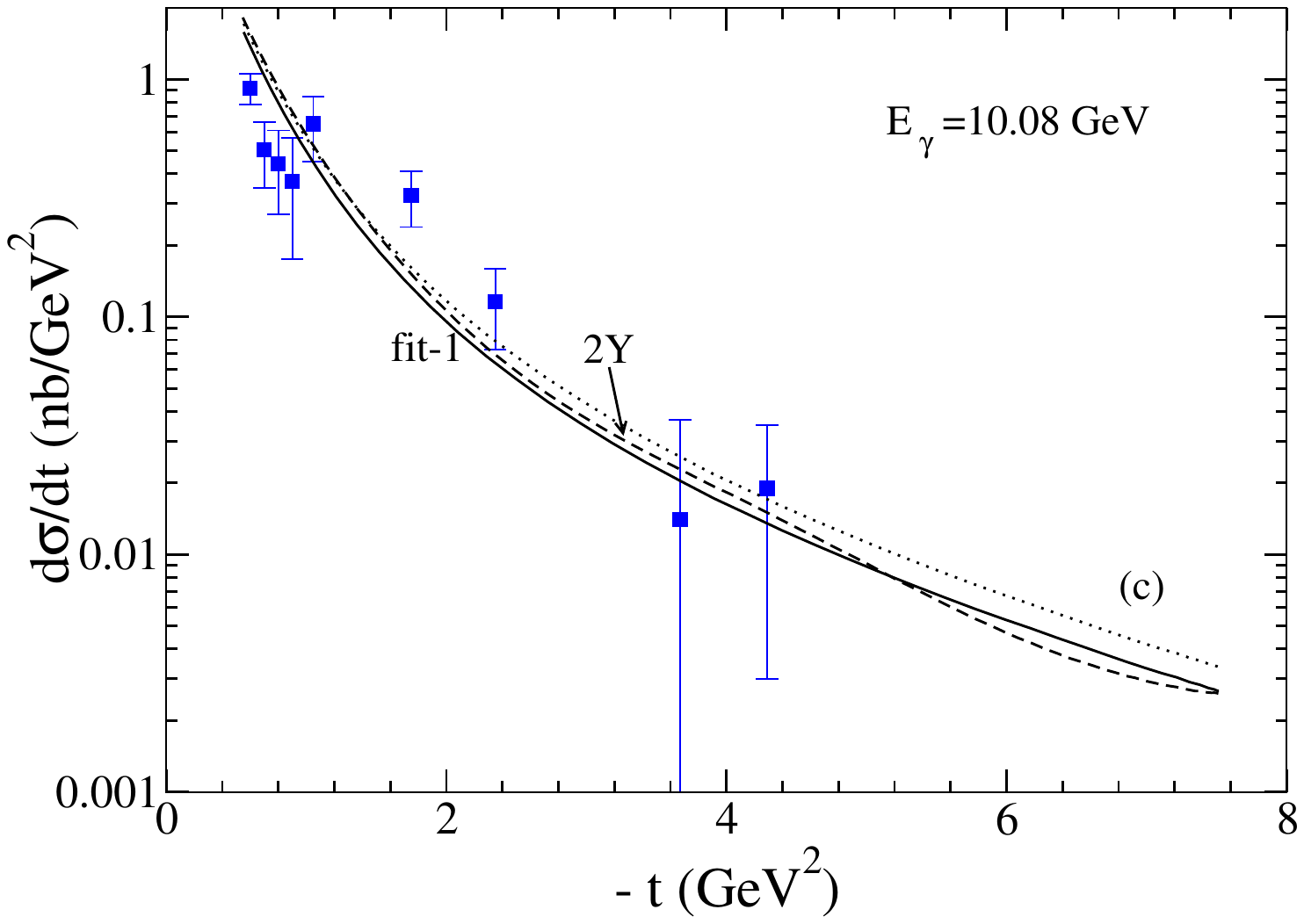}
\includegraphics[width=0.68\columnwidth,angle=0]{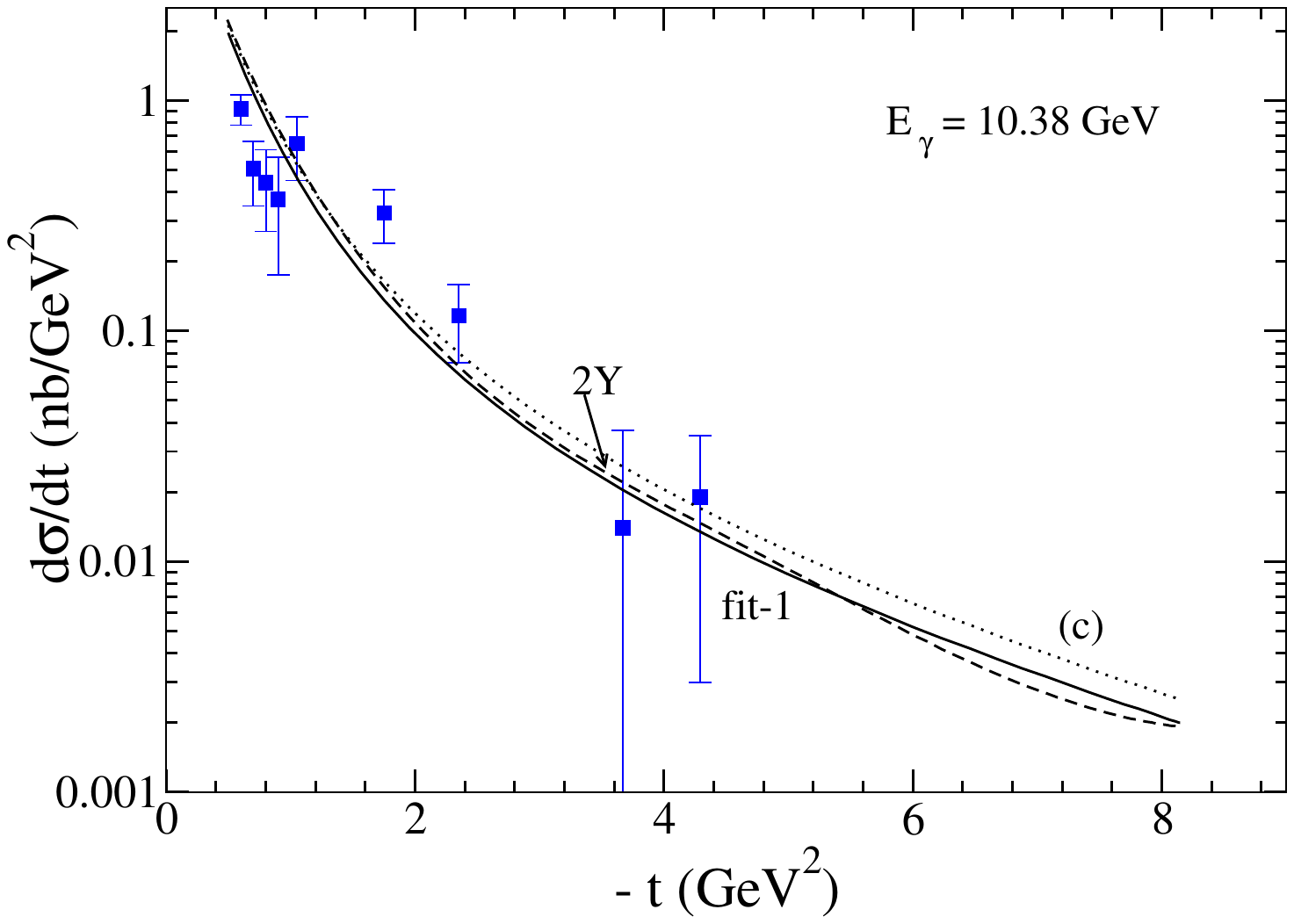}
\includegraphics[width=0.68\columnwidth,angle=0]{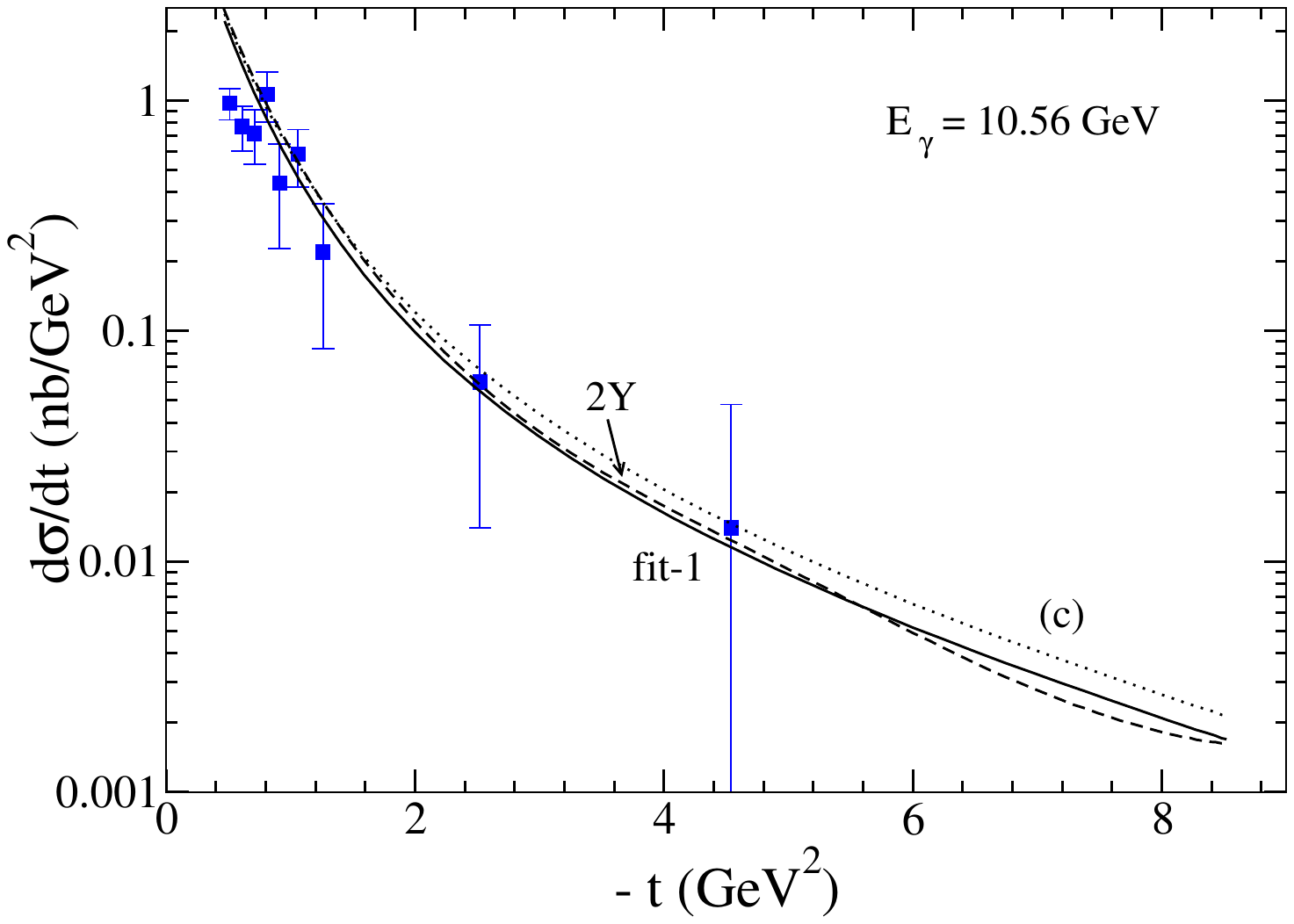}
\caption{Differential  cross sections from the models 2Y, fit-1 and (c).
The data in the top three panels are 
 taken from the  GlueX  Collaboration~\cite{GlueX-23}. The other data  are
taken  from an experiment at JLab's Hall-C~\cite{jlab-hallc}.}. \label{fig:dsdt-low}
\label{fig:dsdt-low-a}
\end{figure*}

\begin{figure}
\centering
\includegraphics[width=0.9\columnwidth,angle=0]{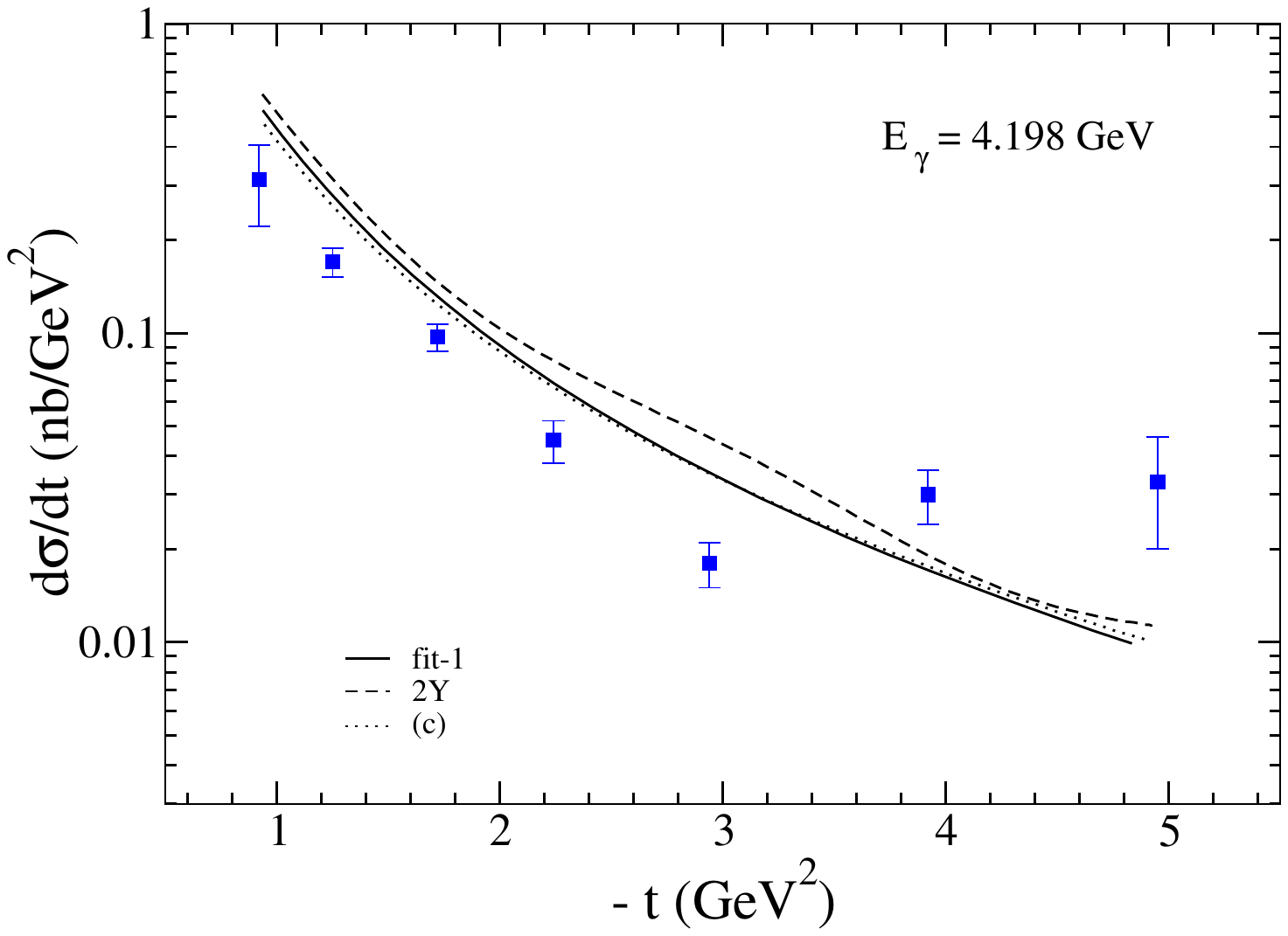}
\includegraphics[width=0.9\columnwidth,angle=0]{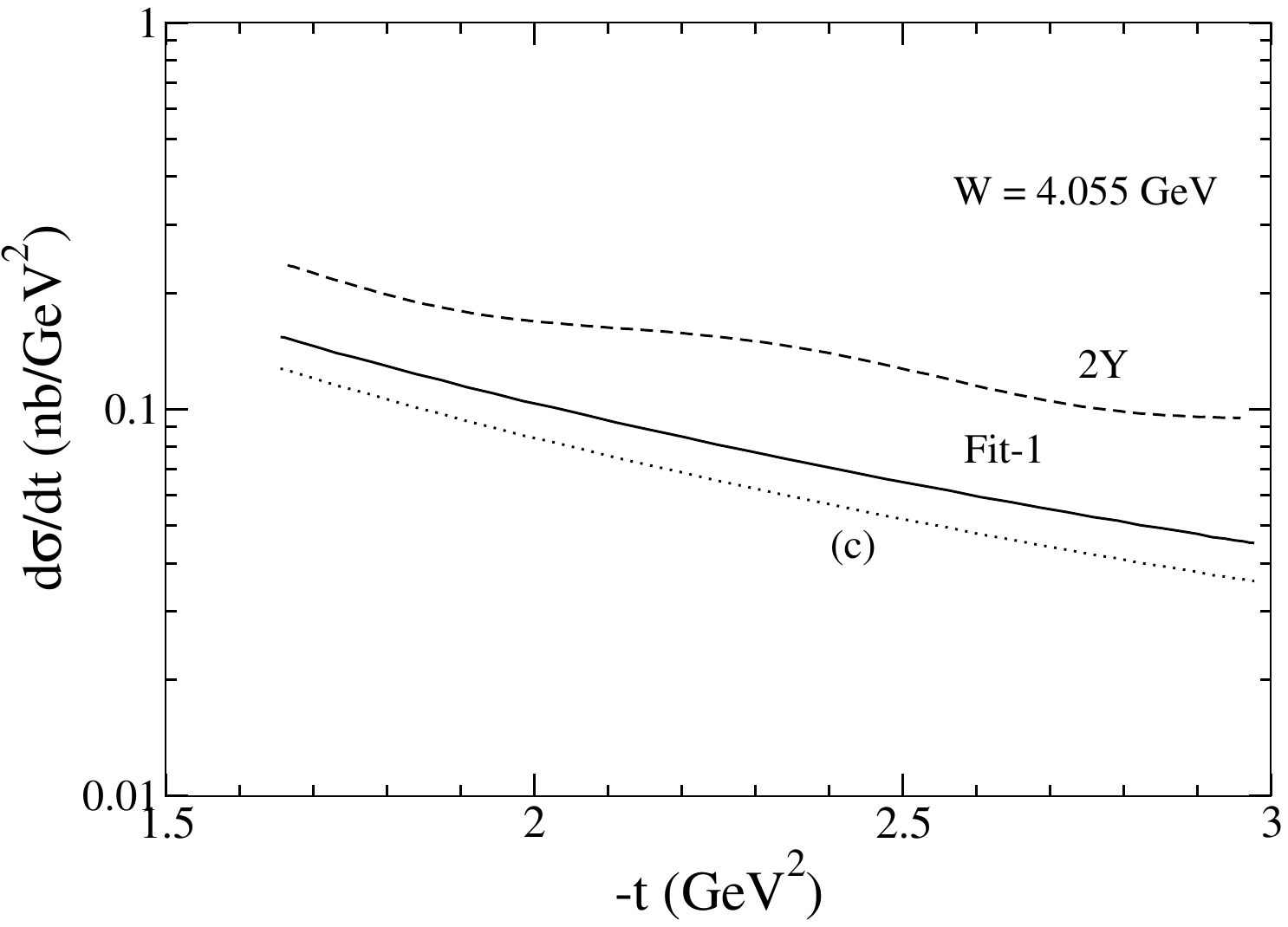}
\caption{Differential cross sections obtained from each  
model listed in Table~\ref{tab:tab-1}. 
The data are taken from~\cite{jlab-hallc}.}
\label{fig:dsdt-lqcd}
\end{figure}

\begin{table}[t]
	\caption{\label{tab:tab-1} {The parameters for the models  (a), (b), (c)  imposing LQCD constraints on FSI.}}
	\begin{ruledtabular}
	\begin{tabular}{lccccr}
	Model & $\alpha_{\rm FSI}$ & $\mu$ (GeV) & $\mu_1$ (GeV)& $a$(fm) &$\alpha_B$\\
	\hline
   (a) &  $-0.03$ & 0.3 &  $-$ & $-0.15$ & $-0.162$\\
 (b)& $-0.055$ & 0.5 & $-$ & $-0.233$ & $-0.152$\\
 (c) & $-0.1$ & 0.9 & 1.8 & $-0.057$ & $-0.163$\\
	\end{tabular}
	\end{ruledtabular}
\end{table}
We also find that  the  differential  cross sections from all models with
LQCD constraints, as given above, can describe the available data of
differential cross sections as well as the 2Y model.
This is illustrated
in Fig.~\ref{fig:dsdt-low-a} for the model fit-1 and (c).
For future measurements at energies very close to the threshold,  we
present predicted differential cross sections  in Fig.~\ref{fig:dsdt-lqcd}.

\begin{figure}[t]
\centering
\includegraphics[width=0.9\columnwidth,angle=0]{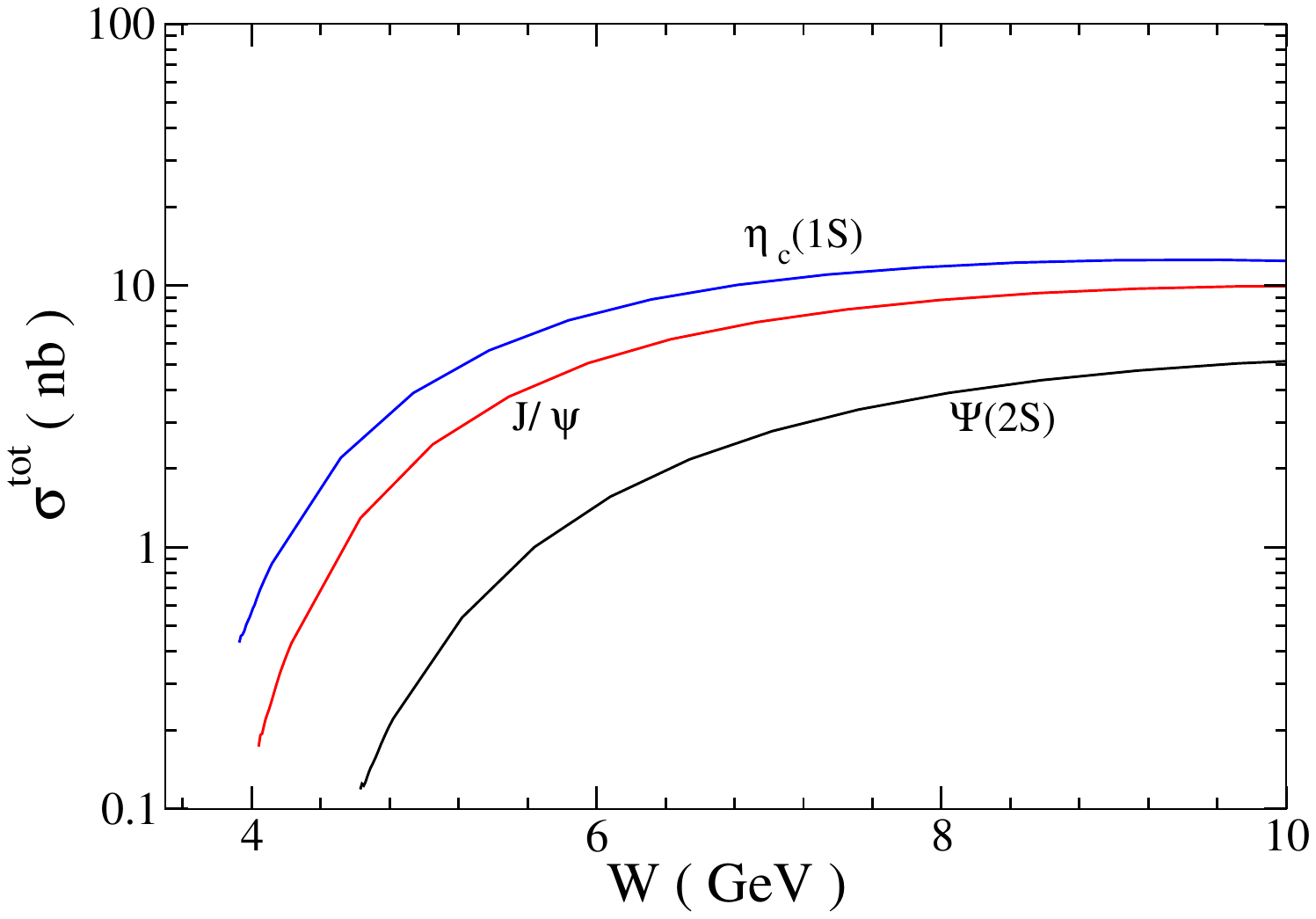}
\includegraphics[width=0.9\columnwidth,angle=0]{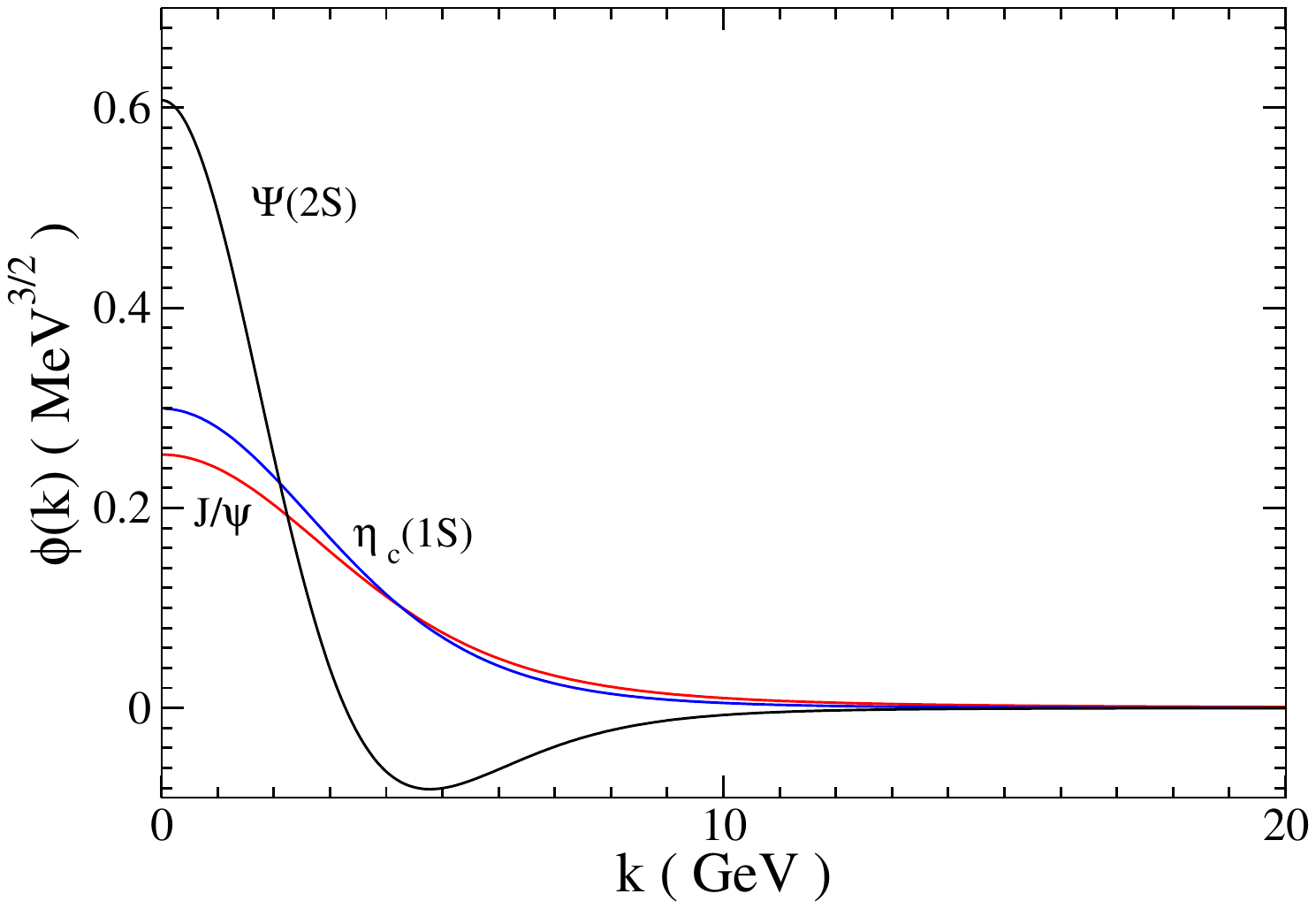}
\caption{Top pandel: predicted total cross sections  of
$\eta_c(1S)$ and $\psi(2S)$ photo-production,
Bottom panel: the wavefunctions of $J/\psi$, $\eta_c(1S)$ and $\psi(2S)$.
}
\label{fig:totcrst-pred}
\end{figure}
\begin{figure*}[t]
\centering
\includegraphics[width=0.8\columnwidth,angle=0]{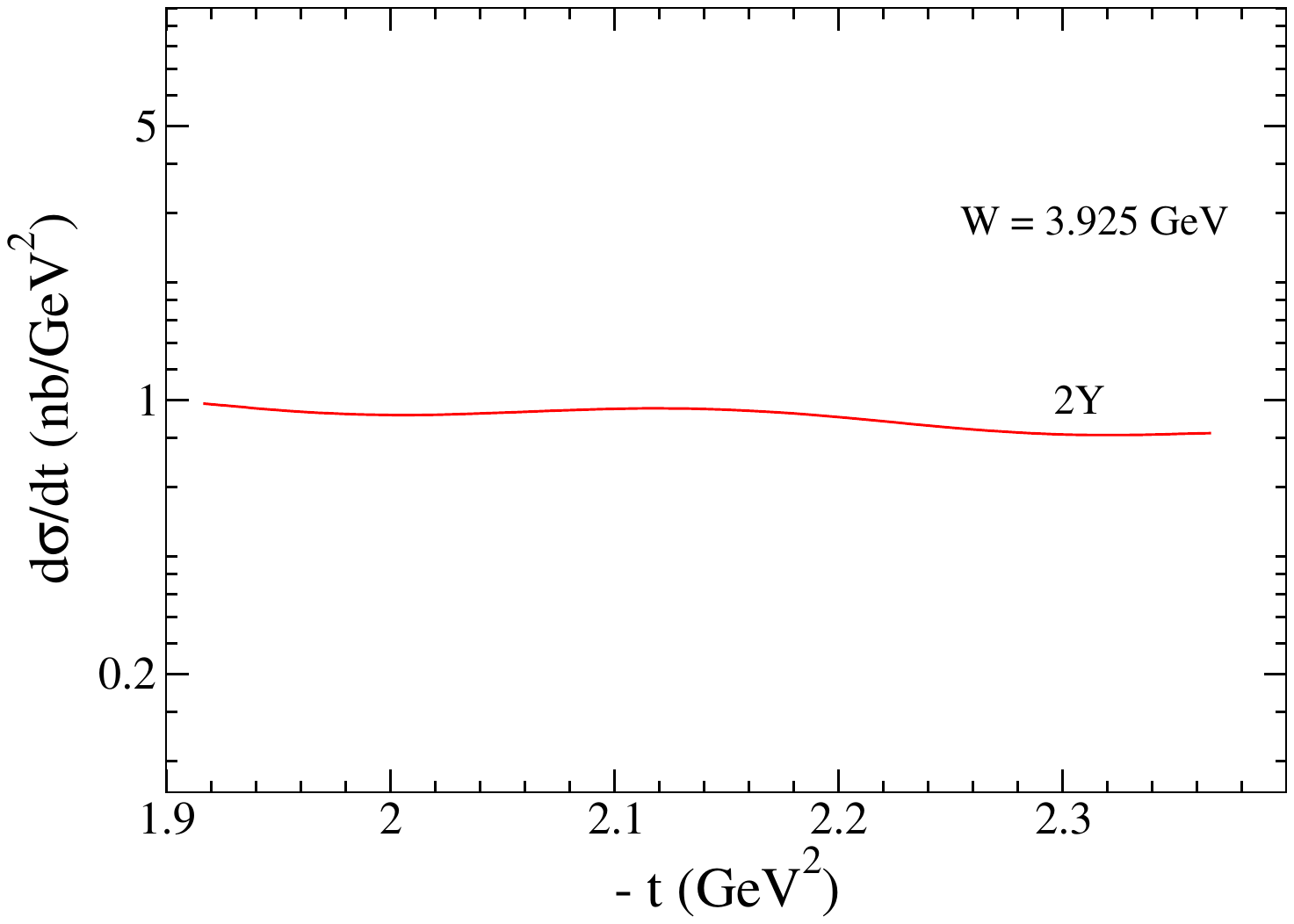}
\includegraphics[width=0.8\columnwidth,angle=0]{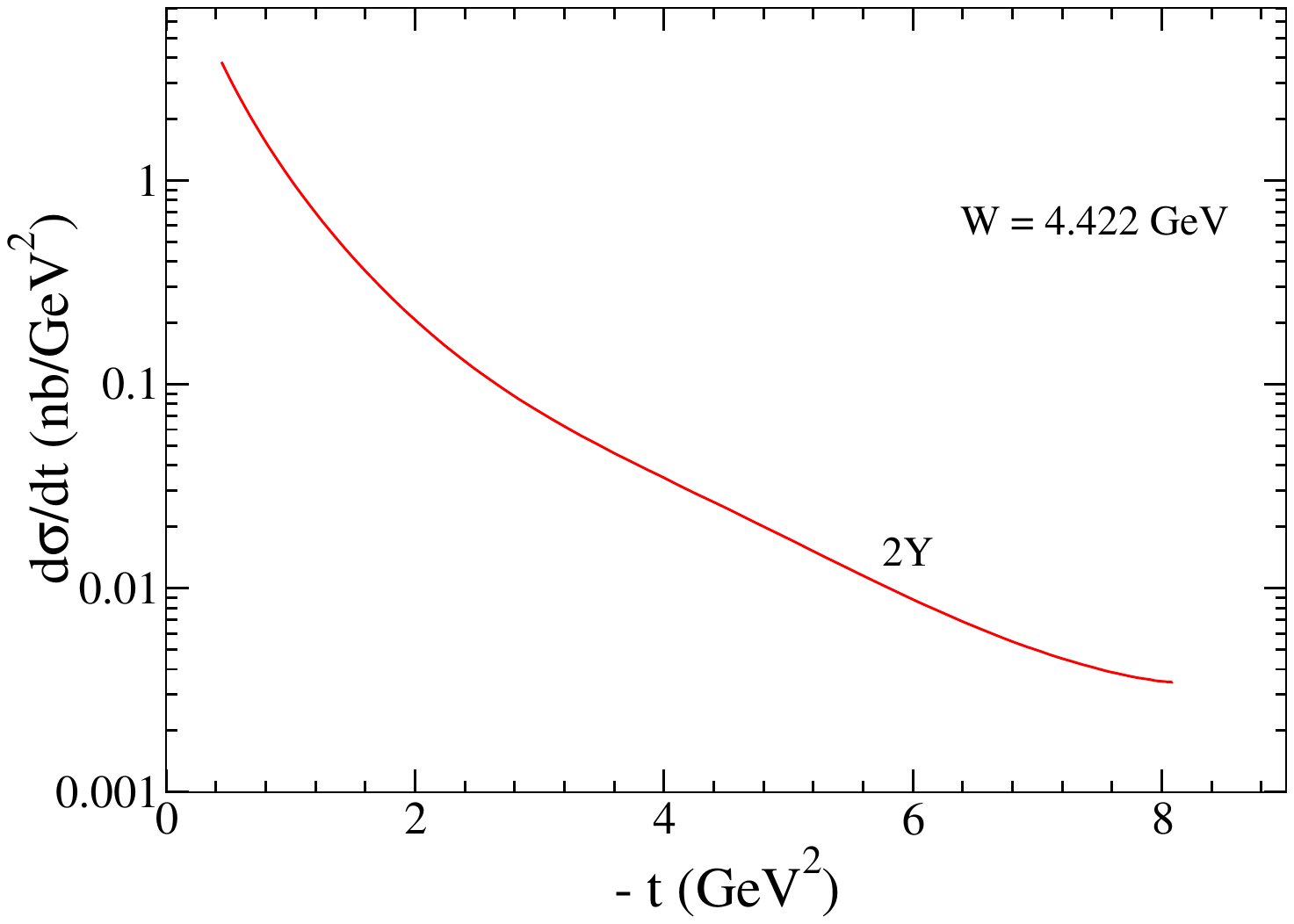}
\includegraphics[width=0.8\columnwidth,angle=0]{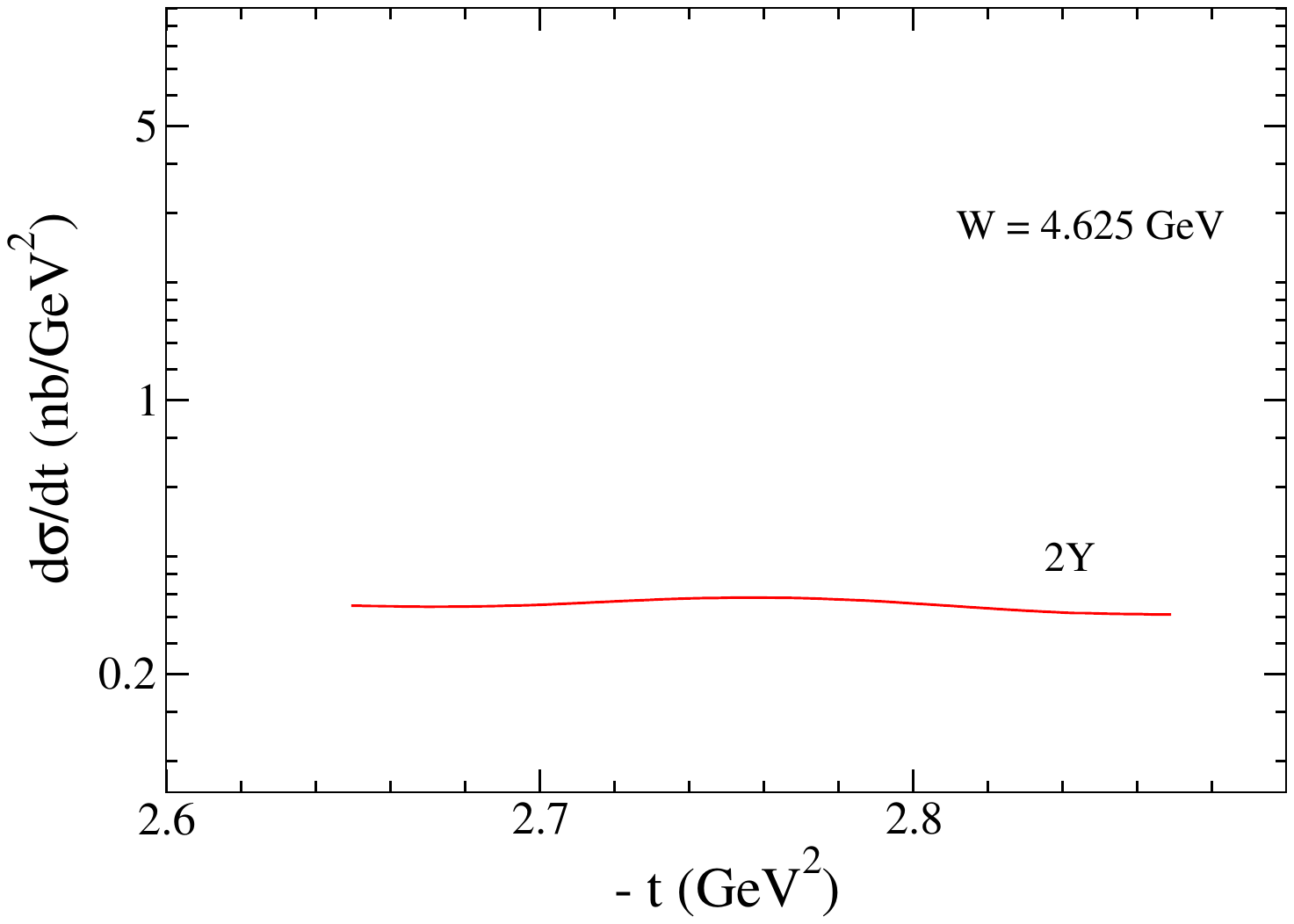}
\includegraphics[width=0.8\columnwidth,angle=0]{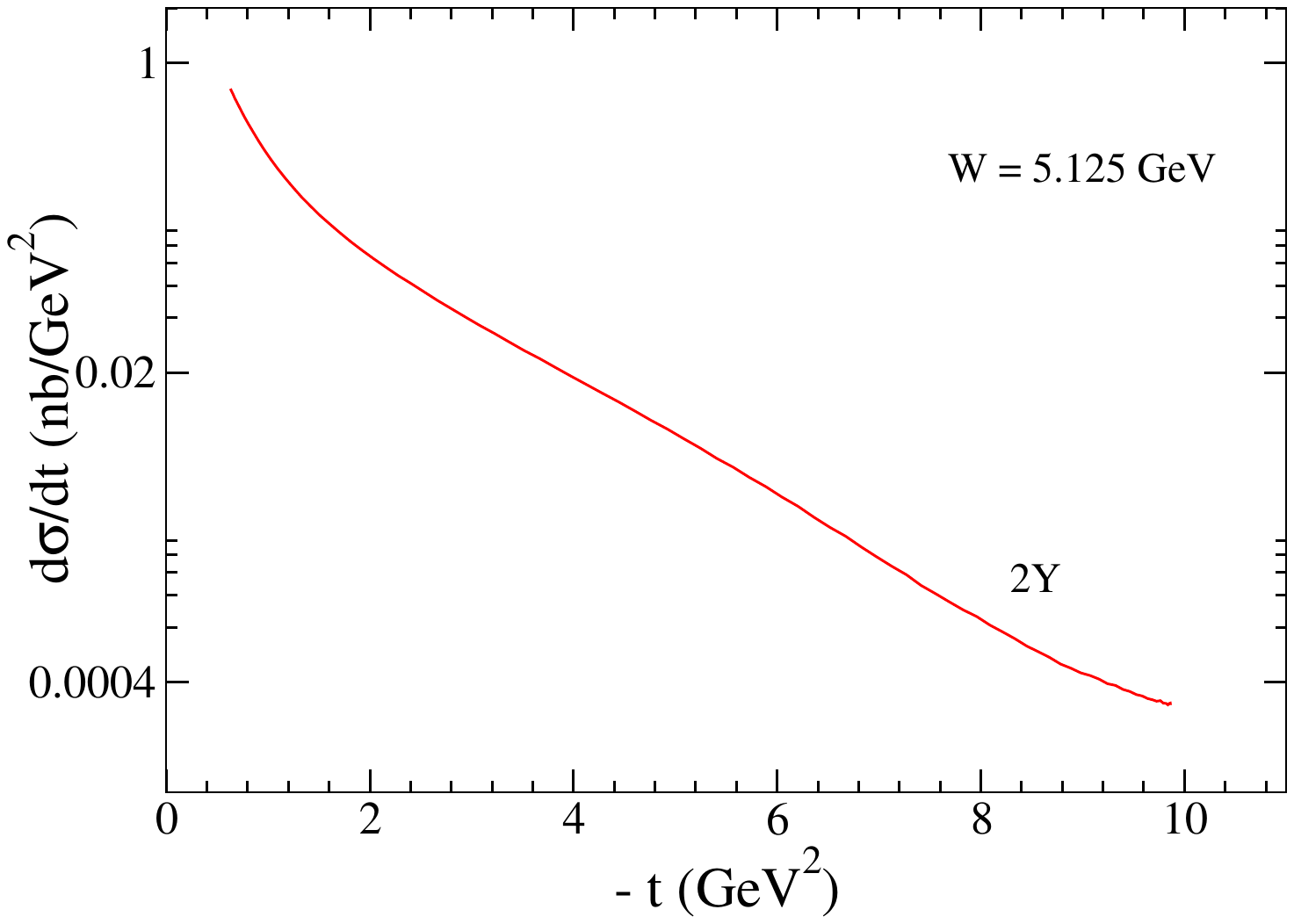}
\caption{Top two panels:
 predicted  differential cross sections of $eta_c(1S)$ photo-production 
at 20 MeV (left) and 500MeV (right) 
from threshold. Bottom two panels: predicted differential cross sections of $eta_c(1S)$
 at 20 MeV (left  two panels) and 500MeV (right  two  panels) from the  thresholds.
}
\label{fig:dsdt-etac}
\end{figure*}

\subsection{Photo-production of $\eta_c(1S)$ and $\psi(2S)$}
With the parameters of $v_{cN}$ determined from fitting the  
JLab data~\cite{GlueX-19,GlueX-23,jlab-hallc}, 
we proceed to predict the  photo-production of 
$\eta_c(1S)$ and $\psi(2s)$ mesons.
Because $v_{cN}$ is   spin independent  and the  wavefunctions
for $\eta_c(1S)$ and $\psi(2S)$  are also of the $s$-wave form  given by
Eq.~(\ref{eq:phi-v}), the calculations
can be done by using the same formula presented in Section  III by
simply  changing the  wavefunctions and the kinematics associated with the masses of mesons.
By  using the  parameters  of the  2Y model presented in the previous
section, the predicted total cross  sections for the
photo-productions of $\eta_c(1S)$, $\psi(2s)$, and $J/\psi$ are  presented in the top panel of Fig.~\ref{fig:totcrst-pred},
along with the comparison of their wavefunctions in the bottom panel. 
In Fig.~\ref{fig:dsdt-etac}, we present the predicted
differential cross sections for $\eta_c(1S)$  (upper  panels) 
and $\psi(2S)$ ( lower panels) at energies 20 MeV (left)  and  500  MeV (right)
above the thresholds. 

As a validation test for our model, it would be intriguing to compare our predictions with forthcoming data expected from experiments at JLab 
and the future EIC.

\section{Summary and Future improvements}
Within a Hamiltonian  formulation~\cite{SL96,MSL06,JLMS07,KNLS13}, 
a dynamical model based on the Constituent Quark Model (CQM)
 and a phenomenological charm quark-nucleon 
potential $v_{cN}(r)$ is constructed to investigate the
 $J/\psi$ photo-production  on the nucleon at energies near threshold.
The main feature of the model 
lies in the quark-N  potential, $v_{cN}$, which
generates a photo-production amplitude defined by
 a  $c\bar{c}$-loop integration over the $\gamma  \rightarrow c\bar{c}$
vertex function and the  $J/\psi$ wavefunction, $\phi_{J/\psi}(c\bar{c})$, from the CQM described in Ref.~\cite{SEFH13}.
In addition, the $J/\psi$-N final state  interaction is
calculated  from a  $J/\psi$-nucleon potential $V_{J/\psi N}(r)$,
which  is constructed  by folding $v_{cN}(r)$ into 
the same wavefunction $\phi_{J/\psi}(c\bar{c})$.
By also incorporating the Pomeron-exchange amplitudes determined in Refs.~\cite{WL13,lso23},
the constructed model can describe the available data from threshold
to  high  energies, up to the invariant mass $W=300$ GeV.

The parametrization   of  $v_{cN}(r)$ is chosen such that
 the constructed $V_{J/\psi N}(r)$ at large distances has the same
Yukawa potential form extracted from a LQCD calculation.
The parameters of  $v_{cN}$  are determined by fitting  the
total cross section data  of JLab~\cite{GlueX-19,GlueX-23} by performing
 calculations  that include $J/\psi$-N  final state interactions,
 as required by the unitarity condition.
The predicted differential cross sections $d\sigma/dt$ are in reasonably
good agreement with  the data from JLab~\cite{jlab-hallc,GlueX-23}.
Furthermore, it is shown that the FSI effects dominate the cross section
in the very near-threshold region.
This indicates that the low energy $J/\psi$-N scattering amplitudes and
the associated models of  $J/\psi$-N  potentials $V_{J/\psi N}(r)$  
can be extracted from the $J/\psi$ photo-production  data with high sensitivity.
The determined $V_{J/\psi N}(r)$ can be  used  to
 understand the nucleon
resonances $N^*(P_c)$  reported by the LHCb
collaboration~\cite{LHCb-15,LHCb-16a,LHCb-19,LHCb-21a},
to extract the gluonic distributions in nuclei, $J/\psi$ production in relativistic
heavy-ion collisions, and to study the
existence of nuclei with hidden charms~\cite{BD88a,BSD90,GLM00,BSFS06,WL12}.

By imposing the constraints of $J/\psi$-N potential 
extracted from the LQCD  calculation of Refs.~\cite{KS10b,KS11,sasaki-1},
 we have obtained  three 
$J/\psi$-N potentials which fit the JLab  data  equally well.
The resulting  $J/\psi$-N scattering lengths
are in the range of $a=(-0.05$ fm $\sim$ $-0.25$ fm), which rule out the existence
of  free $J/\psi$-N bound states. On the other hand, the available
data near threshold is far from sufficient to draw a conclusion.
Clearly, more extensive and   precise experimental
data are needed to make further progress.

With the determined quark-nucleon potential $v_{cN}(r)$ and the wavefunctions
generated from the same CQM of  Ref.~\cite{SEFH13}, the
constructed dynamical model has been used to  predict
 the cross  sections  of photo-production of 
$\eta_c(1S)$  and $\psi(2S)$ mesons. 
It will be interesting  to have  data from 
 experiments at JLab  and  EIC to test our predictions.

The model  presented here needs to be improved  in the future.
Our parametrization of quark-nucleon potential $v_{cN} $ is guided by the Yukawa  form extracted
from a LQCD calculation of Ref.~\cite{KS10b}. This must be improved by using more advanced LQCD calculations of
$J/\psi$-N scattering, in particular the  short-range part of the  potential.
A recent LQCD calculation of $\phi$-N interaction leads to a more complex form than the
simple Yukawa form.
In  addition, the quark sub-structure of the nucleon  
must be considered  in the  future  improvements of our
model as well as most, if not all, of the
previous models (as reviewed in Ref.~\cite{lso23}).
However, this is  an  non-trival many-body problem similar
to that encountered in the investigations of  nuclear reactions~\cite{feshbach}.

The results from the model based on the effective Lagrangian approach~\cite{du}
suggest that we need to extend our  model to include
the coupled-channel effects via the $\bar{D}*\Lambda_c$ channel in order to explain the cusp structure
of the total cross  section  data
near $W\sim 4.2-4.3$ GeV. However, this improvement must also consider other coupled-channel
effects arising from the coupling with $\pi N$ and $\rho N$ channels, as investigated  in Ref.~\cite{WL13}.

It is necessary to improve our model by developing $c\bar{c}$-loop calculations
 of Pomeron-exchange mechanism,  which are needed to fit the high  energy data.
 This requires an extension of our approach to include
relativistic effects.  This can be done 
straightforwardly 
within the Instant Form of relativistic Quantum
Mechanics of Dirac~\cite{KP91}, 
while the Light-front form is also possible with much more works.

\section*{Acknowledgement}
We would like to thank J. Segovia for providing us with 
the CQM wavefunctions used in this work.
The works of S.S and H.-M.C. were supported by the National Research Foundation of Korea (NRF) under Grant No. NRF- 2023R1A2C1004098.
The work of T.-S.H.L. was supported  by the U.S. Department  of Energy, 
Office of Science, Office of Nuclear Physics,  under Contract No.AC02-06CH11357.

\end{document}